\documentclass[12pt,oneside,reqno]{amsart}
\usepackage[T1]{fontenc}
\usepackage[utf8]{inputenc}
\usepackage[a4paper]{geometry}
\usepackage{float}
\usepackage{amstext}
\usepackage{amsthm}
\usepackage{amssymb}
\usepackage{setspace}
\usepackage{graphicx}
\usepackage{booktabs} 
\usepackage{tabu}
\usepackage[authoryear]{natbib}
\usepackage{color}
\usepackage{comment}
\usepackage{enumerate}
\usepackage{url}
\usepackage{amsaddr}
\usepackage{comment}
\usepackage{multirow}
\usepackage{subcaption}
\makeatletter

\numberwithin{equation}{section}
\numberwithin{figure}{section}

\usepackage{amsfonts}
\usepackage{amstext}
\usepackage{amsthm}
\usepackage{hyperref}

\newtheorem{thm}{Theorem}
\newtheorem{lem}{Lemma}

\newtheorem{prop}{Proposition}
\newtheorem{asm}{Assumption}

\theoremstyle{definition}

\makeatother

\begin{document}
\title{Dyadic Regression with Sample Selection}
\author{Kensuke Sakamoto}
\date{\today}
\address{University of Wisconsin-Madison}
\email{kenlfo2080@wisc.edu}
\thanks{The author thanks Jack Porter, Bruce Hansen, Harold Chiang for support and helpful comments from Xiaoxia Shi, Harold Chiang, Kohei Yata, and participants in New York Camp Econometrics XVII. This paper was supported by the Summer Fellowship from the Department of Economics at University of Wisconsin-Madison.}

\begin{abstract}
	This paper addresses the sample selection problem in panel dyadic regression analysis. Dyadic data often include many zeros in the main outcomes due to the underlying network formation process. This not only contaminates popular estimators used in practice but also complicates the inference due to the dyadic dependence structure.  
	We extend \cite{Kyriazidou1997}'s approach to dyadic data and characterize the asymptotic distribution of our proposed estimator. The convergence rates are $\sqrt{n}$ or $\sqrt{n^{2}h_{n}}$, depending on the degeneracy of the H\'{a}jek projection part of the estimator, where $n$ is the number of nodes and $h_{n}$ is a bandwidth. We propose a bias-corrected confidence interval and a variance estimator that adapts to the degeneracy. A Monte Carlo simulation shows the good finite-sample performance of our estimator and highlights the importance of bias correction in both asymptotic regimes when the fraction of zeros in outcomes varies. We illustrate our procedure using data from the paper by \cite{Moretti2017} on migration.
\end{abstract}

\maketitle
\noindent \textbf{\small Keywords:} \small Dyadic Data, Sample Selection, Fixed Effects, Network Formation, Bias Correction.
\newpage
\section{Introduction}
\noindent Dyadic data describe pairwise outcomes, such as trade volume between countries. Numerous applications have analyzed such data using the regression model, referred to as dyadic regression. Examples include gravity equations in trade, migration, and urban economics \citep{Helpman2008, Moretti2017, Monte2018}, and risk-sharing networks in development economics \citep{Fafchamps2007}. One of the prominent features of dyadic data is the non-negligible number of zeros in the outcomes of interest, \footnote{\cite{Helpman2008} document that there was no trade among roughly 50\% of country pairs
from 1970 to 1997. In 2017, there was no migration among about 60\% of country pairs (the author calculated using the data available from the World Bank (\url{https://www.worldbank.org/en/topic/
migrationremittancesdiasporaissues/brief/migration-remittances-data}).} possibly due to economic mechanisms such as prohibitive fixed costs. This paper deals with panel dyadic data, where zeros are prevalent both across cross-sections and over time.

How should we treat zeros in dyadic regression? In applications, zeros are often discarded due to the log-linear specification \citep{Moretti2017}. The Poisson pseudo-maximum-likelihood (PPML) estimator is also frequently used to avoid discarding zeros and address issues related to log-linearization \citep{SantosSilva2006}. These approaches implicitly assume that zeros occur exogenously. Since a zero in a pairwise outcome results from no link between two units, we can associate zeros with the underlying network formation mechanism that determines which pairs appear in a sample. If the network is formed endogenously as a result of an interaction between two agents, the empirical practices mentioned above can be subject to sample selection bias, as in \cite{J.Heckman1979}.

This paper has two primary objectives. First, we aim to jointly model network formation and the outcome generation on such networks.  This joint modeling allows identification of the effects of changes in pair-level or individual-level characteristics, separating them from the effects caused by changes in networks. In contrast, the dyadic regression literature has primarily focused on regression with fixed or exogenous networks. Second, we develop a robust inference method that accounts for the dyadic dependence structure. Pairwise outcomes are likely to be dependent on each other through common shocks to individuals. This dyadic dependence can be especially important in the presence of zeros and the network formation because a few individuals can have significantly more links than others, \footnote{For example, in \cite{Moretti2017}'s migration flow data, star scientists' migration from or to California constituted approximately 14\% of the links in the sample on average. This percentage is much higher than the expected 2\% when considering all potential links in the sample.} which strengthens the influence of shocks to those individuals on the dyadic dependence. At the same time, it is known that with dyadic data, we can have different asymptotic regimes depending on the nature of those individual-level shocks \citep{Menzel2021}. To be practitioner-friendly, our inference method needs to consider the dyadic dependence and ensure adaptivity to different resulting asymptotic regimes.

Our setup will be a linear panel dyadic regression model, featuring the network formation process as a sample selection mechanism that generates both zeros and unobservable outcomes. To capture the dyadic dependence structure, we incorporate two types of unobservable individual heterogeneity into the model: time-invariant fixed effects and time-varying random effects, which is a new modeling strategy in the literature. We extend \cite{Kyriazidou1997}'s identification argument, originally designed for individualistic data, to dyadic data, and correspondingly propose a semiparametric, kernel-based estimator that assigns weights to pairs whose selection index remains stable over time. A significant challenge we face when analyzing our estimator is the need to address the
dependence structure caused by node-level shocks, which is absent in individualistic data models analyzed in \cite{Kyriazidou1997}. To control for this type of dependence, we utilize the U-statistic-like structure of our estimator, which gives us a mutually uncorrelated decomposition into the node-level Hájek projection part and the dyad-level projection error part.

We show that our estimator is asymptotically normal with two different convergence rates depending on the nature of errors. If the H\'{a}jek projection is non-degenerate (i.e., each summand has positive variance), our estimator achieves $\sqrt{n}$-asymptotic normality, where $n$ is the number of nodes. In this case, we not only have zero asymptotic bias but also share the same convergence rates as the usual fixed effect estimator and PPML estimator when its leading term is also non-degenerate. The latter point implies that there is no loss in effective sample sizes with our estimator for using a kernel-based local method compared with the usual non-weighted estimator. If the H\'{a}jek projection is degenerate, our estimator achieves $\sqrt{Nh_{n}}$-asymptotic normality, where $N\sim n^{2}$ is the number of dyads and $h_{n}$ is a bandwidth. While the usual fixed effect estimator and the PPML estimator can be non-Gaussian in the limit \citep{Menzel2021}, our estimator is guaranteed to be asymptotically normal regardless of degeneracy. This result is analogous to \cite{Hall1984}'s central limit theorem for degenerate U-statistics, allowing common statistics of interest, such as confidence intervals, to be constructed in a standard manner. In the degenerate case, our estimator exhibits asymptotic bias, which motivates us to introduce a bias correction.

We propose a variance estimator and bias-corrected confidence intervals that adapt to the degeneracy. Our variance estimator is similar to the one proposed by \cite{Graham2019} for nonparametric dyadic density estimation. We show that our estimator is consistent for the asymptotic variances in both non-degenerate and degenerate cases, after being rescaled by $\sqrt{n}$ or $\sqrt{N h_{n}}$, respectively. For the bias correction, we use a consistent estimator for the asymptotic bias in the degenerate case. We show that the correction term is negligible in the non-degenerate case after being rescaled by $\sqrt{n}$. Combining both bias-corrected estimator and variance estimator, we can construct bias-corrected confidence intervals for our estimator. These intervals have asymptotically correct sizes regardless of the (non-)degeneracy of the leading term in our estimator.

We conduct a simple simulation exercise to demonstrate the performance of our estimators compared to the usual fixed effect estimator and PPML estimator, as we vary the fraction of selected dyads from 10\% to 90\%. Our proposed estimator exhibits better finite sample properties than the other two estimators. Our bias-corrected confidence intervals also outperform the alternatives in coverage probabilities, regardless of degeneracy. This result underscores the importance of bias correction in finite samples, even though the asymptotic bias is zero in the non-degenerate case, which is a new finding in the literature.

We apply our estimator to the regression specification proposed by \cite{Moretti2017}, which estimates the effects of state tax differences on the internal migration flows within the U.S. Comparing our proposed estimator with \cite{Moretti2017}'s, we find that their conclusion, which suggests that state tax differences have a significant impact on internal migration, may not be robust in the presence of a dyadic dependence structure and sample selection biases.

This paper is closely related to the growing literature on dyadic regression 
\citep{Cameron2014, Tabord-Meehan2019, Bonhomme2020, Zeleneev2020, Graham2020, graham2021minimax,sassi2023}. With the exception of \cite{Bonhomme2020} and \cite{Zeleneev2020}, most of these papers do not address non-random sample selection, but instead focus on the consequences of dyadic dependence. \cite{Bonhomme2020} primarily studies cases where selection is conditionally random with random effects, and briefly discusses conditionally non-random selection without providing a theoretical analysis. \cite{Zeleneev2020} investigates identification and estimation in cross-sectional dyadic regression models with more flexible combinations of node-level fixed effects, including fixed selection effects as a special case. However, this flexibility comes at the cost of more complex inference, which is not covered in their paper. In contrast, our paper focuses on models with an additional time dimension, enabling us to develop asymptotic distribution theory and a practical inference method that adapts to degeneracy.

This paper also contributes to the literature on econometric analysis of models with
endogenous network formation. Examples include \cite{Moon2021}, \cite{Auerbach2022}, and \cite{Jochmans2023}. While these papers study social interaction/peer effects type models where outcomes of interest are individualistic, our paper studies the direct
consequence of network formation on dyadic outcomes.

\section{Model}
	\subsection{Setup}\mbox{}\\
	There are $n$ nodes in the data (e.g., states, countries), indexed by $i=1,...,n$. Let $\{(X_{it},Z_{it})_{t=1,...,T})\}_{i=1}^{n}$ be a node-level observation, where $X_{it}\in\mathbb{R}^{q_{x}}$ and $Z_{it}\in\mathbb{R}^{q_{z}}$. For each dyad $ij$ and time $t$, $Y_{ijt}\in\mathbb{R}$ is a main outcome, and we observe a binary variable $d_{ijt}\in\{0,1\}$, which indicates that $Y_{ijt}$ is observable \footnote{Since we focus on a linear model, we can interchange unobservability with zero. Alternatively, we can interpret $Y_{ijt}$ as the logarithm of $\tilde{Y}_{ijt}\geq 0$.} only if $d_{ijt}=1$. We can interpret the adjacency matrix $D_{t}\equiv [d_{ijt}]_{i,j=1,...,n}$ as a network that summarizes the existence of interactions between nodes.  In this paper, we restrict our attention to a model with $T=2$ and an undirected graph where $Y_{ijt}=Y_{jit}$, $d_{ijt}=d_{jit}$ for all $i,j,t$. We also rule out self-loops by convention: $Y_{iit}=d_{iit}=0$ for all $i,t$. An extension to $T>2$ and a directed graph is discussed in Section 4.1.
	
	The data is generated according to the following model:
	\begin{align}
		W_{ijt}&=w(X_{it},X_{jt}), R_{ijt}=r(Z_{it},Z_{jt}), \label{model:regressor}\\
		Y_{ijt}^{*}&=W_{ijt}'\beta + \psi(A_{i},A_{j}) +\epsilon_{ijt},\label{model:structural}\\
		d_{ijt}&=\boldsymbol{1}\{R_{ijt}'\gamma + \varphi(B_{i},B_{j}) -\eta_{ijt}\}, \label{model:firststep}\\
		Y_{ijt}&=\begin{cases}
			Y_{ijt}^{*}\text{ if }d_{ijt}=1\\
			\text{unobserved if }d_{ijt}=0
		\end{cases}.
	\end{align}
	The regressors $W_{ijt}\in\mathbb{R}^{q_{w}}$ and $R_{ijt}\in\mathbb{R}^{q_{r}}$ are constructed from some user-specified symmetric functions $w:\mathbb{R}^{q_{x}}\times\mathbb{R}^{q_{x}}\to\mathbb{R}^{q_{w}}$ and $r:\mathbb{R}^{q_{z}}\times \mathbb{R}^{q_{z}}\to\mathbb{R}^{q_{r}}$ such that $w(x,y)=w(y,x)$ and $r(x',y')=r(y',x')$ for any $x,y\in\mathbb{R}^{q_{x}}$ and $x',y'\in\mathbb{R}^{q_{z}}$. For example, we can specify $w$ to be a pairwise summation $w(x,y)=x+y$. The symmetry in these functions is needed as our graphs are undirected; we can relax this requirement with directed graphs, as discussed in Section 4.1. The node-level fixed effects $A_{i},B_{i}\in\mathbb{R}$ are unobservable, and we allow them to correlate with the regressors, as in the usual fixed effect model. The functions $\psi:\mathbb{R}\times\mathbb{R}\to\mathbb{R}$ and $\varphi:\mathbb{R}\times\mathbb{R}\to\mathbb{R}$ are unknown symmetric functions that capture the interaction between two nodes through their fixed effects. 
	
	We specify the structure of errors $\epsilon_{ijt},\eta_{ijt}$ as follows: For $1\leq i<j\leq n$, 
	\begin{align}
		(\epsilon_{ij1},\epsilon_{ij2},\eta_{ij1},\eta_{ij2})=\tau(U_{i1},U_{i2},U_{j1},U_{j2},U_{ij1},U_{ij2}),
		\label{eq:error}
	\end{align}
	where $U_{i}\equiv (U_{i1},U_{i2})$ and $U_{ij}\equiv (U_{ij1},U_{ij2})$ are node-level and dyad-level random vectors, respectively, and $\tau$ is an unknown multivariate function.\footnote{Here, we need not specify the dimensions of those vectors and the function since the following results do not depend on them as long as those dimensions are fixed.}
	
	Let $\xi_{i}\equiv (X_{i1},X_{i2},Z_{i1},Z_{i2},A_{i},B_{i})$ be a vector that contains observed and unobserved information in the two periods with respect to node $i$. We impose the following distributional assumption:
	\begin{asm}\mbox{}
		\begin{enumerate}
			\item $\xi_{i}$, $i=1,...,n$ are independently and identically distributed.
			\item $(\epsilon_{ijt},\eta_{ijt})_{t=1,2}, 1\leq i<j\leq n$ are generated according to (\ref{eq:error}).
			\item Conditionally on $\{\xi_{i}\}_{i=1}^{n}$, $U_{i}$, $i=1,...,n$ are independent, $U_{ij}$, $1\leq i<j\leq n$ are independent, and both of them are mutually independent.
			\item For $i<j$, $(U_{i},U_{j},U_{ij})$ conditional on $\{\xi_{i}\}_{i=1}^{n}$ has the same distribution as $(U_{i},U_{j},U_{ij})$ conditional on $\xi_{i},\xi_{j}$.
			\item For $i<j<k$, if $\xi_{i}=\xi_{j}=\xi_{k}$, $(U_{i},U_{j},U_{ij})$ and $(U_{i},U_{k},U_{ik})$ has the same distribution conditionally on $\xi_{i},\xi_{j},\xi_{k}$.
		\end{enumerate}
		\label{asm:data}
	\end{asm}
	 Part (1) imposes homogeneity on the node-level data-generating process. Parts (2) and (3) are new to the literature on dyadic regression with fixed effects. While the previous literature assumes conditional independence of dyadic-level errors \citep{Graham2017,Zeleneev2020,Candelaria2020}, our error structure (\ref{eq:error}) allows for the conditional dependence between errors with a common node (e.g., $\epsilon_{ij1}$ and $\epsilon_{ik1}$) through $U_{i}$, but also includes conditional independence as a special case where node-level random vectors $U_{it},U_{jt}$ are degenerate given $\{\xi_{i}\}_{i=1}^{n}$. Part (4) is the standard assumption in the literature and excludes "externalities," where dyad $ij$ can be affected by nodes other than $i$ or $j$. Part (5) ensures the conditional exchangeability of $(\epsilon_{ijt},\eta_{ijt})_{t=1,2}$ across dyads.
	 
	 \subsection{Identification}\mbox{}\\
	 \indent The following two assumptions are crucial for the identification of $\beta$:
	 \begin{asm}
	 	$(\epsilon_{ij1},\epsilon_{ij2},\eta_{ij1},\eta_{ij2})$ and $(\epsilon_{ij2},\epsilon_{ij1},\eta_{ij2},\eta_{ij1})$ are identically distributed conditionally on $\xi_{i},\xi_{j}$.
		\label{asm:exchange}
	 \end{asm}
 	
 	\begin{asm}
 		$E[d_{ij1}d_{ij2}\Delta W_{ij}\Delta W_{ij}'|\Delta R_{ij}'\gamma=0]$ is non-singular where $\Delta W_{ij}=W_{ij1}-W_{ij2}$ and $\Delta R_{ij}=R_{ij1}-R_{ij2}$.
 		\label{asm:inv}
 	\end{asm}
 	Assumption \ref{asm:exchange} excludes cases where, for example, the conditional variance of $\epsilon_{ijt}$ depends only on period $t$'s information: $Var(\epsilon_{ijt}|\xi_{i},\xi_{j})=\sigma^{2}\times W_{ijt}'\beta$. However, it allows time invariant heteroskedasticity such as $Var(\epsilon_{ijt}|\xi_{i},\xi_{j})=\sigma^{2}(W_{ij1}+W_{ij2})'\beta\times A_{i}\times A_{j}$. From (\ref{eq:error}), this assumption is implied by the conditional exchangeability of $U_{it}$ and $U_{ijt}$ with respect to time and the symmetry of function $\tau$ in the sense that $\tau(u_{1},u_{2},v_{1},v_{2},w_{1},w_{2})=\tau(u_{2},u_{1},v_{2},v_{1},w_{2},w_{1})$ for any $u_{1},u_{2},v_{1},v_{2},w_{1},w_{2}$. 
	 Assumption \ref{asm:inv} excludes cases where $W_{ijt}$ is exactly the same as $R_{ijt}$ and implies that some variables in $R_{ijt}$ must be excluded from $W_{ijt}$. Since 
 	\begin{align*}
 		&E[d_{ij1}d_{ij2}\Delta W_{ij}\Delta W_{ij}'|\Delta R_{ij}'\gamma=0]\\
 		&=Pr(d_{ij1}d_{ij2}=1|\Delta R_{ij}'\gamma=0)\times E[\Delta W_{ij}\Delta W_{ij}'|d_{ij1}d_{ij2}=1,\Delta R_{ij}'\gamma=0],
 	\end{align*}
 	this assumption also implies that the networks $D_{1},D_{2}$ are locally dense across time in the sense that $Pr(d_{ij1}d_{ij2}=1|\Delta R_{ij}'\gamma=0)>0$.
 	
 	Our identification argument is summarized in the following two steps, similarly to \cite{Kyriazidou1997}. First, take the time-difference on observed outcomes (dyads with $d_{ij1}=d_{ij2}=1$) to eliminate the fixed effects:
 	\begin{align*}
 		\Delta Y_{ij}=\Delta W_{ij}'\beta +\epsilon_{ij1}-\epsilon_{ij2}.
 	\end{align*}
 	If we take expectation of both sides conditionally on $d_{ij1}=d_{ij2}=1$ and $\xi_{i},\xi_{j}$,
 	\begin{align*}
 		E[\Delta Y_{ij}|d_{ij1}d_{ij2}=1,\xi_{i},\xi_{j}]=\Delta W_{ij}'\beta+\underbrace{E[\epsilon_{ij1}-\epsilon_{ij2}|d_{ij1}d_{ij2}=1,\xi_{i},\xi_{j}]}_{\text{Sample selection effect}}.
 	\end{align*}
 	Note that, in general, the sample selection effect is not $0$.
 	
 	Second, we seek to find conditions to eliminate the selection effect. Assumption \ref{asm:exchange} is equivalent to
 	\begin{align*}
 		F(\epsilon_{ij1},\epsilon_{ij2},\eta_{ij1},\eta_{ij2}|\xi_{i},\xi_{j})=F(\epsilon_{ij2},\epsilon_{ij1},\eta_{ij2},\eta_{ij1}|\xi_{i},\xi_{j}),
 	\end{align*}
 	where $F$ is the conditional distribution of the errors given $\xi_{i},\xi_{j}$. Then, for dyad $ij$ with $\Delta R_{ij}'\gamma=R_{ij1}'\gamma-R_{ij2}'\gamma=0$,
 	\begin{align*}
 		&E[\epsilon_{ij1}|d_{ij1}d_{ij2}=1,\xi_{i},\xi_{j},\Delta R_{ij}'\gamma=0]\\
 		&=E[\epsilon_{ij1}|R_{ij1}'\gamma+\varphi(B_{i},B_{j})\geq \eta_{ij1},R_{ij2}'\gamma+\varphi(B_{i},B_{j})\geq\eta_{ij2},\xi_{i},\xi_{j},\Delta R_{ij}'\gamma=0]\\
 		&=E[\epsilon_{ij2}|R_{ij2}'\gamma+\varphi(B_{i},B_{j})\geq \eta_{ij2},R_{ij1}'\gamma+\varphi(B_{i},B_{j})\geq\eta_{ij1},\xi_{i},\xi_{j},\Delta R_{ij}'\gamma=0]\\
 		&=E[\epsilon_{ij2}|d_{ij1}d_{ij2}=1,\xi_{i},\xi_{j},\Delta R_{ij}'\gamma=0].
 	\end{align*}
 	Hence, the conditional expectation of $\Delta Y_{ij}$ given $d_{ij1}d_{ij2}=1$, $\xi_{i},\xi_{j}$, and $\Delta R_{ij}'\gamma=0$ is
 	\begin{align*}
 		E[\Delta Y_{ij}|d_{ij1}d_{ij2}=1,\xi_{i},\xi_{j},\Delta R_{ij}'\gamma=0]=\Delta W_{ij}'\beta.
 	\end{align*}
 	Multiplying the both sides by $\Delta W_{ij}$ and aggregating $\xi_{i},\xi_{j}$, we get
 	\begin{align*}
 		E[\Delta W_{ij}\Delta Y_{ij}|d_{ij1}d_{ij2}=1,\Delta R_{ij}'\gamma=0]=E[\Delta W_{ij}\Delta W_{ij}'|d_{ij1}d_{ij2}=1,\Delta R_{ij}'\gamma=0]\beta.
 	\end{align*}
 	Then, under Assumption \ref{asm:inv}, $\beta$ is uniquely written as
 	\begin{align}
 		\beta=E[d_{ij1}d_{ij2}\Delta W_{ij}\Delta W_{ij}'|\Delta R_{ij}'\gamma=0]^{-1}E[d_{ij1}d_{ij2}\Delta W_{ij}\Delta Y_{ij}|\Delta R_{ij}'\gamma=0].
 		\label{eq:ident}
 	\end{align}
 	
 	\subsection{Estimation}\mbox{}\\
 	\indent Estimation is done in two steps. In the first step, we estimate $\gamma$ with a consistent estimator $\hat{\gamma}_{n}$, and in the second step we estimate $\beta$ with $\hat{\beta}_{n}$, a sample analogue of the identified $\beta$ with $\gamma$ replaced by $\hat{\gamma}_{n}$. 
 	
 	In the following, we focus on the second step. The sample-analogue of (\ref{eq:ident}) is given by
 	\begin{align*}
 		\hat{\beta}_{n}=\left[\sum_{i<j}d_{ij1}d_{ij2}\Delta W_{ij}\Delta W_{ij}'K_{h_{n}}(\Delta R_{ij}'\hat{\gamma}_{n})\right]^{-1}\left[\sum_{i<j}d_{ij1}d_{ij2}\Delta W_{ij}\Delta Y_{ij}K_{h_{n}}(\Delta R_{ij}'\hat{\gamma}_{n})\right],
 	\end{align*}
 	where $\sum_{i<j}=\sum_{i=1}^{n-1}\sum_{j=i+1}^{n}$, $K_{h_{n}}(v)=h_{n}^{-1}K(v/h_{n})$ is a kernel, and $h_{n} $ is a bandwidth. The weight function is used to smooth the condition $\Delta R_{ij}'\gamma=0$ and puts larger weight on observations with small $\Delta R_{ij}'\hat{\gamma}_{n}$.
 	
 	To evaluate $\hat{\beta}_{n}$ in terms of $\beta$, rewrite the time-differenced model as
 	\begin{align*}
 		\Delta Y_{ij}=\Delta W_{ij}'\beta +\lambda_{ij}+\nu_{ij},
 	\end{align*} 
 	where 
 	\begin{align*}
 		\lambda_{ij}&\equiv E[\epsilon_{ij1}-\epsilon_{ij2}|d_{ij1}d_{ij2}=1, \xi_{i},\xi_{j}]\\
 		\nu_{ij}&\equiv \epsilon_{ij1}-\epsilon_{ij2}-\lambda_{ij}.
 	\end{align*}
 	Note that $E[\nu_{ij}|d_{ij1}d_{ij2}=1,\xi_{i},\xi_{j}]=0$ by construction. Define
 	\begin{align*}
 		\hat{S}_{WW}&\equiv \frac{1}{N}\sum_{i<j}d_{ij1}d_{ij2}\Delta W_{ij}\Delta W_{ij}'K_{h_{n}}(\Delta R_{ij}'\hat{\gamma}_{n}),\\
 		\hat{S}_{W\lambda}&\equiv \frac{1}{N}\sum_{i<j}d_{ij1}d_{ij2}\Delta W_{ij}\lambda_{ij}K_{h_{n}}(\Delta R_{ij}'\hat{\gamma}_{n}),\\
 		\hat{S}_{W\nu}&\equiv \frac{1}{N}\sum_{i<j}d_{ij1}d_{ij2}\Delta W_{ij} \nu_{ij}K_{h_{n}}(\Delta R_{ij}'\hat{\gamma}_{n}).
 	\end{align*}
 	Substituting $\Delta Y_{ij}$ into $\hat{\beta}_{n}$ yields
 	\begin{align*}
 		\hat{\beta}_{n}=\beta +\hat{S}_{WW}^{-1}\hat{S}_{W\lambda}+\hat{S}_{WW}^{-1}\hat{S}_{W\nu}.
 	\end{align*}
 	The terms $\hat{S}_{WW}^{-1}\hat{S}_{W\lambda}$ and $\hat{S}_{WW}^{-1}\hat{S}_{W\nu}$ can be understood as the selection bias term and the stochastic error term of the estimator, respectively.
 
\section{Asymptotic Analysis}
	
	\subsection{Regularity Conditions}\mbox{}\\
	\indent For ease of notation, we write the following conditions in terms of dyads $12$ and $13$, which entails no loss of generality under the undirected graph and Assumption \ref{asm:data}.
	
	Let $f_{R\gamma,2}$ be the joint density of $\Delta R_{12}'\gamma$ and $\Delta R_{13}'\gamma$ when it exists and $f_{R\gamma,2|\xi_{2},U_{2},\xi_{3},U_{3}}$ be the conditional density given $\xi_{2},U_{2},\xi_{3},U_{3}$.
	 Let $f_{R\gamma}$ be the marginal density and $f_{R\gamma|\xi_{1},U_{1}}$ be the conditional density given $\xi_{1},U_{1}$.
	\begin{asm}
		The joint distribution of $\Delta R_{12}'\gamma$ and $\Delta R_{13}'\gamma$ is absolutely continuous, and for some $\kappa_{0}>0$, the following hold in the neighborhoods $(-\kappa_{0},\kappa_{0})^{2}$ or $(-\kappa_{0},\kappa_{0})$ around $(0,0)$ or $(0)$, respectively:
		\begin{enumerate}
			\item The density $f_{R\gamma,2}(\cdot,\cdot)$ is $k\geq 2$ times continuously differentiable, and the derivatives $\frac{\partial^{2}}{\partial x^{p}\partial y^{q}}f_{R\gamma,2}(\cdot,\cdot)$ are bounded for $p+q\leq k,p,q\geq 0$ and bounded away from $0$.
			\item The conditional density $f_{R\gamma,2|\xi_{2},U_{2},\xi_{3},U_{3}}(\cdot,\cdot)$ given $\xi_{2},U_{2},\xi_{3},U_{3}$ is continuous and bounded almost surely.
			\item The marginal density $f_{R\gamma}(\cdot)$ is bounded away from $0$.
			\item The conditional marginal density $f_{R\gamma|\xi_{1},U_{1}}(\cdot)$ given $\xi_{1},U_{1}$ is continuous and bounded almost surely.
		\end{enumerate}
		\label{asm:index}
	\end{asm}
	Part (1) is a smoothness assumption on the density as in the nonparametric regression literature. 
	Part (3) ensures that we observe $\Delta R_{12}'\gamma$ around $0$, which is crucial for identification. Parts (2) and (4) essentially requires well-behaved $r(\cdot,\cdot)$ in (\ref{model:regressor}).
	
	Define $(w_{1},w_{2})\mapsto \Lambda(w_{1},w_{2},\xi_{1},\xi_{2})$ as
	\begin{align*}
		\Lambda(w_{1},w_{2},\xi_{1},\xi_{2})\equiv E[\epsilon_{12t}|\eta_{12t}\leq w_{1},\eta_{12s}\leq w_{2},\xi_{1},\xi_{2}]
	\end{align*}
	with $t,s=1,2,t\neq s$. This $\Lambda$ is the sample selection effect caused by the correlation between errors $\epsilon_{12t},\epsilon_{12s}$ and $\eta_{12t},\eta_{12s}$. Note that the function $\Lambda$ does not depend on time $t$ or $s$ because of Assumption \ref{asm:exchange}.
	\begin{asm}
		The function $(w_{1},w_{2})\mapsto \Lambda(w_{1},w_{2},\xi_{1},\xi_{2})$ is differentiable  in the neighborhoods $(-\kappa_{0},\kappa_{0})^{2}$ around $(0,0)$ for some $\kappa_{0}>0$.
	\label{asm:bias}
	\end{asm}
	This assumption is essential for controlling the sample selection effect and characterizing the asymptotic bias in some cases. An implication of this assumption is that for some $\Lambda_{12}\equiv \tilde{\Lambda}(w_{1},w_{2},\xi_{1},\xi_{2})$, 
	\begin{align*}
		\Lambda(w_{1},w_{2},\xi_{1},\xi_{2})-\Lambda(w_{2},w_{1},\xi_{1},\xi_{2})=\Lambda_{12}\times (w_{1}-w_{2}),
	\end{align*}
	by the multivariate mean-value theorem.
	Note that the function $\Lambda$ does not depend on time $t$ or $s$ because of Assumption \ref{asm:exchange}. This assumption is strong because the difference in $\Lambda$ must be exactly linear in the first and second elements. If we focus on the degenerate case discussed below, since the asymptotic bias is $0$ in that case, we can relax the differentiability to Lipschitz-like continuity on $\Lambda$: $|\Lambda(w_{1},w_{2},\xi_{1},\xi_{2})-\Lambda(w_{2},w_{1},\xi_{1},\xi_{2})|\leq |\Lambda_{12}|\times |w_{1}-w_{2}|$. 
	
	Let $\|\cdot\|$ denote a Euclidian norm of vectors.
	\begin{asm}
		For some $\kappa_{0}>0$, the following hold in the neighborhoods $(-\kappa_{0},\kappa_{0})^{2}$ or $(-\kappa_{0},\kappa_{0})$ around $(0,0)$ or $(0)$, respectively.
		\begin{enumerate}
			\item The following moments are  bounded almost surely:
			\begin{align*}
				&E[\|\Delta W_{12}\|^{8}|\Delta R_{12}'\gamma=\cdot,\xi_{1},U_{1}],
				E[\|\Delta R_{12}\|^{6}|\Delta R_{12}'\gamma=\cdot,\xi_{1},U_{1}],\\
				&E[\nu_{12}^{8}|\Delta R_{12}'\gamma=\cdot,\xi_{1},U_{1}],
				E[\Lambda_{12}^{6}|\Delta R_{12}'\gamma=\cdot,\xi_{1},U_{1}].
			\end{align*}
			\item The following moments are continuous and bounded, and the first two are positive definite:
			\begin{align*}
				&E[d_{121}d_{122}\Delta W_{12}\Delta W_{12}'|\Delta R_{12}'\gamma=\cdot],E[d_{121}d_{122}\Delta W_{12}\Delta W_{12}'\nu_{12}^{2}|\Delta R_{12}'\gamma=\cdot]\\
				&E[d_{121}d_{122}d_{131}d_{132}\Delta W_{12}\Delta W_{13}'\nu_{12}\nu_{13}|\Delta R_{12}'\gamma=\cdot,\Delta R_{13}'\gamma=\cdot].
			\end{align*}
			\item $g(\cdot)\equiv E[d_{121}d_{122}\Delta W_{12}\Lambda_{12}|\Delta R_{12}'\gamma=\cdot]f_{R\gamma}(\cdot)$ is $k$-times continuously differentiable with bounded derivatives.
			\item $g_{\xi_{1},U_{1}}(\cdot)\equiv E[d_{121}d_{122}\Delta W_{12}\nu_{12}|\Delta R_{12}'\gamma=\cdot,\xi_{1},U_{1}]f_{R\gamma|\xi_{1},U_{1}}(\cdot)$ is $k$-times continuously differentiable with bounded derivatives almost surely.
		\end{enumerate}
	\label{asm:moments}
	\end{asm}
	Part (1) assumes the existence of conditional moments for the relevant variables. The conditioning on $\xi_{1}$ and $U_{1}$ is needed for controlling the dyadic dependence structure. Part (2) is crucial for obtaining the convergence results used below, and the positive definiteness is needed for ensuring the non-degeneracy of our estimator in the limit. Part (3) is used for characterizing the asymptotic bias provided below. Part (4) is essential for the negligibility of the approximation error of our variance estimator.
	
	\begin{asm}
		The following moments exist:
		\begin{align*}
			E[\|\Delta W_{12}\|^{8}], E[\|\Delta R_{12}\|^{8}],E[\Lambda_{12}^{6}],E[\nu_{12}^{6}]
		\end{align*}
	\label{asm:unconmoments}
	\end{asm}
	Additionally to Assumption \ref{asm:moments}, which restricts the moments locally around $(0,0)$ or $(0)$, we use the existence of these unconditional moments when bounding error terms coming from the usage of $\hat{\gamma}_{n}$.
	 
	\begin{asm}
		A kernel function $K(\cdot)$ satisfies the following:
		\begin{enumerate}
			\item For some $\kappa>0$, $K$ is $0$ outside of $[-\kappa,\kappa]$, bounded in $[-\kappa,\kappa]$, and three times continuously differentiable with bounded derivatives in $(-\kappa,\kappa)$.
			\item $\int K(s)ds=1$.
			\item $\int s^{i}K(s)ds=0$ for $i=1,...,k$.
		\end{enumerate}
	\label{asm:kernel}
	\end{asm}
	For example, a fourth-order biweight kernel $K(x)=106/64(1-3u^{2})(1-x^{2})^{2}\boldsymbol{1}\{|x|< 1\}$ satisfies this assumption with $\kappa=1$ and $k=3$.
	
	\begin{asm}
		The sequence of bandwidths $\{h_{n}\}$ satisfies $h_{n}\to 0$ and $nh_{n}\to\infty$ as $n\to\infty$.
	\label{asm:band}
	\end{asm}
	This assumption is standard in the nonparametric regression literature. We impose further conditions on $\{h_{n}\}$ in each statement below.
	
	\begin{asm}
		The first-step estimator $\hat{\gamma}_{n}$ satisfies $\sqrt{Nh_{n}}(\hat{\gamma}_{n}-\gamma)=o_{p}(1)$.
		\label{asm:first}
	\end{asm}
	This assumption requires the first-step estimator to be consistent and converge faster than our estimator. For example, if $\eta_{ijt}\sim Logistic(0,1)$ independently across $ij$ and $t$, we can show that \cite{Chamberlain1980}'s conditional logit estimator satisfies  $\hat{\gamma}_{n}-\gamma=O_{p}(1/\sqrt{N})$ so that $\sqrt{Nh_{n}}(\hat{\gamma}_{n}-\gamma)=O_{p}(\sqrt{h_{n}})=o_{p}(1)$. In Section 3.6, we discuss the availability of alternative estimators for $\gamma$. We leave the case where $\hat{\gamma}_{n}$ converges slower than required in this assumption for future research.
	
	\subsection{Asymptotic Normality}\mbox{}\\
	\indent Define the following components that will appear in the asymptotic bias and variance expression:
	\begin{align*}
		\Sigma_{WW}&\equiv f_{R\gamma}(0)E[d_{121}d_{122}\Delta W_{12}\Delta W_{12}'|\Delta R_{12}'\gamma=0]\\
		\Sigma_{W\lambda}&\equiv\frac{1}{k!}\frac{\partial^{k}g(0)}{\partial w^{k}}\int s^{k+1}K(s)ds,\\
		\Sigma_{W\nu,1}&\equiv 4f_{R\gamma,2}(0,0)E[d_{121}d_{122}d_{131}d_{132}\Delta W_{12}\Delta W_{13}'\nu_{12}\nu_{13}|\Delta R_{12}'\gamma=\Delta R_{13}'\gamma=0],\\
		\Sigma_{W\nu,2}&\equiv f_{R\gamma}(0)E[d_{121}d_{122}\Delta W_{12}\Delta W_{12}'\nu_{12}^{2}|\Delta R_{12}'\gamma=0]\int K^{2}(s)ds.
	\end{align*}
	We have the following result:
	\begin{thm}
		Suppose that Assumptions \ref{asm:data}-\ref{asm:first} hold. Fix an arbitrary non-zero vector $c\in\mathbb{R}^{q_{w}}$ and some constant $h\in[0,\infty)$. Let $c_{W}=\Sigma_{WW}^{-1}c$. Then, as $n\to\infty$, we have the following three cases:
		\begin{enumerate}
			\item If $Nh_{n}^{2k+3}\to h$ and $c_{W}'\Sigma_{W\nu,1}c_{W}>0$:
			\begin{align*}
				\sqrt{n}c'(\hat{\beta}_{n}-\beta)\to_{d}\mathcal{N}(0,c_{W}'\Sigma_{W\nu,1}c_{W}).
			\end{align*}
			\item If $Nh_{n}^{2k+3}\to h$ and $c_{W}'\Sigma_{W\nu,1}c_{W}=0$:
			\begin{align*}
			\sqrt{Nh_{n}}c'(\hat{\beta}_{n}-\beta)\to_{d}\mathcal{N}(\sqrt{h}c_{W}'\Sigma_{W\lambda}, c_{W}'\Sigma_{W\nu,2}c_{W}).
			\end{align*}
			\item If $Nh_{n}^{2k+3}\to\infty$ and $nh^{2k}_{n}\to\infty$:
			\begin{align*}
				h_{n}^{-(k+1)}(\hat{\beta}_{n}-\beta)\to_{p}\Sigma_{WW}^{-1}\Sigma_{W\lambda}.
			\end{align*}
		\end{enumerate}
 	\label{thm:normality}
	\end{thm}

	Parts (1) and (2) of Theorem \ref{thm:normality} show that our estimator is asymptotically normal, with different convergence rates depending on $\Sigma_{W\nu,1}$. Part (1) differs from \cite{Kyriazidou1997} in that the convergence rate is parametric and based on the number of nodes $n$, rather than the number of dyads $N$. When $c_{W}'\Sigma_{W\nu,1}c_{W}>0$ in part (1), the covariance between summands sharing a common node (e.g., dyads $ij$ and $ik$) does not vanish asymptotically, reducing the effective sample size to $n$. The leading term is an average of conditional means given $\xi_{i}$ and $U_{i}$, which averages out and eliminates $h_{n}$ from the convergence rate. This $\sqrt{n}$-asymptotic normality matches results in the dyadic nonparametric density estimation literature \citep{Graham2019}. When $\Sigma_{W\nu,1}=0$, as in part (2), our result aligns with \cite{Kyriazidou1997}, with nonparametric convergence rates based on the number of dyads. This corresponds to the degenerate case for dyadic dependence, as discussed in the literature \citep{Graham2019,cattaneo2024}. Part (3) of Theorem \ref{thm:normality} shows that, with suitable normalization, our estimator converges to the asymptotic bias term regardless of degeneracy. This property is used to construct the bias-corrected estimator in the following section.
	
	We can compare our estimator with the usual fixed effect estimator:
	\begin{align}
		\hat{\beta}_{FE}=\left[\sum_{i<j}d_{ij1}d_{ij2}\Delta W_{ij}\Delta W_{ij}'\right]^{-1}\left[\sum_{i<j}d_{ij1}d_{ij2}\Delta W_{ij}\Delta Y_{ij}\right],
		\label{eq:beta_fe}
	\end{align}
	which is biased because of the selection effect $\lambda_{ij}$.
	First, in the case of non-degeneracy, our estimator and the re-centered (infeasible) fixed effect estimator share the same convergence rates of $\sqrt{n}$ \citep{Daveziez2021}. This implies that there is no reduction in the effective sample size for using our kernel-based local estimator, which amends the need for fairly large samples as discussed in \cite{Kyriazidou1997}. Second, in the case of degeneracy, the fixed effect estimator applied to our model can exhibit a non-Gaussian distribution in the limit \citep{Menzel2021}, while our estimator is asymptotically normal regardless of the degeneracy. This guaranteed asymptotic normality is analogous to \cite{Hall1984}'s central limit theorem for degenerate U-statistics, and thus the common statistics of interest, such as confidence intervals, can be constructed in a standard manner.

	If we interpret the structural equation \eqref{model:structural} as the log-linearized version of the canonical gravity model \citep{SantosSilva2006,Head2014} with additive fixed effects,
	\begin{align*}
		\tilde{Y}_{ijt} = exp(W_{ijt}'\beta + A_{i}+A_{j})\times \underbrace{\eta_{ijt}}_{=d_{ijt}exp(\epsilon_{ijt})},
	\end{align*}
	the Poisson pseudo-maximum-likelihood estimator (PPML) for $\beta$ can be compared with our estimator. The PPML estimator $\hat{\beta}_{PPML}$ with two-way fixed effects is defined as the solution to:
	\begin{align}
		\sum_{i<j}\sum_{t=1}^{2}d_{ijt}\left[\tilde{Y}_{ijt}-exp(W_{ijt}'b+ a_{i}+a_{j})\right]W_{ijt}=0,
		\label{eq:beta_ppml}
	\end{align}
	where $a_{1},...,a_{n}$ satisfy 
	\begin{align*}
		\sum_{j=i+1}^{n}\sum_{t=1}^{2}d_{ijt}\left[\tilde{Y}_{ijt}-exp(W_{ijt}'\hat{\beta}_{PPML}+ a_{i}+ a_{j})\right]=0,\quad i=1,...,n-1.
	\end{align*}
	We can make a similar comparison as in the fixed effect estimator based on the results by \cite{Daveziez2021} and \cite{Menzel2021}: $\hat{\beta}_{PPML}$ will be biased because of the misspecified errors, and the re-centered $\hat{\beta}_{PPML}$ is asymptotically normal at the rate of $\sqrt{n}$ in the non-degenerate case and can be non-Gaussian in the degenerate case.

	\subsection{Variance Estimation}\mbox{}\\
	\indent Since our estimator exhibits different asymptotic distributions depending on $\Sigma_{W\nu,1}$, it is desirable to have a variance estimator that adapts to the degeneracy.
	
	First, we estimate $\Sigma_{W\nu,1}$. Define
	\begin{align*}
		\hat{S}_{ij}\equiv 2d_{ij1}d_{ij2}K_{h_{n}}(\Delta R_{ij}'\hat{\gamma}_{n})\Delta W_{ij}\Delta \hat{\epsilon}_{ij},
	\end{align*} 
	where $\Delta\hat{\epsilon}_{ij}$ is a residual $\Delta Y_{ij}-\Delta W_{ij}'\hat{\beta}_{n}$. Then, we propose an estimator for $\Sigma_{W\nu,1}$ as
	\begin{align*}
		\hat{\Sigma}_{W\nu,1}={n\choose 3}^{-1}\Sigma_{i<j<k}\frac{1}{3}(\hat{S}_{ij}\hat{S}_{ik}'+\hat{S}_{ij}\hat{S}_{jk}'+\hat{S}_{ik}\hat{S}_{jk}').
	\end{align*}
	Next, we estimate $\Sigma_{W\nu,2}$ by
	\begin{align*}
		\hat{\Sigma}_{W\nu,2}=\frac{h_{n}}{N}\sum_{i<j}d_{ij1}d_{ij2}K_{h_{n}}(\Delta R_{ij}'\hat{\gamma}_{n})^{2}\Delta W_{ij}\Delta W_{ij}'\Delta \hat{\epsilon}_{ij}.
	\end{align*} 
	The following result shows consistency of these estimators and their usefulness in adaptive variance estimation.
	
	\begin{prop}
		Suppose that Assumptions \ref{asm:data}-\ref{asm:first} hold. Set $h_{n}=hN^{-1/(2k+3)}$ for some $h\in(0,\infty)$. We have
		\begin{align*}
			\hat{\Sigma}_{W\nu,1}&\to_{p}\Sigma_{W\nu,1},\\
			\hat{\Sigma}_{W\nu,2}&\to_{p}\Sigma_{W\nu,2},
		\end{align*}
		as $n\to\infty$. If $c_{W}\Sigma_{W\nu,1}c_{W}=0$ with $c_{W}=\Sigma_{WW}^{-1}c$ for some $c\in\mathbb{R}^{q_{w}}$, we have
		\begin{align*}
			nh_{n}c'\hat{S}_{WW}^{-1}\hat{\Sigma}_{W\nu,1}\hat{S}_{WW}^{-1}c\to_{p}0,
		\end{align*}
		as $n\to\infty$.
		\label{prop:variance}
	\end{prop}

	We now propose our variance estimator as follows:
	\begin{align*}
		\hat{\Sigma}\equiv \hat{S}_{WW}^{-1}\left[\frac{n-2}{n(n-1)}\hat{\Sigma}_{W\nu,1}+\frac{1}{Nh_{n}}\hat{\Sigma}_{W\nu,2}\right]\hat{S}_{WW}^{-1}.
	\end{align*}
	We can see that this estimator is adaptive to the degeneracy: When $\Sigma_{W\nu,1}$ is positive definite, since $n/(Nh_{n})=o(1)$,
	\begin{align*}
		nc'\hat{\Sigma}c=c'\hat{S}_{WW}^{-1}\left[\frac{n-2}{n-1}\hat{\Sigma}_{W\nu,1}+\frac{n}{Nh_{n}}\hat{\Sigma}_{W\nu,2}\right]\hat{S}_{WW}^{-1}c\to_{p}c_{W}'\Sigma_{W\nu,1}c_{W},
	\end{align*}
	as $n\to\infty$ by Proposition \ref{prop:variance} and Lemma \ref{lem:S_WW} in Appendix A. When $c_{W}'\Sigma_{W\nu,1}c_{W}=0$, since $nh_{n}c'\hat{S}_{WW}^{-1}\hat{\Sigma}_{W\nu,1}\hat{S}_{WW}^{-1}c=o_{p}(1)$ by Proposition \ref{prop:variance},
	\begin{align*}
		Nh_{n}c'\hat{\Sigma}c=c'\hat{S}_{WW}^{-1}\left[2(n-2)h_{n}\hat{\Sigma}_{W\nu,1}+\hat{\Sigma}_{W\nu,2}\right]\hat{S}_{WW}^{-1}c\to_{p}c_{W}'\Sigma_{W\nu,2}c_{W},
	\end{align*}
	as $n\to\infty$. 
	
	Our variance estimator is adapted from the one provided in \cite{Graham2019} for a dyadic nonparametric density estimator. They show that this type of estimator can be adaptive to the "knife edge" case, where $nh_{n}$ is bounded from above and below asymptotically so that $Nh_{n}\sim n$. Here, we additionally show that the estimator is adaptive to the degeneracy by showing that the term involving $\hat{\Sigma}_{W\nu,1}$ decays fast enough to be negligible when the convergence rate is $\sqrt{Nh_{n}}$.
	
	\subsection{Bandwidth Selection}\mbox{}\\
	\indent From the asymptotic distributional approximation result in Theorem \ref{thm:normality}, we can write down the mean squared error of our estimator (without negligible parts)
	\begin{align*}
		MSE(c'\hat{\beta}_{n})=h_{n}^{2(k+1)}(c_{W}'\Sigma_{W\lambda})^{2}+\frac{1}{n}c_{W}'\Sigma_{W\nu,1}c_{W}+\frac{1}{Nh_{n}}c_{W}'\Sigma_{W\nu,1}c_{W}.
	\end{align*}
	The optimal solution for minimizing this mean squared error with respect to $h_{n}$ is given by
	\begin{align*}
		h_{n}^{*}&=\left(\frac{c_{W}'\Sigma_{W\nu,2}c_{W}}{2(k+1)N(c_{W}'\Sigma_{W\nu})^{2}}\right)^{\frac{1}{2k+3}}\\
			&=h^{*}N^{-\frac{1}{2k+3}}.
	\end{align*}

	We can estimate $h^{*}$ by the plug-in method. By Proposition \ref{prop:variance}, we have a consistent estimator for the variance part. For the bias part, we use a pilot bandwidth given by
	\begin{align*}
		h_{n,\delta}=hN^{-\delta/(2k+3)},
	\end{align*}
	for some $\delta\in(0,\frac{2k+3}{4k+4})$ and $h>0$. Let $\hat{\beta}_{n,\delta}$ be our estimator calculated with $h_{n,\delta}$. We can check that this bandwidth satisfies $Nh_{n,\delta}^{2k+3}\to\infty$ and $nh_{n,\delta}^{2k+2}\to\infty$. Thus, by Theorem \ref{thm:normality},
	\begin{align*}
		h_{n,\delta}^{-(k+1)}(\hat{\beta}_{n,\delta}-\beta)\to_{p}\Sigma_{WW}^{-1}\Sigma_{W\nu},
	\end{align*}
 	as $n\to\infty$. By replacing $\beta$ by $\hat{\beta}_{n}$, calculated with $h_{n}=hN^{-\frac{1}{2k+3}}$, we have the following result:
 	\begin{prop}
	 		Suppose that Assumptions \ref{asm:data}-\ref{asm:first} hold. Let $\hat{\beta}_{n}$ and $\hat{\beta}_{n,\delta}$ be the proposed estimators with bandwidths $h_{n}=hN^{-1/(2k+3)}$ and $h_{n,\delta}=hN^{-\delta/(2k+3)}$, respectively, for some $h>0$ and $\delta\in(0,\frac{2k+3}{4k+4})$. Then,
 			\begin{align*}
 			h_{n,\delta}^{-(k+1)}(\hat{\beta}_{n,\delta}-\hat{\beta}_{n})\to_{p}\Sigma_{WW}^{-1}\Sigma_{W\lambda},
 		\end{align*}
 		as $n\to\infty$.
 		\label{prop:bias}
 	\end{prop}
 	Thus, 
 	\begin{align*}
 		\hat{h}^{*}=\left(\frac{c'\hat{S}_{WW}^{-1}\hat{\Sigma}_{W\nu,2}\hat{S}_{WW}^{-1}c}{2(k+1)\{h_{n,\delta}^{-(k+1)}c'(\hat{\beta}_{n,\delta}-\hat{\beta}_{n})\}^{2}}\right)^{\frac{1}{2k+3}}
 	\end{align*}
 	is a consistent estimator for $h^{*}$ by Propositions \ref{prop:variance} and \ref{prop:bias}.
 	
 	\subsection{Bias Correction}\mbox{}\\
 	\indent Notice that our estimator has the asymptotic bias of $\sqrt{h}\Sigma_{WW}^{-1}\Sigma_{W\lambda}$ in the case of degeneracy, $\Sigma_{W\nu,1}=0$ from Theorem \ref{thm:normality}. If the bias is non-negligible, it distorts the coverage probability of the confidence interval. Correcting the bias part is desirable as it is generally unknown whether the degeneracy occurs. Fortunately, given the similar asymptotic distributional result as \cite{Kyriazidou1997} in the degenerate case, we can use her bias correction strategy as follows. 
 	 
 	Note that $h_{n,\delta}^{-(k+1)}(\hat{\beta}_{n,\delta}-\beta)$ directly estimates the asymptotic bias from Theorem \ref{thm:normality}. We can construct a bias-corrected estimator $\hat{\beta}_{n,bc}(\beta)$ by subtracting this bias estimator from the original estimator with suitable normalization: Let $r_{n,\delta}=N^{(1-\delta)/(2k+3)}$. The bias-corrected estimator is given by
 	\begin{align*}
 		\hat{\beta}_{n,bc}(\beta)=\hat{\beta}_{n}-r_{n,\delta}^{-(k+1)}(\hat{\beta}_{n,\delta}-\beta).
 	\end{align*}
 	
 	We can check that this estimator is asymptotically unbiased regardless of the degeneracy: When $c_{W}'\Sigma_{W\nu,1}c_{W}>0$,
 	\begin{align*}
 		\sqrt{n}c'(\hat{\beta}_{n,bc}(\beta)-\beta)&=\sqrt{n}c'(\hat{\beta}_{n}-\beta)-\sqrt{n}r_{n,\delta}^{-(k+1)}c'(\hat{\beta}_{n,\delta}-\beta)\\
 		&=\underbrace{\sqrt{n}(\hat{\beta}_{n}-\beta)}_{\to_{d}\mathcal{N}(0,c_{W}'\Sigma_{W\nu,1}c_{W})}-\underbrace{\sqrt{n}h^{k+1}N^{-(k+1)/(2k+3)}}_{\to 0}\underbrace{h_{n,\delta}^{-(k+1)}c'(\hat{\beta}_{n,\delta}-\beta)}_{\to_{p}c_{W}'\Sigma_{W\lambda}}\\
 		&\to_{d}\mathcal{N}(0,c_{W}'\Sigma_{W\nu,1}c_{W}),
 	\end{align*}
 	as $n\to\infty$. When $c_{W}'\Sigma_{W\nu,1}c_{W}=0$, 
 	\begin{align*}
 		\sqrt{Nh_{n}}c'(\hat{\beta}_{n,bc}(\beta)-\beta)&=\sqrt{Nh_{n}}c'(\hat{\beta}_{n}-\beta)-\sqrt{Nh_{n}}h_{n,1-\delta}^{-(k+1)}c'(\hat{\beta}_{n,\delta}-\beta)\\
 		&=\underbrace{\sqrt{Nh_{n}}c'(\hat{\beta}_{n}-\beta)}_{\to_{d}\mathcal{N}(\sqrt{h^{2k+3}}c_{W}'\Sigma_{W\lambda},c_{W}'\Sigma_{W\nu,2}c_{W})}-\underbrace{\sqrt{h}h_{n,\delta}^{-(k+1)}c'(\hat{\beta}_{n,\delta}-\beta)}_{\to_{p}\sqrt{h^{2k+3}}c_{W}\Sigma_{W\lambda}}\\
 		&\to_{d}\mathcal{N}(0,c_{W}'\Sigma_{W\nu,2}c_{W}),
 	\end{align*}
  	as $n\to\infty$. Thus, given the adaptivity of $\hat{\Sigma}$ to the degeneracy, we have
  	\begin{align*}
  		(c'\hat{\Sigma}c)^{-1/2}c'(\hat{\beta}_{n,bc}(\beta)-\beta)\to_{d}\mathcal{N}(0,1),
  	\end{align*}
  	as $n\to\infty$ for an arbitrary non-zero vector $c\in\mathbb{R}^{q_{w}}$.
  	
  	Then, we can construct the bias-corrected confidence interval as follows: Letting $\Phi_{1-\alpha/2}^{-1}$ be $1-\alpha/2$ quantile of the standard normal distribution, we have
  	\begin{align*}
  		&-\Phi_{1-\alpha/2}^{-1}\leq (c'\hat{\Sigma}c)^{-1/2}c'(\hat{\beta}_{n,bc}(\beta)-\beta)\leq \Phi_{1-\alpha/2}^{-1}\\
  		&\Longleftrightarrow -(c'\hat{\Sigma}c)^{-1/2}\Phi_{1-\alpha/2}^{-1}\leq c'\hat{\beta}_{n}-h_{n,1-\delta}^{-(k+1)}c'\hat{\beta}_{n,\delta}-(1-h_{n,1-\delta}^{-(k+1)})c'\beta\leq (c'\hat{\Sigma}c)^{-1/2}\Phi_{1-\alpha/2}^{-1}\\
  		&\Longleftrightarrow CI_{L,\alpha,c}\leq c'\beta\leq CI_{U,\alpha,c},
  	\end{align*}
  	where
  	\begin{align*}
  		CI_{L,\alpha,c}&\equiv (1-h_{n,1-\delta}^{-(k+1)})^{-1}\left[c'\hat{\beta}_{n}-h_{n,1-\delta}^{(-k+1)}c'\hat{\beta}_{n,\delta}-(c'\hat{\Sigma}c)^{-1/2}\Phi_{1-\alpha/2}^{-1}\right],\\
  		CI_{U,\alpha,c}&\equiv (1-h_{n,1-\delta}^{-(k+1)})^{-1}\left[c'\hat{\beta}_{n}-h_{n,1-\delta}^{(-k+1)}c'\hat{\beta}_{n,\delta}+(c'\hat{\Sigma}c)^{-1/2}\Phi_{1-\alpha/2}^{-1}\right].
  	\end{align*}
  	The full inference procedure is summarized as follows:
  	\begin{enumerate}
  		\item Compute the first step estimator $\hat{\gamma}_{n}$.
  		\item Choose $k\geq 2,\delta\in(0,\frac{2k+3}{4k+4})$, and $h>0$ to compute $\hat{\beta}_{n}$ and $\hat{\beta}_{n,\delta}$ with bandwidths $h_{n}=hN^{-1/(2k+3)}$ and $h_{n,\delta}=hN^{-\delta/(2k+3)}$, respectively.
  		\item Compute $\hat{\Sigma}$ and $h_{n,\delta}^{-(k+1)}(\hat{\beta}_{n,\delta}-\hat{\beta}_{n})$ to estimate the asymptotic variance and bias and obtain $\hat{h}^{*}$.
  		\item Update $\hat{\beta}_{n}$ and $\hat{\beta}_{n,\delta}$ with bandwidths $h_{n}=\hat{h}^{*}N^{-1/(2k+3)}$ and $h_{n,\delta}=\hat{h}^{*}N^{-\delta/(2k+3)}$, respectively.
  		\item Construct the confidence interval by computing $CI_{L,\alpha,c}$ and $CI_{U,\alpha,c}$ from $\hat{\beta}_{n},\hat{\beta}_{n,\delta}$, and $c'\hat{\Sigma}c$.
  	\end{enumerate}
  
  	\subsection{First-step Estimator}\mbox{}\\
  	\indent Remember that we want to estimate $\gamma$ from the selection equation or network formation process (\ref{model:firststep}):
  	\begin{align*}
  		d_{ijt}=\boldsymbol{1}\{R_{ijt}'\gamma+\varphi(B_{i},B_{j})-\eta_{ijt}\geq 0\}.
  	\end{align*}
  	 This DGP can be interpreted as a panel discrete choice model as well as a network formation model. Estimators for discrete choice models such as \cite{Chamberlain1980}, \cite{Manski1987}, or \cite{Horowitz1992} can be candidates for estimating $\gamma$. Also, estimators for network formation models such as \cite{Graham2017} or \cite{Candelaria2020} can be applicable under additional conditions.
  	 
  	 Whether those estimators can be used as our first-step estimator $\hat{\gamma}_{n}$ boils down to their convergence rates: Recall that Assumption \ref{asm:first} requires that $\sqrt{Nh_{n}}(\hat{\gamma}_{n}-\gamma)=o_{p}(1)$, which implies that the first-step estimator needs to converge faster than $\hat{\beta}_{n}$. We can conjecture that, without additional conditions on $\eta_{ijt}$, the convergence rates of those estimators are $\sqrt{n}$ in worst cases due to the conditional dependence across dyads. Obviously, $\sqrt{n}$-rate is incompatible with Assumption \ref{asm:first}. In the following, we discuss what kind of additional conditions are needed to ensure Assumption \ref{asm:first}.
  	 
  	 We may assume additive separability for $\eta_{ijt}$: $\eta_{ijt}=V_{it}+V_{jt}+V_{ijt}$, where conditionally on $\{\xi_{i}\}_{i=1}^{n}$, $(V_{i1},V_{i2}),i=1,...,n$ is independent, $(V_{ij1},V_{ij2}),1\leq i<j\leq n$ is independent, and both are mutually independent. This assumption is weaker than assuming $(\eta_{ij1},\eta_{ij2}),1\leq i<j\leq n$ is conditionally independent given $\{\xi_{i}\}_{i=1}^{n}$, where $V_{it}$ is treated as degenerate. With additional conditions, we can directly apply \cite{Graham2017}'s joint maximum likelihood estimator or \cite{Candelaria2020}'s semiparametric estimator, both of which leverage the cross-sectional variation in $d_{ijt}$ and $R_{ijt}$. We can show that in our setting (especially Assumptions \ref{asm:data} and $\ref{asm:moments}$), the limiting networks are dense, which implies that both \cite{Graham2017} and \cite{Candelaria2020}'s estimators satisfy $\sqrt{N}(\hat{\gamma}_{n}-\gamma)=O_{p}(1)$ and Assumption \ref{asm:first}.
  	 
  	 Alternatively, we may assume that $(\eta_{ij1},\eta_{ij2}),1\leq i<j\leq n$ is conditionally independent given $\{\xi_{i}\}_{i=1}^{n}$ and $\varphi(B_{i},B_{j})=B_{i}+B_{j}$.
	  \cite{Graham2017} and \cite{Candelaria2020}'s estimators still satisfy Assumption \ref{asm:first}, but we can also show that \cite{Chamberlain1980}'s conditional logit estimator and \cite{Horowitz1992}'s smoothed maximum score estimator can satisfy Assumption \ref{asm:first}. Under the conditional independence assumption, the latter two estimators can be written in an asymptotically locally linear form where the corresponding influence function is indexed by $ij$ with $0$ covariances. Thus, the convergence rates are based on $N$ and Assumption \ref{asm:first} can be satisfied depending on the tuning parameters.
  	 
\section{Extension}
  	 \subsection{Directed Graph with Multiple Periods}\mbox{}\\
  	 \indent In the above analysis, we restricted our attention to an undirected graph; the variables are all symmetric with respect to nodes (e.g., $Y_{ijt}=Y_{jit}$). Also, there were only two time periods, $t=1,2$. The extension to a directed graph case with $t=1,...,T\, (T\geq 2)$ is straightforward; Letting $\Delta_{st}A\equiv A_{s}-A_{t}$ denote the time difference between $s$ and $t$, we propose the following estimator:
  	 \begin{align*}
  	 	\hat{\beta}_{n}=&\left[\sum_{s<t}\sum_{i=1}^{n}\sum_{j\neq i}d_{ijs}d_{ijt}\Delta_{st}W_{ij}\Delta_{st}W_{ij}'K_{h_{n}}(\Delta_{st}R_{ij}'\hat{\gamma}_{n})\right]^{-1}\\
  	 	&\times\left[\sum_{s<t}\sum_{i=1}^{n}\sum_{j\neq i}d_{ijs}d_{ijt}\Delta_{st}W_{ij}\Delta_{st}Y_{ij}K_{h_{n}}(\Delta_{st}R_{ij}'\hat{\gamma}_{n})\right].
  	 \end{align*}
   	 All the results and their proofs are valid with some modification because we can always rewrite the double sum $\sum_{i=1}^{n}\sum_{j\neq i}A_{ij}$ as $\sum_{i<j}(A_{ij}+A_{ji})$ for any variables $\{A_{ij}\}$. We will use this version of the estimator in our empirical application.
   	 
   	 \subsection{Pairwise Fixed Effects}\mbox{}\\
   	 \indent In the model (\ref{model:structural}) and (\ref{model:firststep}), all the fixed effects are node-wise. Since we are interested in coefficients on time-varying dyadic variables, it is possible to include pairwise fixed effects $A_{ij}$ and $B_{ij}$ in each equation, additionally to $A_{i},A_{j}$ and $B_{i},B_{j}$. Clearly, with pairwise fixed effects, the identification and estimator will be the same as with node-wise fixed effects since we are leveraging the time variation. Thus, a similar asymptotic analysis will also hold as long as $(A_{ij},B_{ij}),1\leq i<j\leq n$ are independently distributed conditionally on $\{\xi_{i}\}_{i=1}^{n}$. 
  		
  	 Alternatively, we can also do away with the additive separability by incorporating node-wise fixed effects into pairwise ones:
  	 \begin{align*}
  	 A_{ij}=\tilde{\tau}\left(\tilde{A}_{i},\tilde{A}_{j},\tilde{A}_{ij}\right),
  	 \end{align*}
   	 where $\tilde{\tau}$ is some unknown function, $\tilde{A}_{i}$ is a node-wise fixed effect, and $\tilde{A}_{ij}$ is a pairwise fixed effect. We can impose a similar structure for $B_{ij}$.  Again, the asymptotic analysis will hold as long as $(\tilde{A}_{ij},\tilde{B}_{ij}),1\leq i<j\leq n$ are conditionally independent. With a more general dependence structure, we could show a similar asymptotic result using \cite{Kojevnikov2020}'s central limit theorem for $\psi$-dependent data.

   	 \subsection{Sparsity}\mbox{}\\
	 \indent Above, we argue that our model and assumptions imply that the limiting networks $D_{1}$ and $D_{2}$ are locally dense around $\Delta R_{ij}'\gamma\sim 0$. Thus, we limit our attention to cases where the number of dyads in the sample must be proportional to $N$. Our modeling is appropriate in some applications, such as trade or migration, where the number of dyads is rather dense. However, ours can be inappropriate for some applications where the networks are sparse such as employee-employer, bank-firm matched data (e.g., \cite{Abowd1999}, \cite{jimenez2014}).
	   	
	 We can accommodate sparse networks by the following modification; let us modify Assumption \ref{asm:data} so that $\xi_{i},i=1,...,n$ are drawn from some distribution that is allowed to depend on $n$. For example, as argued in \cite{Graham2017}, we can consider a distribution where the fixed effects are such that $\liminf_{1\leq i\leq n}B_{i}=-\infty$. Then, we can discuss identification and estimation with fixed $n$, and the moments of interest are all dependent on $n$. Especially, we can consider the sequence of networks such that $Pr(d_{121}d_{122}=1|\Delta R_{12}'\gamma=0)\to 0$ and $r_{n}Pr(d_{121}d_{122}=1|\Delta R_{12}'\gamma=0)=\Omega(1)$ for some $r_{n}\to\infty$ to incorporate sparsity. We do not pursue sparsity in this paper and leave it for future projects.
 
\section{Simulation}
	To see the performance of the estimator, we conduct some simulation exercises. Consider the following data-generating process:
	\begin{align*}
		&W_{ijt}=X_{it}+X_{jt}, R_{it}=(W_{ijt},Z_{it}+Z_{jt})',\\
		&A_{i}=\frac{X_{i1}+ X_{i2}}{2}, B_{i}=\frac{Z_{i1}+ Z_{i2}}{2},\\
		&d_{ijt}=\boldsymbol{1}\{R_{ijt}'(1,1)'+\theta\times (B_{i}+B_{j})-\eta_{ijt}\geq 0\},\\
		&Y_{ijt}=d_{ijt}(W_{ijt}+A_{i}+A_{j}+\epsilon_{ijt})
	\end{align*}
	where 
	\begin{align*}
		&X_{it},Z_{it}\sim \mathcal{N}(2,1), i.i.d. \text{ across }i,t,\\
		&\eta_{ijt}\sim Logistic(0,1), i.i.d.\text{ across }ij,t,\\
		&\epsilon_{ijt}=U_{it}+U_{jt}+\eta_{ijt},\text{ where }U_{it}\sim\mathcal{N}(0,\sigma),i.i.d.\text{ across }i,t.
	\end{align*}
	Note that $\beta=1$ and $\gamma=(1,1)'$. We have $\theta\in\{-0.3,-2.0,-3.0\}$ inside of $d_{ijt}$ to control for the fraction of zeros in the simulated data set:
	\begin{align*}
		Pr(d_{121}\times d_{122}=0)\sim \begin{cases}
			20\%\text{ if }\theta =-0.3\\
			75\%\text{ if }\theta=-2.0\\
			90\%\text{ if }\theta =-3.0
			\end{cases}.
	\end{align*}
	We also change $\sigma\in\{0.0,1.0\}$ for $U_{it}$ so that $\sigma=0.0$ ($\sigma=1.0$) corresponds to the degenerate (non-degenerate) case.
	
	As described above, we can interpret this data-generating process as a log-linearized version of the canonical gravity model (\cite{Head2014}); by writing $\tilde{Y}_{ijt}$ as an observable outcome, we redefine the main equation as
	\begin{align*}
		\tilde{Y}_{ijt}=exp(W_{ijt}+A_{i}+A_{j})\times \underbrace{\eta_{ijt}}_{=d_{ijt}exp(\epsilon_{ijt})}.
	\end{align*}
	We can take a log and recover the original model for a unit with $d_{ijt}=1$. This modeling allows a mass at $\tilde{Y}_{ijt}=0$, one important feature of dyadic data.
	
	We conduct experiments for $n\in\{50,100,150,200\}$, $\theta\in\{-0.3,-2.0,-3.0\}$, and $\sigma\in\{0.0,1.0\}$, and iterate $2000$ times for each one. We calculate $\hat{\gamma}_{n}$ by \cite{Chamberlain1980}'s conditional logit estimator:
	\begin{align*}
		\hat{\gamma}_{n}=\underset{g\in\mathcal{G}}{argmax}\sum_{i<j:d_{ij1}+d_{ij2}=1}M_{ij}(g)
	\end{align*}
	where $\mathcal{G}$ is a compact subset of $\mathbb{R}^{q_{r}}$ and
	\begin{align*}
		M_{ij}(g)=\boldsymbol{1}\{d_{ij1}=1\}\ln \left(\frac{exp(\Delta R_{ij}'g)}{1+exp(\Delta R_{ij}'g)}\right)+\boldsymbol{1}\{d_{ij2}=1\}\ln \left(\frac{1}{1+exp(\Delta R_{ij}'g)}\right).
	\end{align*}
	
	For $\hat{\beta}_{n}$, we use a biweight kernel for $K(\cdot)$, given by $K(x)=15/16(1-x^{2})^{2}\boldsymbol{1}\{|x|\leq 1\}$. This choice implies that we assume that the smoothness of the model is given by $k=2$. We set $\delta=0.4$ and $h=3.0$ and calculate each estimator and confidence interval according to the inference procedure discussed above.
	
	For comparison, we calculate the fixed effect estimator $\hat{\beta}_{FE}$ given by (\ref{eq:beta_fe}).
	The standard error is calculated by $\hat{\Sigma}$, with $K_{h_{n}}(\cdot)$ replaced by $1$. We also calculate the Poisson pseudo-maximum-likelihood (PPML) estimator $\hat{\beta}_{PPML}$ given by (\ref{eq:beta_ppml}).
	We compute $\hat{\beta}_{PPML}$ and its standard error by the \textit{penppml} package in \textit{R} \citep{penppml}. The standard error is clustered at the node level, which is close to $\hat{\Sigma}_{WW}^{-2}\hat{\Sigma}_{W\nu,1}$ in our setting \citep{Graham2020}.
	
	The result is summarized in the following TABLE 1 and 2. In TABLE 1, we evaluate the three estimators by mean and median biases (MeanBias), root mean square error (RMSE) for $\sigma=0,1$. In TABLE 2, we compute 95\% coverage probabilities (Coverage) of four different confidence intervals: $CI_{conv}$ (conventional CI from $\hat{\hat{\beta}}_{n}$ and $\hat{\Sigma}$), $CI_{bc}$ (bias-corrected CI given by $CI_{L,0.05}$ and $CI_{U,0.05}$), $CI_{FE}$ (conventional CI from $\hat{\beta}_{FE}$ and $\hat{\Sigma}$ with a flat kernel.), and $CI_{PPML}$ (conventional CI from $\hat{\beta}_{PPML}$ and its node-level clustered standard error).
	
	From TABLE \ref{table:1} and \ref{table:2}, we can see that our estimator performs better than the fixed effect estimator and the PPML estimator in terms of bias, which shows that the weights given by the first step estimator work well in eliminating the bias. Our estimator also outperforms the competitors regarding RMSE, which implies that the loss in precision is not severe. Our estimator also performs well even when there is a large fraction of zeros in $Y$ ($Pr(D_{ij1}\times D_{ij2})\sim 90\%$ when $\theta=-3.0$). There is little difference between $\sigma=0$ and $\sigma=1$ other than added variances in the estimators.
	
	From TABLE \ref{table:3} and \ref{table:4}, we can see that $CI_{bc}$ is close to 95\% regardless of the degeneracy ($\Sigma_{W\nu,1}=0$ or $>0$) while the others are off from the targeted nominal coverage. This result confirms the effectiveness of the bias correction strategy as well as the adaptivity of our variance estimator, as claimed in Section 3.3. Also, it is notable to see that the bias correction is important for obtaining correct coverage probabilities even though the asymptotic bias is $0$ in the case of $\sigma=1.0$ so that $\Sigma_{W\nu,1}>0$ (Theorem \ref{thm:normality}) and  $CI_{conv}$ would return an asymptotically correct coverage.
	
	\begin{table}[htbp]\centering
		\caption{Finite sample properties of $\hat{\beta}_{n}$, $\hat{\beta}_{FE}$, and $\hat{\beta}_{PPML}$ }
		\begin{subtable}{\linewidth}\centering
		\caption{$\sigma=1.0$}
		\begin{tabular}{cc|ccc|ccc}
			\toprule
			&  & \multicolumn{3}{l}{\hfil MeanBias}  &\multicolumn{3}{l}{\hfil RMSE}  \\
			$\theta$ &  $n$   & $\hat{\beta}_{n}$ & $\hat{\beta}_{FE}$ &$\hat{\beta}_{PPML}$  &   $\hat{\beta}_{n}$ & $\hat{\beta}_{FE}$ & $\hat{\beta}_{PPML}$ \\
			\midrule
			-0.3 &  50 &      0.045 &      0.133 & 0.185 &    0.122 &     0.160 &0.430 \\
			-2.0 &  50 &      0.141 &      0.352 & 0.467 &     0.210 &  0.377 &  0.617\\
			-3.0 &  50 &      0.162 &      0.369 & 0.582 & 0.273 &     0.415 & 0.752 \\
			\midrule
			-0.3 & 100 &      0.038 &       0.136 & 0.195 &  0.087  &      0.148 & 0.376\\
			-2.0 & 100 &      0.099 &       0.349 & 0.438 &      0.142 &      0.359 & 0.536\\
			-3.0 & 100 &      0.117 &     0.359 & 0.542 &     0.184  &      0.378 &0.657 \\
			\midrule
			-0.3 & 150 &      0.028 &       0.135 & 0.193&      0.070  &      0.143 & 0.327\\
			-2.0 & 150 &      0.075 &       0.346 & 0.427 &      0.112 &    0.353 &0.496\\
			-3.0 & 150 &      0.095 &       0.356 & 0.527 &    0.145 &     0.367 & 0.607\\
			\midrule
			-0.3 & 200 &      0.024 &     0.134 &  0.193  &    0.060 &     0.140  & 0.305\\
			-2.0 & 200 &      0.061 &     0.344 & 0.417&      0.091 &    0.348 & 0.471\\
			-3.0 & 200 &      0.076 &     0.352 &  0.510 &   0.118 &    0.360 & 0.572\\
			\bottomrule
			\label{table:1}
		\end{tabular}
		\end{subtable}\\

		\begin{subtable}{\linewidth}\centering
		\caption{$\sigma=0.0$}
		\begin{tabular}{cc|ccc|ccc}
			\toprule
			&  & \multicolumn{3}{l}{\hfil MeanBias} & \multicolumn{3}{l}{\hfil RMSE}  \\
			$\theta$ &  $n$   & $\hat{\beta}_{n}$ & $\hat{\beta}_{FE}$ & $\hat{\beta}_{PPML}$ & $\hat{\beta}_{n}$ & $\hat{\beta}_{FE}$ & $\hat{\beta}_{PPML}$\\
			\midrule
			-0.3 &  50 &      0.048 &     0.134 & 0.194 &     0.082 &   0.142 & 0.399\\
			-2.0 &  50 &      0.140 &     0.352 & 0.468  &     0.176 &     0.365 &0.586\\
			-3.0 &  50 &      0.161 &      0.369 &  0.581 &      0.229 &   0.397 & 0.714\\
			\midrule
			-0.3 & 100 &      0.037 &      0.135 &  0.193 &    0.053 &    0.138 &0.332\\
			-2.0 & 100 &      0.093 &      0.348 &  0.438 &    0.110 &   0.352 & 0.508\\
			-3.0 & 100 &      0.113 &      0.359 &  0.546 &     0.145 &     0.368 & 0.630\\
			\midrule
			-0.3 & 150 &      0.028 &     0.135 &   0.191 &     0.039 &     0.136 &0.301\\
			-2.0 & 150 &      0.071 &      0.345 &  0.427 &     0.082 &     0.348 &0.477\\
			-3.0 & 150 &      0.089 &      0.354 & 0.529  &   0.108 &   0.359 &0.592\\
			\midrule
			-0.3 & 200 &      0.024 &       0.135 &  0.191  &    0.031 &    0.136 & 0.278\\
			-2.0 & 200 &      0.058 &     0.345 &  0.415 &     0.067 &      0.347 & 0.451\\
			-3.0 & 200 &      0.074 &        0.355 & 0.508 &   0.089 &       0.358 &0.553\\
			\bottomrule
			\label{table:2}
		\end{tabular}
	\end{subtable}
	\caption*{\footnotesize{Note: MeanBias reports the mean of the difference between the corresponding estimator and the true value $\beta=1$. RMSE reports the root mean square error. $\hat{\beta}_{n}$ is our proposed estimator, $\hat{\beta}_{FE}$ is the fixed effect estimator, and $\hat{\beta}_{PPML}$ is the Poisson pseudo-maximum-likelihood estimator.}}
	\end{table}

	\begin{table}[htbp]
		\centering
		\caption{$95\%$ coverage probabilities of $CI_{conv},CI_{bc}$, $CI_{FE}$, and $CI_{PPML}$}
		\begin{subtable}{\textwidth}
		\centering
		\caption{$\sigma=1.0$}
		\begin{tabular}{cc|cccc}
			\toprule
			& &\multicolumn{3}{l}{\hfil Coverage} \\
			$\theta$ &   n & $CI_{conv}$ & $CI_{bc}$ &$CI_{FE}$ & $CI_{PPML}$ \\
			\midrule
			 -0.3 &  50 &      0.790 &       0.961 &       0.498 & 0.537\\
			-2.0 &  50 &      0.646 &       0.963 &       0.150 &0.236\\
			-3.0 &  50 &      0.640 &       0.901 &       0.311 & 0.211\\
			\midrule
			-0.3 & 100 &      0.785 &       0.978 &       0.233 &0.498\\
			-2.0 & 100 &      0.668 &       0.970 &       0.011 &0.173\\
			-3.0 & 100 &      0.674 &       0.953 &       0.072 &0.143\\
			\midrule
			-0.3 & 150 &      0.790 &       0.971 &       0.103 &0.472\\
			-2.0 & 150 &      0.689 &       0.949 &       0.001 &0.117\\
			-3.0 & 150 &      0.688 &       0.944 &       0.016 &0.09\\
			\midrule
			-0.3 & 200 &      0.817 &       0.964 &       0.040&0.426 \\
			-2.0 & 200 &      0.730 &       0.947 &       0.000 & 0.08\\
			-3.0 & 200 &      0.720 &       0.946 &       0.004 & 0.08\\
			\bottomrule
			\label{table:3}
		\end{tabular}
	\end{subtable}
	\begin{subtable}{\textwidth}
		\centering
		\caption{$\sigma=0.0$}
		\begin{tabular}{cc|cccc}
			\toprule
			& &\multicolumn{3}{l}{\hfil Coverage} \\
			$\theta$ &   n & $CI_{conv}$ & $CI_{bc}$ &$CI_{FE}$ & $CI_{PPML}$ \\
			\midrule
			 -0.3 &  50 &      0.698 &       0.918 &       0.141 &0.547\\
			-2.0 &  50 &      0.535 &       0.935 &       0.026 &0.204\\
			-3.0 &  50 &      0.592 &       0.869 &       0.168 &0.182\\
			\midrule
			-0.3 & 100 &      0.655 &       0.960 &       0.001 &0.515\\
			-2.0 & 100 &      0.482 &       0.958 &       0.000 &0.12\\
			-3.0 & 100 &      0.571 &       0.944 &       0.004 &0.106\\
			\midrule
			-0.3 & 150 &      0.673 &       0.977 &       0.000 &0.45\\
			-2.0 & 150 &      0.471 &       0.945 &       0.000 &0.073\\
			-3.0 & 150 &      0.532 &       0.949 &       0.001 &0.065\\
			\midrule
			-0.3 & 200 &      0.660 &       0.970 &       0.000 &0.407\\
			-2.0 & 200 &      0.444 &       0.939 &       0.000 &0.052\\
			-3.0 & 200 &      0.520 &       0.933 &       0.000 &0.047\\
			\bottomrule
			\label{table:4}
		\end{tabular}
	\end{subtable}
	\caption*{\footnotesize{Note: Coverage reports the coverage probabilities of the corresponding confidence intervals. $CI_{conv}$ is the proposed confidence interval without bias correction, $CI_{bc}$ is the bias-corrected confidence interval given by $CI_{L,0.05}$ and $CI_{U,0.05}$. $CI_{FE}$ is the conventional confidence interval from $\hat{\beta}_{FE}$ and $\hat{\Sigma}$ with a flat kernel. $CI_{PPML}$ is the conventional confidence interval from $\hat{\beta}_{PPML}$ and its node-level clustered standard error.}}
	\end{table}

\section{Empirical example}
	\subsection{Background}\mbox{}\\
	\indent As a leading application of our model, consider \cite{Moretti2017}. They study how state-level tax differences affect migration by top scientists in 
	the U.S. Specifically, they estimate the following model implied by their economic theory:
	\begin{align*}
		log(P_{ijt}/P_{iit})&=\eta \left[log(1-\tau_{jt})-log(1-\tau_{it})\right]\\
		&+\eta'\left[log(1-\tau_{jt}')-log(1-\tau_{it}')\right]+\gamma_{j}+\gamma_{i}+u_{ijt},
	\end{align*}
	where $P_{ijt}$ is the number of scientists migrating to state $j$ from state $i$ at year $t$, $\tau_{it}$ and $\tau_{it}'$ are 
	personal and corporate taxes imposed in state $i$ at year $t$, $\gamma_{i}$ is a state fixed effect, and 
	$u_{ijt}$ is an error term.
	
	Note that if there is no migration from $i$ to $j$ at year $t$, $P_{ijt}=0$ and $log(P_{ijt}/P_{iit})$ is undefined. In \cite{Moretti2017}'s dataset,
	more than 70\% of state-pairs exhibit no migration flow:
	\begin{figure}[htbp]
		\centering
		\caption{Fraction of positive migration flows in \cite{Moretti2017}'s dataset.}
		\includegraphics[width=16.0cm]{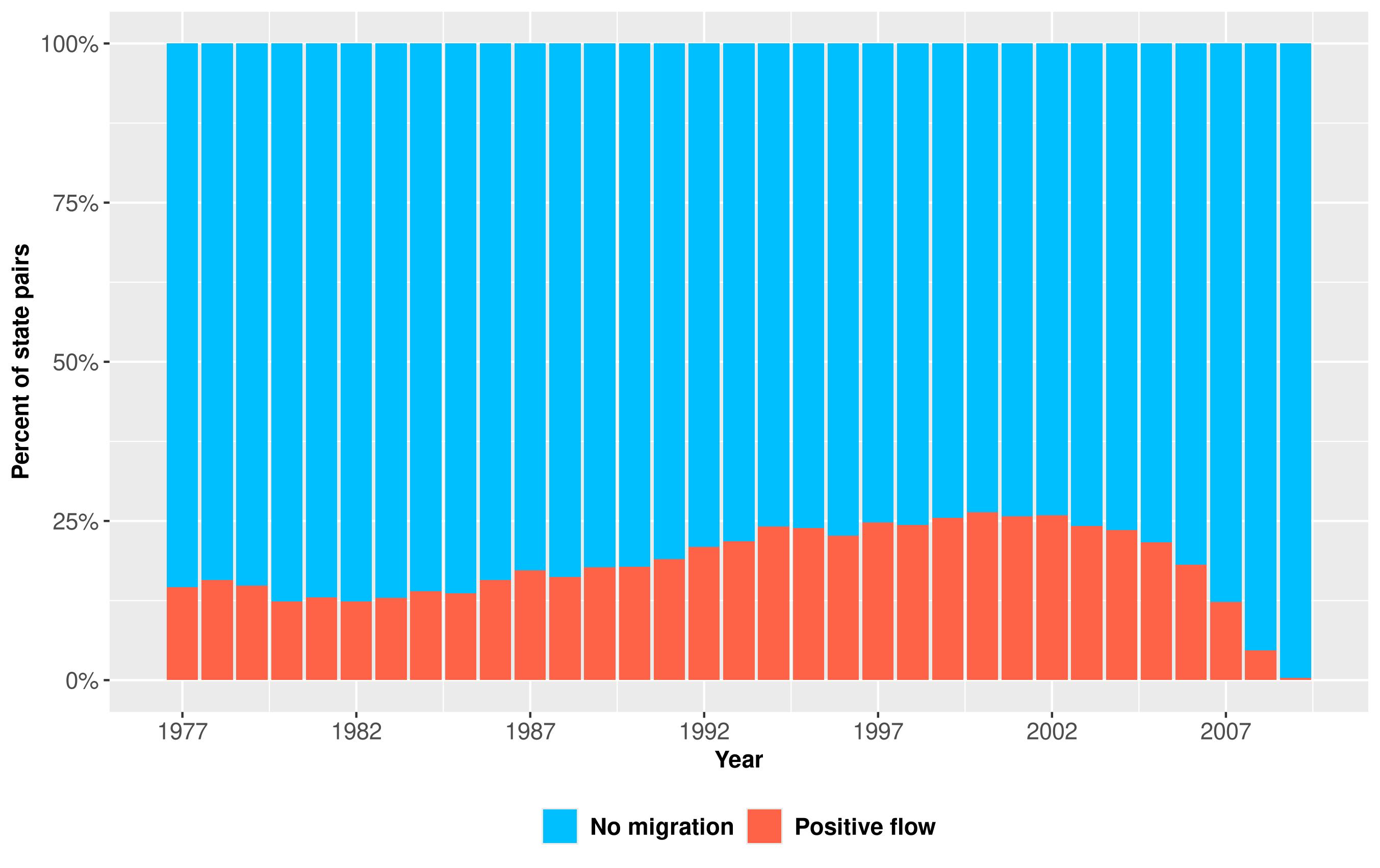}
		\label{fig:zero}
		\caption*{\footnotesize{Note: The migration flow is positive in a given year if there is at least one scientist moving 
			from state $i$ to $j$ (scientists are "star"; They are at or above 95\% quantile in 
			number of patents over the past ten years)}}
	\end{figure}
	When running a regression, they are concerned with a potential sample selection bias stemming from these undefined outcomes. They argue that if the main regressors 
	are not systematically associated with the probability of positive migration flows, the selection bias should be minimal. Running OLS on the linear probability model,
	they find little correlation between the main regressors and no flow.  Re-estimating their model with our method provides a check on the validity of their argument and the appropriateness of using the linear probability model.
	
	When applying our model to their context, we must consider what $R_{ijt}$ should be. Since \cite{Moretti2017}'s underlying theory is based on scientists' and firms' discrete choice,
	one consistent way to generate zero migration flows between some states is to consider endogenous choice sets as in \cite{Dube2021}. Formally, we can write the choice set of representative scientists in state $i$ as $C_{it}=\{j\in\{1,...,51\}:d_{ijt}=1\}$. Here, $\{d_{ijt}\}$ represents the job-market network; if $d_{ijt}=1$, it is possible to move from $i$ to $j$, and vice versa. We can attribute the determinants of the network to the utilities and profits of scientists and firms, as well as the matching costs between the two parties. Such costs are not present in the structural equation if those costs are not compensated through wages; The structural equation consists of the determinants of log wage differences between two states. Thus in the selection equation (\ref{model:firststep}), in addition to $W_{ijt}$, we can include variables in $R_{ijt}$ that capture non-monetary matching costs between two states $i$ and $j$, which does not violate Assumption \ref{asm:inv} as $R_{ijt}$ satisfies the exclusion restriction. 
	
	\subsection{Implementation}\mbox{}\\
	\indent For $W_{ijt}$, as in \cite{Moretti2017}, we include the state-to-state differences in (i) an individual income average income tax rate (ATR) faced by a hypothetical taxpayer at 99\% quantile of the national income distribution, (ii) 
	the corporate tax rate (CIT), (iii) the investment tax credit (ITC), and (iv) the R\&D tax credit (R\&D credit). This is the same set of regressors as \cite{Moretti2017}'s baseline regression. For $R_{ijt}$, we use $W_{ijt}$ plus state-to-state difference in the logarithm of population (POP) and a dummy variable that indicates whether $i$ and $j$ share their governors' political parties (GOV). The additional variables in $R_{ijt}$ arguably measure non-monetary costs of connecting firms and workers in two states. 
	
	We implement the first step estimation as follows. We use the conditional logit estimator extended to a directed graph with multi-periods case, which is given by
	\begin{align*}
		\hat{\gamma}_{n}=\underset{g\in\mathcal{G}}{argmax}\sum_{s<t}\sum_{i,j}M_{ij,st}(g),
	\end{align*}
	where $\mathcal{G}$ is a compact subset of $\mathbb{R}^{q_{r}}$ and
	\begin{align*}
		M_{ij,st}(g)&=\boldsymbol{1}\{d_{ijs}+d_{ijt}=1\}\times \left[\boldsymbol{1}\{d_{ijs}=1\}ln (e_{ij,st})+\boldsymbol{1}\{d_{ijt}=1\}ln(1-e_{ij,st})\right],\\
		e_{ij,st}&=\frac{exp(\Delta_{st}R_{ij}'g)}{1+exp(\Delta_{st}R_{ij}'g)}.
	\end{align*}
	TABLE \ref{table:5} reports the first step estimation result. We can see that the coefficients on the newly added variables $GOV$ and $POP$ deviate from zero, which implies that a part of the identification assumptions (Assumption \ref{asm:inv}) is satisfied. Also, for these two variables, the estimated coefficients imply that the job market network exhibits homophily; similar states are more likely to be connected.
	\begin{table}[htbp]
		\centering
			\caption{First Step Estimation Result}
			\begin{tabular}{cc}
				\toprule
				Variable & $\hat{\gamma}_{n}$\\
				\midrule
				GOV & 0.162 \\
				POP & -13.144\\
				ATR & 8.561 \\
				CIT & 5.850 \\
				ITC & 5.350 \\
				R$\&$D credit & -0.230 \\
				\bottomrule
			\end{tabular}
		\label{table:5}
		\caption*{\footnotesize{Note: The first column reports the variable names included in $R_{ijt}$. The second column reports the estimated coefficients corresponding to each variable.}}
	\end{table}
	
	For the second step estimator, we use our $\hat{\beta}_{n}$ defined above and extend it to the directed graph with multiple period cases, as discussed in Section 4.
	We use a biweight kernel for $K(\cdot)$ (so $k=2$), choose $h=3.0$ as an initial constant for pilot bandwidths, and
	use $\delta=0.4$ for calculating $h_{n,\delta}$. We extend and use $\hat{\Sigma}$ to calculate the standard error while taking into account the correlation across time (case 1). We also calculate the bias-corrected $95\%$-confidence interval by computing $CI_{L,0.05}$ and $CI_{U,0.05}$ as defined above. Also, we list $\hat{\beta}_{MW}$, an estimate from \cite{Moretti2017} (page 1883, TABLE 2A, specification (3)) and calculate the conventional $95\%$-confidence interval based on their standard errors. Note that $\hat{\beta}_{MW}$ is the fixed estimator, where its standard error is calculated by clustering across time and the origin, destination, and origin-destination pairs. 
	
	We summarize the result in TABLE \ref{table:6}. We can see that $\hat{\beta}_{n}$ returns similar values as $\hat{\beta}_{WM}$, which claims the robust positive effect of 
	income and corporate-related tax differences on migration. Thus, \cite{Moretti2017}'s estimates are not likely to be qualitatively affected by the sample selection effects. However, while \cite{Moretti2017}'s estimates are statistically significant at $5\%$ level, our confidence intervals show that all of the estimates are no longer statistically significant at that level except for ITC. Our insignificance result is driven by both the increase in standard errors \footnote{Our standard error is from $\hat{\Sigma}$, which takes fully into account the dependence among pairs that share origin and destination, such as California$\to$Wisconsin and New York$\to$California. \cite{Moretti2017}'s standard error calculation ignores such dependence structure.} and the asymptotic bias correction. Thus, our exercise shows that some of the results in \cite{Moretti2017} may not be robust to the presence of sample selection due to the endogeneity of the job market network.

	\begin{table}[htbp]
		\centering
		\caption{Comparison of our estimator and \cite{Moretti2017}}
		\begin{subtable}{0.50\textwidth}
			\centering
			\caption{This paper}
			\begin{tabular}{ccc}
				\toprule
				Variable & Estimator (s.e) & CI \\
				\midrule
				ATR & 1.634 & $[-2.656,5.872]$ \\
					& (1.886) &  \\
				CIT & 1.666 & $[-2.948,6.276]$ \\
					& (2.040) & \\
				ITC & 1.980 & $[0.651, 3.290]$ \\
					& (0.584) &     \\
				R$\&$D credit & 0.429 & $[-1.352,2.117]$ \\
					& (0.780) &      \\
				\bottomrule
			\end{tabular}
		\end{subtable}%
		\begin{subtable}{0.50\textwidth}
			\centering
			\caption{ \cite{Moretti2017}}
			\begin{tabular}{cc}
				\toprule
				 Estimator (s.e) & CI\\
				\midrule
				 1.926 & $[0.918,2.933]$ \\
				 (0.514) & \\
				1.840 & $[0.687,2.992]$ \\
				(0.588) & \\
				1.793 & $[0.987,2.598]$ \\
				(0.411) & \\
				 0.368 & $[0.011, 0.724]$ \\
				 (0.182) & \\
				\bottomrule
			\end{tabular}
		\end{subtable}
		\label{table:6}
		\caption*{\footnotesize{Note: The first column reports the variable names included in $W_{ijt}$. The second column reports the estimates and standard errors (in parentheses) corresponding to each variable. The third column reports the $95\%$-confidence interval. The left table is our result, and the right table is \cite{Moretti2017}'s result (see page 1883, TABLE 2A, specification (3)).}}
	\end{table}
\section{Conclusion}
This paper studies identification and inference of a panel dyadic data sample selection model. We show that \cite{Kyriazidou1997}'s identification strategy can be extended to our dyadic data setting, and we prove asymptotic normality of the proposed estimator.

Our estimator has some appealing properties. The distributional result implies that our estimator has the same convergence rates as the usual estimators used in practice in the non-degenerate case, and there is no loss of effective sample size for using our nonparametric type estimator. Also, our estimator is guaranteed to be asymptotically normal, while others can be non-Gaussian in the limit.

We also provide consistent estimators for asymptotic bias and variance that adapts to the degeneracy. Specifically, the bias-corrected confidence interval has an asymptotically correct size. Our simple simulation exercise confirms the validity of these estimators and highlights the importance of bias correction in both degenerate and non-degenerate cases.

\newpage

\bibliographystyle{ecta}
\bibliography{citation}
\newpage
\appendix
\section*{Appendix A. Proofs}
 	\subsection*{Proof of Theorem \ref{thm:normality}}
 	\begin{proof}
 		First, consider the infeasible version of $\beta_{n}$, where $\hat{\gamma}_{n}$ is replaced by the true $\gamma$:
 		\begin{align*}
 			\tilde{\beta}_{n}=\beta+S_{WW}^{-1}S_{W\lambda}+S_{WW}^{-1}S_{W\nu},
 		\end{align*}
 		where $S_{WW},S_{W\lambda},$ and $S_{W\nu}$ are the same as $\hat{S}_{WW},\hat{S}_{W\lambda},$ and $\hat{S}_{W\nu}$ except $\hat{\gamma}_{n}$ replaced by $\gamma$. 
 		We use the following lemmas
 		\begin{lem}
 			Suppose Assumptions \ref{asm:data}-\ref{asm:band} hold. Then, 
 			\begin{align*}
 				S_{WW}\to_{p}\Sigma_{WW},
 			\end{align*}
 			as $n\to\infty$.
 			\label{lem:S_WW}
 		\end{lem}
 		\begin{lem}
 			Suppose Assumptions \ref{asm:data}-\ref{asm:band} hold. Fix some $h\in(0,\infty)$. If $Nh_{n}^{2k+3}\to h$, then
 			\begin{align*}
 				\sqrt{Nh_{n}}S_{W\lambda}\to_{p}\sqrt{h}\Sigma_{W\lambda},
 			\end{align*}
 			as $n\to\infty$. If $Nh_{n}^{2k+3}\to\infty$ and $nh_{n}^{2k}\to\infty$, then
 			\begin{align*}
 				h_{n}^{-(k+1)}S_{W\lambda}\to_{p}\Sigma_{W\nu},
 			\end{align*}
 			as $n\to\infty$.
 			\label{lem:S_Wl}
 		\end{lem}
 		\begin{lem}
 			Suppose Assumptions \ref{asm:data}-\ref{asm:band} hold. Fix an arbitrary non-zero vector $c\in\mathbb{R}^{q_{w}}$ and some constant $h\in(0,\infty]$. Let $c_{W}=\Sigma_{WW}^{-1}c$. If $c_{W}'\Sigma_{W\nu,1}c_{W}>0$ and $Nh_{n}^{2k+3}\to h$, then
 			\begin{align*}
 				\sqrt{n}c'S_{WW}^{-1}S_{W\nu }\to_{d}\mathcal{N}(0,c_{W}'\Sigma_{W\nu,1}c_{W}),
 			\end{align*}
 			as $n\to\infty$. If $c_{W}'\Sigma_{W\nu,1}c_{W}=0$ and $Nh_{n}^{2k+3}\to h$, then
 			\begin{align*}
 				\sqrt{Nh_{n}}c'S_{WW}^{-1}S_{W\nu} &\to_{d}\mathcal{N}(0,c_{W}'\Sigma_{W\nu,2}c_{W}),
 			\end{align*}
 			as $n\to\infty$.
 			\label{lem:S_Wnu}
 		\end{lem}
 		
 		By combining Lemmas \ref{lem:S_WW}-\ref{lem:S_Wnu}, the statement of Theorem \ref{thm:normality} follows for $\tilde{\beta}_{n}$. The following lemmas are used to show the negligibility of $\hat{\beta}_{n}-\tilde{\beta}_{n}$:
 		\begin{lem}
 			Suppose Assumptions \ref{asm:data}-\ref{asm:first} hold. Fix some constant $h\in(0,\infty]$. If $Nh_{n}^{2k+3}\to h$, then,
 			\begin{align*}
 				\hat{S}_{WW}=S_{WW}+o_{p}(1).
 			\end{align*}
 			\label{lem:hS_WW}
 		\end{lem}
 		\begin{lem}
 			Suppose Assumptions \ref{asm:data}-\ref{asm:first} hold. Fix some constant $h\in(0,\infty]$. If $Nh_{n}^{2k+3}\to h$, then
 			\begin{align*}
 				\hat{S}_{W\lambda}=S_{W\lambda}+o_{p}\left(\frac{1}{\sqrt{Nh_{n}}}\right).
 			\end{align*}
 			\label{lem:hs_Wl}
 		\end{lem}
 		\begin{lem}
 			Suppose Assumptions \ref{asm:data}-\ref{asm:first} hold. Fix some constant $h\in(0,\infty]$. If $Nh_{n}^{2k+3}\to h$, then
 			\begin{align*}
 				\hat{S}_{W\nu}=S_{W\nu} +o_{p}\left(\frac{1}{\sqrt{Nh_{n}}}\right).
 			\end{align*}
 			\label{lem:hS_Wnu}
 		\end{lem}
 		
 		By combining Lemmas \ref{lem:hS_WW}-\ref{lem:hS_Wnu}, we have
 		\begin{align*}
 			\hat{\beta}_{n}-\beta
 			=\tilde{\beta}_{n}-\beta+o_{p}\left(\frac{1}{\sqrt{Nh_{n}}}\right).
 		\end{align*}
 		 Thus, the normalization $r_{n}\in\{\sqrt{n},\sqrt{Nh_{n}},h_{n}^{-(k+1)}\}$ corresponding to each case results in
 		\begin{align*}
 			r_{n}(\hat{\beta}_{n}-\beta)=r_{n}(\tilde{\beta}_{n}-\beta)+o_{p}(1).
 		\end{align*}
 		Since $\tilde{\beta}_{n}$ satisfies the statement of Theorem \ref{thm:normality}, this completes the proof.
 	\end{proof}
 
 	\subsection*{Proof of Proposition \ref{prop:variance}}
 	\begin{proof}
 		We show the claim by the following steps.
 		
 	\subsubsection*{Step 1: $\hat{\Sigma}_{W\nu,2}\to_{p}\Sigma_{W\nu,2}$}\mbox{}\\
 		\indent By expanding $K^{2}(\Delta R_{ij}'\hat{\gamma}_{n}/h_{n})$ around $\Delta R_{ij}'\gamma$, we get
 	\begin{align*}
 		K^{2}(\Delta R_{ij}'\hat{\gamma}_{n}/h_{n}) = K^{2}(\Delta R_{ij}'\gamma/h_{n})+2\Delta R_{ij}'(\hat{\gamma}_{n}-\gamma)/h_{n}k(c_{ij,n}^{*}/h_{n})K(c_{12n}^{*}/h_{n}),
 	\end{align*}
 	where $c_{ij,n}^{*}$ is between $\Delta R_{ij}'\gamma$ and $\Delta R_{ij}'\hat{\gamma}_{n}$ and $k(\cdot)$ is the derivative of $K(\cdot)$.
 	Then, 
 	\begin{align*}
 		\hat{\Sigma}_{W\nu} &=\underbrace{\frac{1}{Nh_{n}}\sum_{i<j} K^{2}(\Delta R_{12}'\gamma/h_{n})d_{ij1}d_{ij2}\Delta W_{ij}\Delta W_{ij}'\Delta \hat{\varepsilon}^{2}_{ij}}_{D_{p1,1}}\\
 		&+\underbrace{\frac{2}{Nh_{n}^{2}}\sum_{i<j}\Delta R_{ij}'(\hat{\gamma}_{n}-\gamma)k(c_{ijn}^{*}/h_{n})K(c_{ijn}^{*}/h_{n})\Delta W_{ij}\Delta W_{ij}'\Delta \hat{\varepsilon}^{2}_{ij}}_{D_{p1,2}}.
 	\end{align*}
 	\subsubsection*{Sub-Step 1: $D_{p1,1}\to_{p}\Sigma_{W\nu,2}$}\mbox{}\\
 	Observe that
 	\begin{align*}
 		\Delta\hat{\varepsilon}_{ij}^{2}-\nu_{ij}^{2}&=\left(\Delta W_{ij}'(\beta-\hat{\beta}_{n})\right)^{2}+\lambda_{ij}^{2}+2\Delta W_{ij}'(\beta-\hat{\beta}_{n})\lambda_{ij}\\
 		&+2\Delta W_{ij}'(\beta-\hat{\beta}_{n})\nu_{ij}+2\lambda_{ij}\nu_{ij}.
 	\end{align*}
 	Thus,
 	\begin{align*}
 		D_{p1,1}&=\frac{1}{Nh_{n}}\sum_{i<j} K^{2}(\Delta R_{ij}'\gamma/h_{n})d_{ij1}d_{ij2}\Delta W_{ij}\Delta W_{ij}'\nu^{2}_{ij}\\
 		&+\frac{1}{Nh_{n}}\sum_{i<j} K^{2}(\Delta R_{ij}'\gamma/h_{n})d_{ij1}d_{ij2}\Delta W_{ij}\Delta W_{ij}'\left(\Delta W_{ij}'(\beta-\hat{\beta}_{n})\right)^{2}\\
 		&+\frac{1}{Nh_{n}}\sum_{i<j} K^{2}(\Delta R_{ij}'\gamma/h_{n})d_{ij1}d_{ij2}\Delta W_{ij}\Delta W_{ij}'\lambda_{ij}^{2}\\
 		&+\frac{2}{Nh_{n}}\sum_{i<j} K^{2}(\Delta R_{ij}'\gamma/h_{n})d_{ij1}d_{ij2}\Delta W_{ij}\Delta W_{ij}'\Delta W_{ij}'(\beta-\hat{\beta}_{n})\lambda_{ij}\\
 		&+\frac{2}{Nh_{n}}\sum_{i<j} K^{2}(\Delta R_{ij}'\gamma/h_{n})d_{ij1}d_{ij2}\Delta W_{ij}\Delta W_{ij}'\Delta W_{ij}'(\beta-\hat{\beta}_{n})\nu_{ij}\\
 		&+\frac{2}{Nh_{n}}\sum_{i<j} K^{2}(\Delta R_{ij}'\gamma/h_{n})d_{ij1}d_{ij2}\Delta W_{ij}\Delta W_{ij}'\lambda_{ij}\nu_{ij}.
 	\end{align*}
 	We call each term by $D_{p1,1}^{i}$ for $i=1,...,6$ that is corresponding to each row.
 	
 	The first term $D_{p1,1}^{1}$ converges to $\Sigma_{W\nu}$. Its expectation coincides with $\Sigma_{W\nu}$ in the limit as
 	\begin{align*}
 		E[D_{p1,1}^{1}]&=\frac{1}{h_{n}}\int E[d_{121}d_{122}\Delta W_{12}\Delta W_{12}'\nu_{12}^{2}|\Delta R_{12}'\gamma=r]K^{2}(r/h_{n})f_{R\gamma}(r)dr\\
 		&=\int E[d_{121}d_{122}\Delta W_{12}\Delta W_{12}'\nu_{12}^{2}|\Delta R_{12}'\gamma=rh_{n}]K^{2}(r)f_{R\gamma}(rh_{n})dr\\
 		&=\Sigma_{W\nu,2}+o(1),
 	\end{align*}
 	where the last line holds by the dominated convergence theorem under Assumptions \ref{asm:index}, \ref{asm:moments}, and \ref{asm:kernel}. For the variance, denoting each summand by $D_{p1,1,ij}^{1}$ and for any vector $a$ with $\|a\|=1$, we have
 	\begin{align*}
 		Var[\|D_{p1,1}^{1}\|]&\leq \frac{1}{Nh_{n}^{2}}E\left[\|D_{p1,1,12}^{1}\|^{2}\right]\\
 		&+\frac{2(n-2)}{Nh_{n}^{2}}E[\|D_{p1,1,12}^{1}\|\times \|D_{p1,1,13}^{1}\|].
 	\end{align*}
 	The first term in the right hand side is $O(1/(Nh_{n}))$ because,
 	\begin{align*}
 		E\left[\|D_{p1,1,12}^{1}\|^{2}\right]&\leq \int E[\|\Delta W_{12}\|^{4}\nu_{12}^{4}|\Delta R_{12}'\gamma=r]K^{2}(r/h_{n})f_{R\gamma}(r)dr\\
 		&= h_{n}\int E[\|\Delta W_{12}\|^{4}\nu_{12}^{4}|\Delta R_{12}'\gamma=rh_{n}]K^{2}(r)f_{R\gamma}(rh_{n})dr\\
 		&=O(h_{n}),
 	\end{align*}
 	where the last line holds from Assumptions \ref{asm:index}, \ref{asm:moments}, and \ref{asm:kernel}. The second term on the right-hand side is $O(1/n$) because
 	\begin{align*}
 		E[\|D_{p1,1,12}^{1}\|\times \|D_{p1,1,13}^{1}\|]
 		&\leq E\Big[\int E[\|\Delta W_{12}\|^{2}\nu_{12}^{2}|\Delta R_{12}'\gamma=r_{1},\xi_{1},U_{1}]\\
 		&\times E[\|\Delta W_{13}\|^{2}\nu_{13}^{2}|\Delta R_{13}'\gamma=r_{1},\xi_{1},U_{1}]\\
 		&\times K^{2}(r_{1}/h_{n})K^{2}(r_{2}/h_{n})f_{R\gamma|\xi_{1},U_{1}}(r_{1})f_{R\gamma|\xi_{1},U_{1}}(r_{2})dr_{1}dr_{2}\Big]\\
 		&=h_{n}^{2}E\Big[\int E[\|\Delta W_{12}\|^{2}\nu_{12}^{2}|\Delta R_{12}'\gamma=r_{1}h_{n},\xi_{1},U_{1}]\\
 		&\times E[\|\Delta W_{13}\|^{2}\nu_{13}^{2}|\Delta R_{13}'\gamma=r_{1}h_{n},\xi_{1},U_{1}]\\
 		&\times K^{2}(r_{1})K^{2}(r_{2})f_{R\gamma|\xi_{1},U_{1}}(r_{1}h_{n})f_{R\gamma|\xi_{1},U_{1}}(r_{1}h_{n})dr_{1}dr_{2}\Big]\\
 		&=O(h_{n}^{2}), 
 	\end{align*}
 	where the first line follows from Assumptions \ref{asm:index}, \ref{asm:moments}, and \ref{asm:kernel}. Thus, 
 	\begin{align*}
 		Var[\|D_{p1,1}^{1}\|]=O\left(\frac{1}{Nh_{n}}\right)+O\left(\frac{1}{n}\right)=o(1).
 	\end{align*}
 	This implies that $D_{p1,1}^{1}\to_{p}\Sigma_{W\nu,2}$ as $n\to\infty$.
 	
 	The second term $D_{p1,1}^{2}$ converges to $0$. Observe that, as $K$ is bounded by Assumption \ref{asm:kernel}, for some absolute constant $C>0$,
 	\begin{align*}
 		\|D_{p1,1}^{2}\|\leq \frac{\|\beta-\beta_{n}\|^{2}}{h_{n}}\times\frac{C}{N}\sum_{i<j}\|\Delta W_{ij}\|^{4}.
 	\end{align*}
 	Since $E[\|\Delta W_{ij}\|^{4}]<\infty$ by Assumption \ref{asm:unconmoments}, we can apply the law of large numbers for U-statistics (\cite{Hoeffding1961}) to $C/N\sum_{i<j}\|\Delta W_{ij}\|^{4}$, which is $O_{p}(1)$. Also, since $\|\beta-\beta_{n}\|=O_{p}(1/\sqrt{n})$ (which is the worst-case rate for the specified $h_{n}$ by Theorem \ref{thm:normality}), we have $\|\beta-\beta_{n}\|^{2}/h_{n}=O_{p}(1/(nh_{n}^{2}))=o_{p}(1)$ as $nh_{n}^{2}\sim n\times n^{-2/(2k+3)}=n^{(2k+1)/(2k+3)}$ diverges. Thus, 
 	\begin{align*}
 		\|D_{p1,1}^{2}\|=o_{p}(1)\times O_{p}(1)=o_{p}(1),
 	\end{align*}
 	and $D_{p1,1}^{2}\to_{p} 0$ as $n\to\infty$.
 	
 	The third term $D_{p1,1}^{3}$ converges to $0$. Observe that,
 	\begin{align*}
 		E[\|D_{p1,1}^{3}\|]&\leq \int E[\|\Delta W_{12}^{2}\Lambda_{12}^{2}|\Delta R_{12}'\gamma=r]r^{2}K^{2}(r/h_{n})f_{R\gamma}(r)dr\\
 		&=h_{n}^{3}E[\|\Delta W_{12}^{2}\Lambda_{12}^{2}|\Delta R_{12}'\gamma=rh_{n}]r^{2}K^{2}(r)f_{R\gamma}(rh_{n})dr\\
 		&=O(h_{n}^{3})
 	\end{align*}
 	where the last line follows from Assumptions \ref{asm:index}, \ref{asm:moments}, and \ref{asm:kernel}. Thus, $E[\|D_{p1,1}^{3}\|]=o(1)$. Observe that, by writing each summand of $D_{p1,1}^{3}$ as $D_{p1,1,ij}^{3}$,
 	\begin{align*}
 		Var[\|D_{p1,1}^{3}\|]\leq\frac{1}{Nh_{n}^{2}}E[\|D_{p1,1,12}^{3}\|^{2}]+\frac{2(n-2)}{Nh_{n}^{2}}E[\|D_{p1,1,12}^{3}\|\times \|D_{p1,1,13}^{3}\|].
 	\end{align*}
 	The first term on the right hand is $O(h_{n}^{3}/N)$ because
 	\begin{align*}
 		E[\|D_{p,1,12}^{3}\|^{2}]&\leq \int E[\|\Delta W_{12}\|^{4}\Lambda_{12}^{4}|\Delta R_{12}'\gamma=r]r^{2}K^{2}(r/h_{n})f_{R\gamma}(r)dr\\
 		&=h_{n}^{5}E[\|\Delta W_{12}\|^{4}\Lambda_{12}^{4}|\Delta R_{12}'\gamma=rh_{n}]r^{4}K^{2}(r)f_{R\gamma}(rh_{n})dr\\
 		&=O(h_{n}^{5}),
 	\end{align*}
 	where the last line holds from Assumptions \ref{asm:index}, \ref{asm:moments}, and \ref{asm:kernel}. The second term on the right hand side is $O(h_{n}^{4}/n)$ because
 	\begin{align*}
 		E[\|D_{p1,1,12}^{3}\|\times \|D_{p1,1,13}^{3}\|]
 		&\leq E\Big[\int E[\|\Delta W_{12}\|^{2}\Lambda_{12}^{2}|\Delta R_{12}'\gamma=r_{1},\xi_{1},U_{1}]\\
 		&\times E[\|\Delta W_{13}\|^{2}\Lambda_{13}^{2}|\Delta R_{13}'\gamma=r_{1},\xi_{1},U_{1}]\\
 		&\times r_{1}^{2}r_{2}^{2}K^{2}(r_{1}/h_{n})K^{2}(r_{2}/h_{n})f_{R\gamma|\xi_{1},U_{1}}(r_{1})f_{R\gamma|\xi_{1},U_{1}}(r_{2})dr_{1}dr_{2}\Big]\\
 		&=h_{n}^{6} E\Big[\int E[\|\Delta W_{12}\|^{2}\Lambda_{12}^{2}|\Delta R_{12}'\gamma=r_{1}h_{n},\xi_{1},U_{1}]\\
 		&\times E[\|\Delta W_{13}\|^{2}\Lambda_{13}^{2}|\Delta R_{13}'\gamma=r_{2}h_{n},\xi_{1},U_{1}]\\
 		&\times r_{1}^{2}r_{2}^{2}K^{2}(r_{1})K^{2}(r_{2})f_{R\gamma|\xi_{1},U_{1}}(r_{1}h_{n})f_{R\gamma|\xi_{1},U_{1}}(r_{2}h_{n})dr_{1}dr_{2}\Big]\\
 		&=O(h_{n}^{6}), 
 	\end{align*}
 	where the last line follows from Assumptions \ref{asm:index}, \ref{asm:moments}, and \ref{asm:kernel}. Hence, we have
 	\begin{align*}
 		Var[\|D_{p1,1}^{3}\|]=O\left(\frac{h_{n}^{3}}{N}\right)+O\left(\frac{h_{n}^{4}}{n}\right)=o(1).
 	\end{align*}
 	This implies that $D_{p1,1}^{3}\to_{p}0$ as $n\to \infty$.
 	
 	The fourth term $D_{p1,1}^{4}$ converges to $0$. Observe that, since $K$ is bounded by Assumption \ref{asm:kernel} and $\|\gamma\|<\infty$, for some constant $C>0$,
 	\begin{align*}
 		\|D_{p1,1}^{4}\|\leq \frac{C\|\beta-\hat{\beta}_{n}\|}{h_{n}}\times\frac{1}{N}\sum_{i<j}\|\Delta W_{ij}\|^{3}\|\Delta R_{ij}\|||\Lambda_{12}|
 	\end{align*}
 	The sum part converges to the expectation of summand by the law of large numbers for U-statistics (\cite{Hoeffding1961}) as $E[\|\Delta W_{12}\|^{3}\|\Delta R_{12}\||\Lambda_{12}|]<\infty$ is bounded by Cauchy-Schwartz and Assumption \ref{asm:unconmoments}. Thus, this part is $O_{p}(1)$. Also note that
 	\begin{align*}
 		\frac{\|\beta-\hat{\beta}_{n}\|}{h_{n}}=O_{p}\left(\frac{1}{nh_{n}}\right)=o_{p}(1),
 	\end{align*}
 	by Assumption \ref{asm:band}. Hence,,
 	\begin{align*}
 		\|D_{p1,1}^{4}\|=o_{p}(1).
 	\end{align*} 
 	This shows that $D_{p1,1}^{4}\to_{p}0$ as $n\to\infty$.
 	
 	The fifth term $D_{p1,1}^{5}$ converges to $0$. Observe that, since $K$ is bounded by Assumption \ref{asm:kernel}, for some constant $C>0$
 	\begin{align*}
 		\|D_{p1,1}^{5}\|\leq \frac{C}{N}\sum_{i<j}\|\Delta W_{ij}\|^{3}|\nu_{ij}|\times \frac{\|\beta-\hat{\beta}_{n}\|}{h_{n}}.
 	\end{align*}
 	The sum part is $O_{p}(1)$ because
 	\begin{align*}
 		E[\|\Delta W_{12}\|^{3}|\nu_{12}|]<\infty,
 	\end{align*}
 	by Assumption \ref{asm:unconmoments} and
 	\begin{align*}
 		Var\left[\frac{1}{N}\sum\|\Delta W_{ij}\||\nu_{ij}|\right]\leq \frac{E[\|\Delta W_{12}\|^{6}\nu_{12}^{2}]}{N}+\frac{2(n-2)}{N}E[\|\Delta W_{12}\|^{3}\Delta W_{13}\|^{3}|\nu_{12}||\nu_{13}|]=o(1),
 	\end{align*}
 	as these two moments are bounded by Assumption \ref{asm:unconmoments}. Thus,
 	\begin{align*}
 		\|D_{p1,1}^{5}\|=o_{p}(1),
 	\end{align*}
 	by the previous calculation for the term involving $\hat{\beta}_{n}-\beta$. This shows that $D_{p1,1}^{5}\to_{p}0$ as $n\to\infty$.
 	
 	The sixth term $D_{p1,1}^{6}$ converges to $0$. Its expectation is exactly $0$ by the conditional mean independence of $\nu_{ij}$. Also, by repeating the similar calculation as $Var[\|D_{p1,1}^{2}\|]$ (by replacing $\nu_{ij}^{2}$ by $\lambda_{ij}\nu_{ij}$), we have 
 	\begin{align*}
 		Var[\|D_{p1,1}^{6}\|]=O\left(\frac{1}{N}\right)+O\left(\frac{h_{n}^{2}}{n}\right)=o(1).
 	\end{align*}
 	This shows that $\tilde{D}_{1,6}\to_{p}0$ as $n\to\infty$.
 	
 	\subsubsection*{Sub-Step 2: $D_{p1,2}\to_{p}0$}\mbox{}\\
 	\indent As before, we can decompose $D_{p1,2}$ into $D_{p1,2}^{i}$ for $i=1,...,6$. Unlike in $D_{p1,1}$, we can no longer have the moments scaled by $h_{n}^{\alpha}$ because the middle values $c_{ij,n}^{*}$ are in the kernels. Thus, by the previous calculation for $D_{p1,1}$, the $D_{p1,2}^{i}$ that involves $\nu_{ij}^{2}$,$\lambda_{ij}^{2}$, or $\lambda_{ij}\nu_{ij}$ will have the slowest convergence rate. So, it suffices to show that those terms converge to $0$ in probability.
 	
 	Pick up such $D_{p1,2}^{i}$ with $\nu_{ij}^{2}$, which is $D_{p1,2}^{1}$ and given by
 	\begin{align*}
 		D_{p1,2}^{1}=\frac{2}{Nh_{n}^{2}}\sum_{i<j}d_{ij1}d_{ij2}\Delta W_{ij}\Delta W_{ij}'\Delta R_{ij}'\nu_{ij}^{2}k(c_{ij,n}^{*}/h_{n})K(c_{ij,n}^{*})(\hat{\gamma}_{n}-\gamma)
 	\end{align*} 
 	Observe that, for some constant $C>0$
 	\begin{align*}
 		\|D_{p1,2}^{1}\|\leq \frac{C}{N}\sum_{i<j}\|\Delta W_{ij}\|^{2}\|\Delta R_{ij}\|\nu_{ij}^{2}\times \frac{\|\hat{\gamma}_{n}-\gamma\|}{h_{n}^{2}}
 	\end{align*}
 	The sum part is $O_{p}(1)$ because
 	\begin{align*}
 		E[\|\Delta W_{12}\|^{2}\|\Delta R_{12}\|\nu_{12}^{2}]<\infty,
 	\end{align*}
 	by Assumption \ref{asm:unconmoments}, and
 	\begin{align*}
 	&Var\left[\frac{1}{N}\sum_{i<j}\|\Delta W_{ij}\|^{2}\|\Delta R_{ij}\|\nu_{ij}^{2}\right]\\
 	&\leq \frac{E[\|\Delta W_{12}\|^{4}\|\Delta R_{12}\|^{2}\nu_{12}^{4}]}{N}+\frac{2(n-2)}{N}E[\|\Delta W_{12}\|^{2}\|\Delta W_{13}\|^{2}\|\Delta R_{12}\|\|\Delta R_{13}\|\nu_{12}^{2}\nu_{13}^{2}]\\
 	&=o(1),
 	\end{align*}
 	as these moments are bounded by Assumption \ref{asm:unconmoments}. The term involving $\hat{\gamma}_{n}$ is $o_{p}(1)$ because
 	\begin{align*}
 		\frac{\|\hat{\gamma}_{n}-\gamma\|}{h_{n}^{2}}&=\frac{\sqrt{Nh_{n}}\|\hat{\gamma}_{n}-\gamma\|}{\sqrt{Nh_{n}^{5}}}=o_{p}(1),
 	\end{align*}
    by Assumption \ref{asm:first} and $Nh_{n}^{5}=Nh_{n}^{2k+3}\times h_{n}^{-2k+2}$ diverges for $k\geq 2$.
 	Hence,
 	\begin{align*}
 		\|D_{p1,2}^{1}\|=O_{p}(1)\times o_{p}(1)=o_{p}(1).
 	\end{align*}
 	This shows that $D_{p1,2}^{1}\to_{p} 0$ as $n\to\infty$. Thus, by the above argument, it follows that $D_{p1,2}\to_{p}0$ as $n\to\infty$.
 	
 	These two sub-steps conclude that \begin{align*}
 		\hat{\Sigma}_{W\nu,2}\to_{p}\Sigma_{W\nu,2},
 	\end{align*}
 	as $n\to\infty$. This finishes Step 1.
 	
 	\subsubsection*{Step 2: $\hat{\Sigma}_{W\nu,1}\to_{p}\Sigma_{W\nu,1}$}\mbox{}\\
 	\indent Define
 	\begin{align*}
 		S_{ij}\equiv 2d_{ij1}d_{ij2}K_{h_{n}}(\Delta R_{ij}'\gamma)\Delta W_{ij}\Delta\hat{\epsilon}_{ij},
 	\end{align*}
 	and let $\tilde{\Sigma}_{W\nu,1}$ be $\hat{\Sigma}_{W\nu,1}$ with $\hat{S}_{ij}$ replaced by $S_{ij}$. First, we use the following result:
 	\begin{lem}
 		Suppose that Assumptions \ref{asm:data}-\ref{asm:first} hold. If $h_{n}=hN^{-1/(2k+3)}$ for some $h>0$, we have
 		\begin{align*}
 			\tilde{\Sigma}_{W\nu,1}\to_{p}\Sigma_{W\nu,1},
 		\end{align*}
 		as $n\to\infty$.
 		\label{lem:tSigma1}
 	\end{lem}
 	Then, it is enough to show that $\hat{\Sigma}_{W\nu,1}$ is well approximated by $\tilde{\Sigma}_{W\nu,1}$:
 	\begin{lem}
 		Suppose that Assumptions \ref{asm:data}-\ref{asm:first} hold. If $h_{n}=hN^{-1/(2k+3)}$ for some $h>0$, we have
 		\begin{align*}
 			\|\hat{\Sigma}_{W\nu,1}-\tilde{\Sigma}_{W\nu,1}\|=o_{p}(1).
 		\end{align*}
 		\label{lem:Sigma1er}
 	\end{lem}
	Lemmas \ref{lem:tSigma1} and \ref{lem:Sigma1er} imply that
	\begin{align*}
		\|\hat{\Sigma}_{W\nu,1}-\Sigma_{W\nu,1}\|\leq \|\hat{\Sigma}_{W\nu,1}-\tilde{\Sigma}_{W\nu,1}\|+\|\tilde{\Sigma}_{W\nu,1}-\Sigma_{W\nu,1}\|=o_{p}(1),
	\end{align*}
 	which shows the consistency of $\hat{\Sigma}_{W\nu,1}$ for $\Sigma_{W\nu,1}$. This finishes Step 2. 
 	
 	\subsubsection*{Step 3: $c_{W}'\Sigma_{W\nu,1}c_{W}=0$ case}\mbox{}\\
 	Observe that, by some algebra,
 	\begin{align*}
 		&nh_{n}c'\hat{S}_{WW}^{-1}\hat{\Sigma}_{W\nu,1}\hat{S}_{WW}^{-1}c\\
 		&=nh_{n}c'(\hat{S}_{WW}^{-1}-\Sigma_{WW}^{-1})\hat{\Sigma}_{W\nu,1}(\hat{S}_{WW}^{-1}-\Sigma_{WW}^{-1})c\\
 		&+nh_{n}c'\Sigma_{WW}^{-1}\hat{\Sigma}_{W\nu,1}(\hat{S}_{WW}^{-1}-\Sigma_{WW}^{-1})c
 		+nh_{n}c'(\hat{S}_{WW}^{-1})\hat{\Sigma}_{W\nu,1}\Sigma_{WW}^{-1}c\\
 		&+3nh_{n}c'\Sigma_{WW}^{-1}\hat{\Sigma}_{W\nu,1}\Sigma_{WW}^{-1}c.
 	\end{align*}
 	
 	We show the negligibility of the first line in the right hand side of this decomposition. By the proof of Lemma \ref{lem:S_WW}, we have that
 	\begin{align*}
 		\hat{S}_{WW}^{-1}-\Sigma_{WW}^{-1}=o_{p}(n^{-\alpha/2})
 	\end{align*}
 	for any $\alpha\in(0,1)$. Thus,
 	\begin{align*}
 		nh_{n}c'(\hat{S}_{WW}^{-1}-\Sigma_{WW}^{-1})\hat{\Sigma}_{W\nu,1}(\hat{S}_{WW}^{-1}-\Sigma_{WW}^{-1})=n^{1-\alpha}h_{n}o_{p}(1)\hat{\Sigma}_{W\nu,1}o_{p}(1)=o_{p}(1),
 	\end{align*}
 	for $\alpha\in[(2k+1)/(2k+3),1)$ as $n^{1-\alpha}h_{n}=n^{(2k+1-\alpha(2k+3))/(2k+3)}=o(1)$ and $\hat{\Sigma}_{W\nu,1}=O_{p}(1)$ by the above Step 2.
 	
 	The remaining terms are shown to be negligible by applying the following lemmas:
 	\begin{lem}
 		Suppose that Assumptions \ref{asm:data}-\ref{asm:first} hold. If $c_{W}'\Sigma_{W\nu,1}c_{W}=0$ and $h_{n}=hN^{-1/(2k+3)}$ for some $h\in(0,\infty)$, we have
 		\begin{align*}
 			n^{1-\alpha/2}h_{n}\hat{\Sigma}_{W\nu,1}c_{W}&\to_{p}0,\\
 			n^{1-\alpha/2}h_{n}c_{W}'\hat{\Sigma}_{W\nu,1}&\to_{p}0,
 		\end{align*}
 		as $n\to\infty$ for $\alpha\in[6/(2k+3),1)$.
 		\label{lem:negligible_partial}
 	\end{lem}
 	\begin{lem}
 		Suppose that Assumptions \ref{asm:data}-\ref{asm:first} hold. If $c_{W}'\Sigma_{W\nu,1}c_{W}=0$ and $h_{n}=hN^{-1/(2k+3)}$ for some $h\in(0,\infty)$, we have
 		\begin{align*}
 			nh_{n}c_{W}'\hat{\Sigma}_{W\nu,1}c_{W}\to_{p}0,
 		\end{align*}
 		as $n\to\infty$.
 		\label{lem:negligible_full}
 	\end{lem}
 	Then, by Lemmas \ref{lem:negligible_partial} and \ref{lem:negligible_full}, the last two lines are shown to be
 	\begin{align*}
 		&nh_{n}c'\Sigma_{WW}^{-1}\hat{\Sigma}_{W\nu,1}(\hat{S}_{WW}^{-1}-\Sigma_{WW}^{-1})c
 		+nh_{n}c'(\hat{S}_{WW}^{-1}-\Sigma_{WW}^{-1})\hat{\Sigma}_{W\nu,1}\Sigma_{WW}^{-1}c\\
 		&+3nh_{n}c'\Sigma_{WW}^{-1}\hat{\Sigma}_{W\nu,1}\Sigma_{WW}^{-1}c\\
 		&=n^{1-\alpha/2}h_{n}c'\Sigma_{WW}^{-1}\hat{\Sigma}_{W\nu,1}o_{p}(1)
 		+c'o_{p}(1)n^{1-\alpha/2}h_{n}\hat{\Sigma}_{W\nu,1}\Sigma_{WW}^{-1}c
 		+o_{p}(1)\\
 		&=o_{p}(1).
 	\end{align*}
 	Hence,
 	\begin{align*}
 		nh_{n}c'\hat{S}_{WW}^{-1}\hat{\Sigma}_{W\nu,1}\hat{S}_{WW}c=o_{p}(1).
 	\end{align*}
 
 	Steps 1-3 finish the proof of Proposition \ref{prop:variance}.
 	\end{proof}
 	\subsection*{Proof of Proposition \ref{prop:bias}}
 	\begin{proof}
 		Since 
 		\begin{align*}
 			h_{n,\delta}^{-(k+1)}(\hat{\beta}_{n,\delta}-\hat{\beta}_{n})=h_{n,\delta}^{-(k+1)}(\hat{\beta}_{n,\delta}-\beta)-h_{n,\delta}^{-(k+1)}(\hat{\beta}_{n}-\beta),
 		\end{align*}
 		where the first term on the right hand side converges to $\Sigma_{WW}^{-1}\Sigma_{W\lambda}$ by Theorem \ref{thm:normality} as $Nh_{n,\delta}^{2k+3}\to\infty$, it suffices to show that
 		\begin{align*}
 			h_{n,\delta}^{-(k+1)}(\hat{\beta}_{n}-\beta)=o_{p}(1).
 		\end{align*}
 		Take an arbitrary non-zero vector $c\in\mathbb{R}^{q_{w}}$. Since $\hat{\beta}_{n}$ is calculated based on $h_{n}=hN^{-1/(2k+3)}$ such that $Nh_{n}^{2k+3}\to h$, by Theorem \ref{thm:normality},
 		\begin{align*}
 			h_{n,\delta}^{-(k+1)}c'(\hat{\beta}_{n}-\beta)&=\frac{1}{\sqrt{nh_{n,\delta}^{2(k+1)}}}\times \underbrace{\sqrt{n}c'(\hat{\beta}_{n}-\beta)}_{=O_{p}(1)}=o_{p}(1)
 		\end{align*}
 		since 
 		\begin{align*}
 			nh_{n,\delta}^{2(k+1)}\sim n\times n^{-4\delta(k+1)/(2k+3)}=n^{\frac{2k+3-4\delta(k+1)}{2k+3}}
 		\end{align*}
 		diverges for $\delta\in(0,\frac{2k+3}{4k+4})$, which is assumed by the hypothesis. Since $c$ is arbitrary, $h_{n,\delta}^{-(k+1)}(\hat{\beta}_{n}-\beta)=o_{p}(1)$, which completes the proof.
 	\end{proof}
 
 	\subsection*{Proofs of Lemmas}
 	 	\subsubsection*{Proof of Lemma \ref{lem:S_WW}}
 	\begin{proof}
 		Write each summand of $S_{WW}$ as $S_{WW,ij}$. Since it suffices to show the element-wise convergence of $S_{WW}$ to $\Sigma_{WW}$, we use a unit vector $e\in\mathbb{R}^{q_{w}}$ with the arbitrary element being $1$ and $0$ elsewhere.
 		Observe that 
 		\begin{align*}
 			E[e'S_{WW}e]=E[e'S_{WW,ij}e]&=\frac{1}{h_{n}}\int E[d_{121}d_{122}e'\Delta W_{12}\Delta W_{12}'e|\Delta R_{12}'\gamma=r]K(r/h_{n})f_{R\gamma}(r)dr\\
 			&=\int E[d_{121}d_{122}e'\Delta W_{12}\Delta W_{12}'e|\Delta R_{12}'\gamma=rh_{n}]K(r)f_{R\gamma}(rh_{n})dr\\
 			&=e'\Sigma_{WW}e+o_{p}(1),
 		\end{align*}
 		where the last line holds from the dominated convergence theorem under Assumptions \ref{asm:index}, \ref{asm:moments} and \ref{asm:kernel}. 
 		Since $S_{WW,ij}$ and $S_{WW,kl}$ are independent if $i\neq k,l$ and $j\neq k,l$ by Assumption \ref{asm:data}, observe that
 		\begin{align*}
 			Var(e'S_{WW}e)&=\frac{Var(S_{WW,12})}{N}+\frac{2(n-2)}{N}Cov(e'S_{WW,12}e,e'S_{WW,13}e).
 		\end{align*}
 		For the variance, we have
 		\begin{align*}
 			Var(S_{WW,12})&\leq E[(e'S_{WW,12}e)^{2}]\\
 			&\leq \frac{1}{h_{n}^{2}}\int E[\|\Delta W_{12}\|^{4}|\Delta R_{12}'\gamma=r]K^{2}(r/h_{n})f_{R\gamma}(r)dr\\
 			&=\frac{1}{h_{n}}\int E[\|\Delta W_{12}\|^{4}|\Delta R_{12}'\gamma=rh_{n}]K^{2}(r)f_{R\gamma}(rh_{n})dr\\
 			&=O\left(\frac{1}{h_{n}}\right),
 		\end{align*}
 		where the last line holds from Assumptions \ref{asm:index}, \ref{asm:moments} and \ref{asm:kernel}. For the covariance, by the conditional independence of $\Delta W_{12}$ and $\Delta W_{13}$, we have
 		\begin{align*}
 			&Cov(e'S_{WW,12}e,e'S_{WW,13}e)\\
 			&\leq E[|e'S_{WW,12}e\times e'S_{WW,13}e|]\\
			&\leq\frac{1}{h_{n}^{2}}\int \mathbb{E}[\|\Delta W_{12}'\|^{2}\|\Delta W_{13}\|^{2}|\Delta R_{12}'\gamma=r_{1},\Delta R_{13}'\gamma=r_{2}]|K(r_{1}/h_{n})||K(r_{2}/h_{n})|f_{R\gamma,2}(r_{1},r_{2})dr_{1}dr_{2}\\
 			&=\int \mathbb{E}[\|\Delta W_{12}'\|^{2}\|\Delta W_{13}\|^{2}|\Delta R_{12}'\gamma=h_{n}r_{1},\Delta R_{13}'\gamma=h_{n}r_{2}]|K(r_{1})||K(r_{2})|f_{R\gamma,2}(h_{n}r_{1},h_{n}r_{2})dr_{1}dr_{2}\\
 			&=O(1),
 		\end{align*}
 		where the last line holds from Assumptions \ref{asm:index}, \ref{asm:moments} and \ref{asm:kernel}. Thus,
 		\begin{align*}
 			Var(e'S_{WW}e)=O\left(\frac{1}{Nh_{n}}\right)+O\left(\frac{1}{n}\right)=o(1).
 		\end{align*}
 		By Chebychev's inequality, $e'S_{WW}e\to_{p}e'\Sigma_{WW}e$ as $n\to\infty$. Since $e$ is arbitrary, this completes the proof.
 		
 	\end{proof}
 	
 	\subsubsection*{Proof of Lemma \ref{lem:S_Wl}}
 	\begin{proof}
 		Write each summand of $S_{W\lambda}$ as $S_{W\lambda,ij}$. We use a unit vector $e\in\mathbb{R}^{q_{w}}$ with an arbitrary element being $1$ and $0$ elsewhere. Observe that, for large enough $n$,
 		\begin{align*}
 			E[e'S_{W\lambda}]=E[e'S_{W\lambda,ij}]&=\frac{1}{h_{n}}\int E[d_{121}d_{122}e'\Delta W_{12}\lambda_{12}|\Delta R_{12}'\gamma=r]K(r/h_{n})f_{R\gamma}(r)dr\\
 			&=h_{n}\int e'g(rh_{n})rK(r)dr\\
 			&=\frac{h_{n}^{k+1}}{k!}\int \left(e'\frac{\partial^{k}g(rh_{n})}{\partial r^{k}}+o(1)\right)r^{k+1}K(r)dr\\
 			&=h_{n}^{k+1}e'\Sigma_{W\lambda} +o(h_{n}^{k+1}),
 		\end{align*}
 		where the second line holds from $\lambda_{12}=\Lambda_{12}\times \Delta R_{12}'\gamma$, the third line holds from Assumption \ref{asm:kernel} eliminating $\int s^{i}K(s)$ for $i=1,...,k$, and the last line holds from the dominated convergence theorem under Assumptions \ref{asm:moments} and \ref{asm:kernel}. Observe that
 		\begin{align*}
 			Var[e'S_{W\lambda}]=\frac{Var[e'S_{W\lambda,12}]}{N}+\frac{2(n-2)}{N}Cov[e'S_{W\lambda,12},e'S_{W\lambda,13}].
 		\end{align*}
 		For the variance, we have
 		\begin{align*}
 			Var[e'S_{W\lambda,12}]&\leq E[(e'S_{W\lambda,12})^{2}]\\
 			&\leq \frac{1}{h_{n}^{2}}\int E[\|\Delta W_{12}\|^{2}\lambda_{12}^{2}|\Delta R_{12}'\gamma=r]K^{2}(r/h_{n})f_{R\gamma}(r)dr\\
 			&=h_{n}\int E[\|\Delta W_{12}\|^{2}\Lambda_{12}^{2}|\Delta R_{12}'\gamma=rh_{n}]r^{2}K^{2}(r)f_{R\gamma}(rh_{n})dr\\
 			&=O(h_{n}),
 		\end{align*}
 		where the last line holds from Cauchy-Schwartz and Assumptions \ref{asm:index}, \ref{asm:moments}, and \ref{asm:kernel}. For the covariance, we have
 		\begin{align*}
 			&Cov[e'S_{W\lambda,12},e'S_{W\lambda,13}]\\
 			&\leq E[e'S_{W\lambda,12}\times e'S_{W\lambda,13}]\\
 			&=\frac{1}{h_{n}^{2}}E\Big[\int E[d_{121}d_{122}e'\Delta W_{12}\Lambda_{12}|\Delta R_{12}'\gamma=r_{1},\xi_{1},U_{1}]E[d_{131}d_{132}e'\Delta W_{13}\Lambda_{13}|\Delta R_{13}'\gamma=r_{1},\xi_{1},U_{1}]\\
 			&\times r_{1}r_{2}K(r_{1}/h_{n})K(r_{2}/h_{n})f_{R\gamma,\xi_{1},U_{1}}(r_{1})f_{R\gamma,\xi_{1},U_{1}}(r_{2})dr_{1}dr_{2}\Big]\\
 			&\leq h_{n}^{2}E\Big[\int E[\|\Delta W_{12}\||\Lambda_{12}||\Delta R_{12}'\gamma=r_{1}h_{n},\xi_{1},U_{1}]E[\|\Delta W_{13}\||\Lambda_{13}||\Delta R_{13}'\gamma=r_{2}h_{n},\xi_{1},U_{1}]\\
 			&\times r_{1}r_{2}K(r_{1})K(r_{2})f_{R\gamma,\xi_{1},U_{1}}(r_{1}h_{n})f_{R\gamma,\xi_{1},U_{1}}(r_{2}h_{n})dr_{1}dr_{2}\Big]\\
 			&=O(h_{n}^{2}),
 		\end{align*}
 		where the last line holds from Cauchy-Schwartz and Assumptions \ref{asm:index}, \ref{asm:moments}, and \ref{asm:kernel}. Thus,
 		\begin{align*}
 			Var[e'S_{W\lambda}]=O\left(\frac{h_{n}}{N}\right)+O\left(\frac{h_{n}^{2}}{n}\right)=O\left(\frac{h_{n}^{2}}{n}\right),
 		\end{align*}
 		since $O(h_{n}/N)=O(h_{n}^{2}/n)\times O(1/(nh_{n}))=o(h_{n}^{2}/n)$ under Assumption \ref{asm:band}.
 		
 		If $Nh_{n}^{2k+3}\to h$ for some $0<h<\infty$, note that
 		\begin{align*}
 			\sqrt{Nh_{n}}E[e'S_{W\lambda}]=\sqrt{Nh_{n}^{2k+3}}e'\Sigma_{W\lambda}+o(\sqrt{Nh_{n}^{2k+3}})\to \sqrt{h}\Sigma_{W\lambda},
 		\end{align*}
 		as $n\to\infty$. Also,
 		\begin{align*}
 			Var[\sqrt{Nh_{n}}e'S_{W\lambda}]=O\left(\frac{Nh_{n}^{3}}{n}\right)=O(nh_{n})\times O(h_{n}^{2})=o(1),
 		\end{align*}
 		by Assumption \ref{asm:band}. Thus, by Chebyshev's inequality, we have 
 		\begin{align*}
 			\sqrt{Nh_{n}}e'S_{W\lambda}\to_{p}\sqrt{h}e'\Sigma_{W\lambda},
 		\end{align*}
 		as $n\to\infty$. 
 		
 		If $Nh_{n}^{2k+3}\to\infty$, note that
 		\begin{align*}
 			h_{n}^{-(k+1)}E[e'S_{W\lambda}]=e'\Sigma_{W\lambda}+o(1)\to e'\Sigma_{W\lambda},
 		\end{align*}
 		as $n\to\infty$. Also,
 		\begin{align*}
 			Var[h_{n}^{-(k+1)}e'S_{W\lambda}]=O\left(\frac{h_{n}^{2}}{nh_{n}^{2k}}\right)=o(1),
 		\end{align*}
 		as $nh_{n}^{2k}\to\infty$ by the hypothesis. Thus, by Chebyshev's inequality, we have
 		\begin{align*}
 			h_{n}^{-(k+1)}e'S_{W\lambda}\to_{p}e'\Sigma_{W\lambda},
 		\end{align*}
 		as $n\to\infty$. Since $e$ is arbitrary, this completes the proof.
 	\end{proof}
 	
 	\subsubsection*{Proof of Lemma \ref{lem:S_Wnu}}
 	\begin{proof}
 		 The proof is done in the following steps:
 		 \subsubsection*{Step 0: Decomposition}\mbox{}\\
 		 \indent Observe that
 		 \begin{align*}
 		 	c'S_{WW}^{-1}S_{W\nu}=c'(S_{WW}^{-1}-\Sigma_{WW}^{-1})S_{W\nu}+c'\Sigma_{WW}^{-1}S_{W\nu}
 		 \end{align*}
 	 	In Steps 1-2, we verify the asymptotic normality of $S_{W\nu}$, with the worst-case convergence rate being $\sqrt{n}$. Given that result, the first term on the right-hand side is shown to be negligible even when normalized by $\sqrt{Nh_{n}}$:
 	 	 \begin{align*}
 	 	 	\sqrt{Nh_{n}}c'(S_{WW}^{-1}-\Sigma_{WW}^{-1})S_{W\nu}&=\sqrt{Nh_{n}}o_{p}(n^{-\alpha/2}) O_{p}(1/\sqrt{n})\\
 	 	 	&=\sqrt{n^{1-\alpha}h_{n}}o_{p}(1)=o_{p}(1)
 	 	 \end{align*}
  	 	 because by Lemma \ref{lem:S_WW}, $S_{WW}^{-1}-\Sigma_{WW}^{-1}=o_{p}(n^{-\alpha/2})$ for any $\alpha\in(0,1)$ and $n^{1-\alpha}h_{n}=o(1)$ for sufficiently large $\alpha$ under the hypothesis. Thus,
  	 	 \begin{align*}
  	 	 	c'S_{WW}^{-1}S_{W\nu}=c'\Sigma_{WW}^{-1}S_{W\nu}+o_{p}(1/\sqrt{Nh_{n}}),
  	 	 \end{align*}
   	 	and it suffices to establish the asymptotic normality of $S_{W\nu}$.
 		 Write $c=c_{W}$ for short. Observe that, since $E[S_{W\nu}]=0$ by the definition of $\nu_{ij}$, $c'S_{W\nu}$ can be decomposed as
 		\begin{align*}
 			c'S_{W\nu}=\underbrace{\frac{1}{n}\sum_{i=1}^{n}L_{i,W\nu}}_{L_{W\nu}}+ \underbrace{\frac{1}{N}\sum_{i<j}P_{ij,W\nu}}_{P_{W\nu}}
			+\underbrace{\frac{1}{N}\sum_{i<j}Q_{ij,W\nu}}_{Q_{W\nu}}
 		\end{align*}
 		where
 		\begin{align*}
 			L_{i,W\nu}&\equiv 2E[d_{ij1}d_{ij2}c'\Delta W_{ij}\nu_{ij}K_{h_{n}}(\Delta R_{ij}'\gamma)|\xi_{i},U_{i}]\\
			P_{ij,W\nu}&=\mathbb{E}[d_{ij1}d_{ij2}c'\Delta W_{ij}\nu_{ij}K_{h_{n}}(\Delta R_{ij}'\gamma)|\xi_{i},U_{i},\xi_{j},U_{j}]-E[d_{ij1}d_{ij2}c'\Delta W_{ij}\nu_{ij}K_{h_{n}}(\Delta R_{ij}'\gamma)|\xi_{i},U_{i}]\\
 			&-E[d_{ij1}d_{ij2}\Delta W_{ij}\nu_{ij}K_{h_{n}}(\Delta R_{ij}'\gamma)|\xi_{j},U_{j}]\\
 			Q_{ij,W\nu}&= d_{ij1}d_{ij2}c'\Delta W_{ij}\nu_{ij}K_{h_{n}}(\Delta R_{ij}'\gamma)-E[d_{ij1}d_{ij2}c'\Delta W_{ij}\nu_{ij}K_{h_{n}}(\Delta R_{ij}'\gamma)|\xi_{i},U_{i},\xi_{j},U_{j}]
 		\end{align*}
 		By design, we have that $Cov[L_{i,W\nu},L_{j,W\nu}]=Cov[L_{i,W\nu},P_{kl,W\nu}]=Cov[L_{i,W\nu},Q_{kl,W\nu}]=0$, $Cov[P_{ij,W\nu},P_{kl,W\nu}]=Cov[P_{ij,W\nu},Q_{kl,W\nu}]=0$, and $Cov[Q_{ij,W\nu},Q_{kl,W\nu}]=0$ for any $i\neq j$, $k\neq l$, and $ij\neq kl$. We show the asymptotic normality of $c'S_{W\nu}$ in the following.
 		
 		\subsubsection*{Step 1: Asymptotic Normality of $L_{W\nu}$}\mbox{}\\
 		\indent Define $V_{L}$ by
 		\begin{align*}
 			V_{L}=\sqrt{n}L_{W\nu}=\sum_{i=1}\underbrace{\frac{L_{i,W\nu}}{\sqrt{n}}}_{V_{i,L}}
 		\end{align*} 
 		Note that $E[V_{i,W\nu}]=0$ by the mean independence of $\nu_{ij}$. Observe that
 		\begin{align*}
 			Var[V_{L}]&=Var[L_{i,W\nu}]\\
 			&=4E\left[E[d_{121}d_{122}c'\Delta W_{12}\nu_{12}K_{h_{n}}(\Delta R_{12}'\gamma)|\xi_{1},U_{1}]^{2}\right]\\
 			&=4E[d_{121}d_{122}d_{131}d_{132}c'\Delta W_{12}c'\Delta W_{13}\nu_{12}\nu_{13}K_{h_{n}}(\Delta R_{12}'\gamma)K_{h_{n}}(\Delta R_{13}'\gamma)]\\
 			&=\frac{4}{h_{n}^{2}}\int E[d_{121}d_{122}d_{131}d_{132}c'\Delta W_{12}c'\Delta W_{13}\nu_{12}\nu_{13}|\Delta R_{12}'\gamma=r_{1},\Delta R_{13}'\gamma=r_{2}]\\
 			&\times K(r_{1}/h_{n})K(r_{2}/h_{n})f_{R\gamma,2}(r_{1},r_{2})dr_{1}dr_{2}\\
 			&=4\int E[d_{121}d_{122}d_{131}d_{132}c'\Delta W_{12}c'\Delta W_{13}\nu_{12}\nu_{13}|\Delta R_{12}'\gamma=r_{1}h_{n},\Delta R_{13}'\gamma=r_{2}h_{n}]\\
 			&\times K(r_{1})K(r_{2})f_{R\gamma,2}(r_{1}h_{n},r_{2}h_{n})dr_{1}dr_{2}\\
 			&=c'\Sigma_{W\nu,1}c + o_{p}(1),
 		\end{align*}
 		where the last line holds from the dominated convergence theorem under Assumptions \ref{asm:index}, \ref{asm:moments}, and \ref{asm:kernel}. Furthermore, note that
 		\begin{align*}
 			&E[d_{121}d_{122}c'\Delta W_{12}\nu_{12}K_{h_{n}}(\Delta R_{12}'\gamma)|\xi_{1},U_{1}]\\
 			&=\frac{1}{h_{n}}\int E[d_{121}d_{122}c'\Delta W_{12}\nu_{12}|\Delta R_{12}'\gamma=r,\xi_{1},U_{1}]K(r/h_{n})f_{R\gamma|\xi_{1},U_{1}}(r)dr\\
 			&=\int E[d_{121}d_{122}c'\Delta W_{12}\nu_{12}|\Delta R_{12}'\gamma=r,\xi_{1},U_{1}]K(r/h_{n})f_{R\gamma|\xi_{1},U_{1}}(r)dr\\
 			&=O(1),
 		\end{align*}
 		almost surely for sufficiently large $n$ by Assumptions \ref{asm:index}, \ref{asm:moments}, and \ref{asm:kernel}. Thus, we have
 		\begin{align*}
 			\sum_{i=1}^{n}E[|V_{i,L}|^{3}]=n\times O\left(\frac{1}{n\sqrt{n}}\right)=o(1).
 		\end{align*}
 		
 		If $\Sigma_{W\nu,1}$ is positive definite, we have $c'\Sigma_{W\nu,1}c>0$. Thus, by Lyapunov CLT, we have
 		\begin{align*}
 			V_{L}/\sqrt{Var[V_{L}]}\to_{d}\mathcal{N}(0,1),
 		\end{align*}
 		as $n\to\infty$. Thus, $V_{L}=\sqrt{n} L_{W\nu}\to_{d}\mathcal{N}(0,c'\Sigma_{W\nu,1}c)$ as $n\to\infty$. 
 		
 		If $c'\Sigma_{W\nu,1}c=0$, observe that, for some constant $C>0$
 		\begin{align*}
 			&Var[L_{i,W\nu}]\\
 			&=4E\left[\left(\int h_{n}^{-1}E[d_{121}d_{122}c'\Delta W_{12}\nu_{12}|\Delta R_{12}'\gamma=r,\xi_{1},U_{1}] K(r/h_{n})f_{R\gamma|\xi_{1},U_{1}}(r)dr\right)^{2}\right]\\
 			&=4E\left[\left(\int g_{\xi_{1},U_{1}}(rh_{n}) K(r)dr\right)^{2}\right]\\
 			&\leq 4E\left[\left(g_{\xi_{1},U_{1}}(0)+Ch_{n}^{k}\right)^{2}\right]\\
 			&=O(h_{n}^{2k}),
 		\end{align*}
 		where the first inequality holds from the Taylor expansion of $g_{\xi_{1},U_{1}}(rh_{n})$ and $K$ eliminating $\int r^{i}K(r)dr=0$ for $i=1,...,k$, and the last line holds since
 		\begin{align*}
 			&E\left[\left(g_{\xi_{1},U_{1}}(0)\right)^{2}\right]\\
 			&=E\left[\left(E[d_{121}d_{122}c'\Delta W_{12}\nu_{12}f_{R\gamma|\xi_{1},U_{1}}(0)|\Delta R_{12}'\gamma=0,\xi_{1},U_{1}]\right)^{2}\right]\\
 			&=E\left[E[d_{121}d_{122}d_{131}d_{132}c'\Delta W_{12}c'\Delta W_{13}\nu_{12}\nu_{13}f_{R\gamma,2}(0,0)|\Delta R_{12}'\gamma=\Delta R_{13}'\gamma=0,\xi_{1},U_{1}]\right]\\
 			&=f_{R\gamma,2}(0,0)E[d_{121}d_{122}d_{131}d_{132}c'\Delta W_{12}c'\Delta W_{13}\nu_{12}\nu_{13}|\Delta R_{12}'\gamma=\Delta R_{13}'\gamma=0]\\
 			&=c'\Sigma_{W\nu,1}c/4 = 0,
 		\end{align*}
 		and $E[g_{\xi_{1},U_{1}}]=0$ by the conditional mean independence of $\nu_{12}$. Thus,
 		\begin{align*}
 			Var[\sqrt{Nh_{n}}L_{W\nu}]=nh_{n}\times O(h_{n}^{2k})=h_{n}^{k-1/2}\times O(\sqrt{Nh_{n}^{2k+3}})=o(1),
 		\end{align*}
 		since $\sqrt{Nh_{n}^{2k+3}}=O(1)$ by the hypothesis and $k\geq 2$ so that $h_{n}^{k-1/2}=o(1)$. Hence, by Chebyshev's inequality,
 		\begin{align*}
 			\sqrt{Nh_{n}}L_{W\nu}\to_{p}0,
 		\end{align*}
 		as $n\to\infty$ when $\Sigma_{W\nu,1}=0$.
		\subsubsection*{Step 2: Asymptotic Normality of $P_{W\nu}$}\mbox{}\\
		\indent Notice that $P_{ij,W\nu}$ is a degenerate U-statistic of order 2. We use the CLT for degenrate U-statistics by \cite{Hall1984}.
		Note that $P_{ij,W\nu}$ is symmetric in $i$ and $j$, $\mathbb{E}[P_{ij,W\nu}^{2}]<\infty$ by Assumption \ref{asm:moments}, and $E[P_{ij,W\nu}|\xi_{i},U_{i}]=E[P_{ij,W\nu}|\xi_{j},U_{j}]=0$ by the definition of $P_{ij,W\nu}$. Also, we can verify that 
		\begin{align*}
			&\mathbb{E}\left[\left(\mathbb{E}\left[P_{12,W\nu}\times P_{13,W\nu}|\xi_{2},U_{2},\xi_{3},U_{3}\right]\right)^{2}\right]/\left[\mathbb{E}[P_{12,W\nu}^{2}]\right]^{2}\to 0\\
			&\frac{1}{n}\times \mathbb{E}\left[P_{12,W\nu}^{4}\right]/\left[\mathbb{E}[P_{12,W\nu}^{2}]\right]^{2}\to 0.
		\end{align*}
		To see this, first note that, 
		\begin{align*}
			&\mathbb{E}\left[\left(\mathbb{E}\left[P_{12,W\nu}\times P_{13,W\nu}|\xi_{2},U_{2},\xi_{3},U_{3}\right]\right)^{2}\right]\\
			&=\mathbb{E}\Big[\Big(\mathbb{E}\Big[\mathbb{E}[d_{121}d_{122}\nu_{12}|\xi_{1},U_{1},\xi_{2},U_{2}]\mathbb{E}[d_{131}d_{132}\nu_{13}|\xi_{1},U_{1},\xi_{3},U_{3}]\\
			&\times c'\Delta W_{12}c'\Delta W_{13}\nu_{12}\nu_{13}K_{h_{n}}(\Delta R_{12}'\gamma)K_{h_{n}}(\Delta R_{13}'\gamma)|\xi_{2},U_{2},\xi_{3},U_{3}]\Big]\Big)^{2}\Big]+O(1)\\
			&=\mathbb{E}\Big[\Big(h_{n}^{-2}\int \mathbb{E}[\mathbb{E}[d_{121}d_{122}\nu_{12}|\xi_{1},U_{1},\xi_{2},U_{2}]\mathbb{E}[d_{131}d_{132}\nu_{13}|\xi_{1},U_{1},\xi_{3},U_{3}]\\
			&\times c'\Delta W_{12}c'\Delta W_{13}\nu_{12}\nu_{13}|\Delta R_{12}'\gamma=r_{1},\Delta R_{13}'\gamma=r_{2},\xi_{2},U_{2},\xi_{3},U_{3}]\\
			&\times K(r_{1}/h_{n})K(r_{2}/h_{n})f_{R\gamma,2|\xi_{2},U_{2},\xi_{3},U_{3}}(r_{1},r_{2})dr_{1}dr_{2}\Big)^{2}\Big]+O(1)\\
			&=\mathbb{E}\Big[\Big(\int \mathbb{E}[\mathbb{E}[d_{121}d_{122}\nu_{12}|\xi_{1},U_{1},\xi_{2},U_{2}]\mathbb{E}[d_{131}d_{132}\nu_{13}|\xi_{1},U_{1},\xi_{3},U_{3}]\\
			&\times c'\Delta W_{12}c'\Delta W_{13}\nu_{12}\nu_{13}|\Delta R_{12}'\gamma=h_{n}r_{1},\Delta R_{13}'\gamma=h_{n}r_{2},\xi_{2},U_{2},\xi_{3},U_{3}]\\
			&\times K(r_{1})K(r_{2})f_{R\gamma,2|\xi_{2},U_{2},\xi_{3},U_{3}}(h_{n}r_{1},h_{n}r_{2})dr_{1}dr_{2}\Big)^{2}\Big]+O(1)\\
			&=O(1),
		\end{align*}
		where the first equality holds from the calculation in Step 1, and the last line holds from Assumptions \ref{asm:index}, \ref{asm:moments}, and \ref{asm:kernel}. Next, note that
		\begin{align*}
			\mathbb{E}[P_{12,W\nu}^{2}]\leq &\mathbb{E}[(d_{121}d_{122}c'\Delta W_{12}\nu_{12}K_{h_{n}}(\Delta R_{12}'\gamma))^{2}]\\
			&=h_{n}^{-2}\int \mathbb{E}[d_{121}d_{122}(c'\Delta W_{12})^{2}\nu_{12}^{2}|\Delta R_{12}'\gamma=r]K^{2}(r/h_{n})f_{R\gamma}(r)dr\\
			&=h_{n}^{-1}\int \mathbb{E}[d_{121}d_{122}(c'\Delta W_{12})^{2}\nu_{12}^{2}|\Delta R_{12}'\gamma=h_{n}r]K^{2}(r)f_{R\gamma}(h_{n}r)dr\\
			&=O(h_{n}^{-1}),
		\end{align*}
		where the last line holds from Assumptions \ref{asm:index}, \ref{asm:moments}, and \ref{asm:kernel}. Similarly, we can verify that $\mathbb{E}[P_{12,W\nu}^{4}]=O(h_{n}^{-3})$. Thus, we have
		\begin{align*}
		&\mathbb{E}\left[\left(\mathbb{E}\left[P_{12,W\nu}\times P_{13,W\nu}|\xi_{2},U_{2},\xi_{3},U_{3}\right]\right)^{2}\right]/\left[\mathbb{E}[P_{12,W\nu}^{2}]\right]^{2}=O(1)/O(h_{n}^{-1})=O(h_{n})\to 0,\\
			&\frac{1}{n}\times \mathbb{E}\left[P_{12,W\nu}^{4}\right]/\left[\mathbb{E}[P_{12,W\nu}^{2}]\right]^{2}=O(h_{n}^{-3}n^{-1})/O(h_{n})=O(n^{-1}h_{n}^{-2})\to 0,
		\end{align*}
		by the hypothesis on the bandwidth.

		Thus, we can apply Theorem 1 in \cite{Hall1984} to have 
		\begin{align*}
			\sqrt{Nh_{n}}P_{W\nu}\to_{d}\mathcal{N}(0,c'\Sigma_{P}c),
		\end{align*}
		as $n\to\infty$, where
		\begin{align*}
			\Sigma_{P} =\lim_{n\to\infty}h_{n}\mathbb{E}[P_{12,W\nu}^{2}].
		\end{align*}

		To complete the characterization of $\Sigma_{P}$, from Step 1, we have
		\begin{align*}
			(Nh_{n})^{-1}\mathbb{E}[P_{12,W\nu}^{2}]&=h_{n}\mathbb{E}\left[\left(\mathbb{E}\left[d_{121}d_{122}c'\Delta W_{12}\nu_{12}K_{h_{n}}(\Delta R_{12}'\gamma)|\xi_{1},U_{1},\xi_{2},U_{2}\right]\right)^{2}\right]+o(1)
		\end{align*}
		so that 
		\begin{align*}
			\Sigma_{P}&=\lim_{n\to\infty}h_{n}\mathbb{E}\left[\left(\mathbb{E}\left[d_{121}d_{122}c'\Delta W_{12}\nu_{12}K_{h_{n}}(\Delta R_{12}'\gamma)|\xi_{1},U_{1},\xi_{2},U_{2}\right]\right)^{2}\right].
		\end{align*}

 		\subsubsection*{Step 3: Asymptotic Normality of $Q_{W\nu}$}\mbox{}\\ 
 		\indent We use the CLT for martingale differences (Theorem 5.24 and Corollary 5.26 in \cite{White2001}). Define $V_{n,t} (1\leq t\leq N)$, a triangular array, as
 		\begin{align*}
 			V_{n,1}&=\frac{1}{N}Q_{12,W\nu},\\
 			V_{n,2}&=\frac{1}{N}Q_{13,W\nu},\\
 			&\vdots\\
 			V_{n,n-1}&=\frac{1}{N}Q_{1n,W\nu},\\
 			&\vdots\\
 			V_{n,N}&=\frac{1}{N}Q_{n-1n,W\nu}.
 		\end{align*}
 		Notice that $Q_{ij,W\nu}$ is independent of $Q_{km,W\nu}$ if $i\neq k,m$ and $j\neq k,m$. Also, $Q_{ij,W\nu}$ is conditionally independent of $Q_{km,W\nu}$ even if $i=k$ or $m$, or $j=k$ or $m$; Note that $(\epsilon_{ijt},\eta_{ijt})_{t=1,2}$ and $(\epsilon_{imt},\eta_{imt})_{t=1,2}$ are conditionally independent given $\xi_{i},U_{i}$ by Assumption \ref{asm:data}. Since $(\xi_{i},U_{i}),i=1,...n$ is i.i.d., this implies that, for $1<t\leq N$ such that $ij$ corresponds to $t$,
 		\begin{align*}
 			E[V_{n,t}|\{V_{n,s};s<t\}]&=E\left[E[V_{n,t}|\{V_{n,s};s<t\},\{\xi_{i},U_{i}\}_{i=1}^{n}|]\{V_{n,s};s<t\}\right]\\
 			&=E\left[E[V_{n,t}|\xi_{i},U_{i},\xi_{j},U_{j}]|\{V_{n,s};s <t\}\right]\\
 			&=0,
 		\end{align*}
 		as $E[V_{n,t}|\xi_{i},U_{i},\xi_{j},U_{j}]=\mathbb{E}[Q_{ij,W\nu}|\xi_{i},U_{i},\xi_{j},U_{j}]=0$ by construction.  Thus, letting $\mathcal{F}_{t}\equiv \sigma(V_{s}|1\leq s\leq t)$ be a sigma algebra generated by $V_{1},...,V_{t-1}$ ($\mathcal{F}_{1}$ is set to be a trivial $\sigma$-algebra) and $\mathbb{F}\equiv (\mathcal{F}_{t})_{1\leq t\leq N}$ be a filtration, we have
 		\begin{align*}
 			E[V_{n,t}|\mathcal{F}_{t-1}]=0
 		\end{align*}
 		for $1\leq  t\leq N$. Also, for each $t$, for some constant $C>0$,
 		\begin{align*}
 			E[|V_{n,t}|]\leq \frac{3}{N}E[|\Delta W_{12}\||\nu_{12}||K_{h_{n}}(\Delta R_{12}'\gamma)|]\leq \frac{C}{Nh_{n
 			}}E[\|\Delta W_{12}\|^{2}]^{1/2}E[\nu_{12}^{2}]^{1/2}<\infty,
 		\end{align*}
 		by Assumptions \ref{asm:unconmoments} and \ref{asm:kernel}. This shows that $\{V_{n,t}\}$ is a martingale difference sequence.
 		
 		Let $V_{n}=\sum_{t=1}^{N}V_{n,t}$. Define the variance of this sequence by
 		\begin{align*}
 			v_{n}^{2}=Var\left[\sum^{N}_{t=1}V_{n,t}\right]=NVar[V_{n,1}]=\frac{1}{N}E[Q_{12,W\nu}^{2}].
 		\end{align*}
 		We can calculate that
 		\begin{align*}
 			E[Q_{12,W\nu}^{2}]&=E\left[d_{121}d_{122}(c'\Delta W_{12})^{2}\nu_{12}^{2}K^{2}(\Delta R_{12}'\gamma)\right]\\
 			&-E\left[E[d_{121}d_{122}c'\Delta W_{12}\nu_{12}K_{h_{n}}(\Delta R_{12}'\gamma)|\xi_{1},U_{1},\xi_{2},U_{2}]^{2}\right].
 		\end{align*}
 		Observe that
 		\begin{align*}
 			&E\left[d_{121}d_{122}(c'\Delta W_{12})^{2}\nu_{12}^{2}K^{2}(\Delta R_{12}'\gamma)\right]\\
 			&=\frac{1}{h_{n}^{2}}\int E[d_{121}d_{122}(c'\Delta W_{12})^{2}\nu_{12}^{2}|\Delta R_{12}'\gamma=r]K(r/h_{n})f_{R\gamma}(r)dr\\
 			&=\frac{1}{h_{n}}\int E[d_{121}d_{122}(c'\Delta W_{12})^{2}\nu_{12}^{2}|\Delta R_{12}'\gamma=rh_{n}]K(r)f_{R\gamma}(rh_{n})dr\\
 			&=\frac{1}{h_{n}}c'\Sigma_{W\nu,2}c+o\left(\frac{1}{h_{n}}\right),
 		\end{align*}
 		where the last line holds from the dominated convergence theorem under Assumptions \ref{asm:index}, \ref{asm:moments}, and \ref{asm:kernel}. Then,
 		\begin{align*}
 			E[Q_{12,W\nu}^{2}]=\frac{1}{h_{n}}c'\Sigma_{W\nu,2}c-\frac{1}{h_{n}}c'\Sigma_{P}c+o\left(\frac{1}{h_{n}}\right).
 		\end{align*}
 		Hence, we have
 		\begin{align*}
 			v_{n}^{2}=\frac{1}{Nh_{n}}c'\Sigma_{W\nu,2}c-\frac{1}{Nh_{n}}c'\Sigma_{P}c+o\left(\frac{1}{Nh_{n}}\right).
 		\end{align*}
 		
 		The CLT for martingale differences holds if we can show the following two conditions:
 		\begin{align*}
 			\sum_{t=1}^{N}E\left[\left(\frac{V_{n,t}}{v_{n}}\right)^{2+\delta}\right]\to 0\text{ (Lyapunov)},
 		\end{align*}
 		for some $\delta>0$ as $n\to\infty$ and 
 		\begin{align*}
 			\sum_{t=1}^{N}\left(\frac{V_{n,t}}{v_{n}}\right)^{2}\to_{p} 1\text{ (Stability)},
 		\end{align*}
 		as $n\to\infty$. If these conditions are met, we can apply Theorem 5.24 and Corollary 5.26 in \cite{White2001} to show that
 		\begin{align*}
 			\frac{V_{n}}{v_{n}}\to_{d}\mathcal{N}(0,1),
 		\end{align*}
 		as $n\to\infty$. Since $\sqrt{Nh_{n}v_{n}}\to\sqrt{c'(\Sigma_{W\nu,2}-\Sigma_{P})c}$, by Slutsky's lemma,
 		\begin{align*}
 			\sqrt{Nh_{n}}V_{n}\to_{d}\mathcal{N}(0,c'(\Sigma_{W\nu,2}-\Sigma_{P})c),
 		\end{align*}
 		which is equivalent to
 		\begin{align*}
 			\sqrt{Nh_{n}}Q_{W\nu}\to_{d}\mathcal{N}(0,c'(\Sigma_{W\nu,2}-\Sigma_{P})c),
 		\end{align*}
 		as $n\to\infty$. 
 		
 		For Lyapunov's condition, observe that for some constant $C>0$,
 		\begin{align*}
 			E\left[|V_{n,1}|^{3}\right]&\leq \frac{C}{(Nh_{n})^{3}}\int E[\|\Delta W_{12}\|^{3}|\nu_{12}|^{3}|\Delta R_{12}'\gamma=r]|K(r/h_{n})|^{3}f_{R\gamma}(r)dr\\
 			&= \frac{Ch_{n}}{(Nh_{n})^{3}}\int E[\|\Delta W_{12}\|^{3}|\nu_{12}|^{3}|\Delta R_{12}'\gamma=rh_{n}]|K(r)|^{3}f_{R\gamma}(rh_{n})dr\\
 			&=O\left(\frac{1}{N^{3}h_{n}^{2}}\right),
 		\end{align*}
 		where the first inequality follows from Jensen's inequality, the last line follows from Cauchy-Schwartz and Assumptions \ref{asm:index}, \ref{asm:moments}, and \ref{asm:kernel}. Since $v_{n}=O(1/\sqrt{Nh_{n}})$, we have
 		\begin{align*}
 			\sum_{t=1}^{N}E\left[\left|\frac{V_{n,t}}{v_{n}}\right|^{3}\right]=NO\left(\frac{\sqrt{Nh_{n}}}{N^{3}h_{n}^{2}}\right)=O\left(\frac{1}{(Nh_{n})^{3/2}}\right)=o(1)
 		\end{align*}
 		by Assumption \ref{asm:band}. Thus Lyapunov's condition holds.
 		
 		For the stability condition, we can alternatively show that
 		\begin{align*}
 			\frac{1}{v_{n}^{2}}\sum_{t=1}^{N}\left(V_{n,t}^{2}-E[V_{n,t}^{2}]\right)\to_{p} 0,
 		\end{align*}
 		as $n\to\infty$. Note that
 		\begin{align*}
 			\frac{1}{v_{n}^{2}}\sum_{t=1}^{N}\left(V_{n,t}^{2}-E[V_{n,t}^{2}]\right)=\frac{1}{Nv_{n}^{2}}\left(\frac{1}{N}\sum_{i<j}Q_{ij,W\nu}^{2}-E[Q_{12,W\nu}^{2}]\right).
 		\end{align*}
 		Since $Nv_{n}^{2}=O(1/h_{n})$, we need to show that the remaining term is $o_{p}(1/h_{n})$. Since $Q_{ij,W\nu}$ is independent fron $Q_{km,W\nu}$ if there is no common node,
 		\begin{align*}
 			&E\left[\left(\frac{1}{N}\sum_{i<j}Q_{ij,W\nu}^{2}-E[Q_{12,W\nu}^{2}]\right)^{2}\right] \\
 			&=\frac{Var[Q_{12,W\nu}^{2}]}{N}+\frac{2(n-2)}{N}Cov[Q_{12,W\nu}^{2},Q_{13,W\nu}^{2}]\\
 			&\leq \frac{E[Q_{12,W\nu}^{4}]}{N}+\frac{2(n-2)}{N}E[Q_{12,W\nu}^{2}\times Q_{13,W\nu}^{2}].
 		\end{align*}
 		The first term in the far right-hand side is bounded as follows: For some constant $C>0$,
 		\begin{align*}
 			\frac{E[Q_{12,W\nu}^{4}]}{N}
 			&\leq \frac{C}{Nh_{n}^{4}}\int E[\|\Delta W_{12}\|^{4}\nu_{12}^{4}|\Delta R_{12}'\gamma=r]K^{4}(r)f_{R\gamma}(r)dr\\
 			&= \frac{C}{Nh_{n}^{3}}\int E[\|\Delta W_{12}\|^{4}\nu_{12}^{4}|\Delta R_{12}'\gamma=rh_{n}]K^{4}(r)f_{R\gamma}(rh_{n})dr\\
 			&=O\left(\frac{1}{Nh_{n}^{3}}\right),
 		\end{align*}
 		where the first inequality follows from Jensen's inequality, and the last line follows from Cauchy-Schwartz and Assumptions \ref{asm:index}, \ref{asm:moments}, and \ref{asm:kernel}. The second term on the far right-hand side is bounded as follows: For some constant $C>0$,
 		\begin{align*}
 			&\frac{2(n-2)}{N}E[Q_{12,W\nu}^{2}\times Q_{13,W\nu}^{2}]\\
 			&\leq \frac{C(n-2)}{Nh_{n}^{4}}\mathbb{E}\Big[\int E[\|\Delta W_{12}\|^{2}\nu_{12}^{2}|\Delta R_{12}'\gamma=r_{1},\xi_{1},U_{1}]\times E[\|\Delta W_{13}\|^{2}\nu_{13}^{2}|\Delta R_{13}'\gamma=r_{2},\xi_{1},U_{1}]\\
 			&\times K^{2}(r_{1}/h_{n})K^{2}(r_{2}/h_{n})f_{R\gamma|\xi_{1},U_{1}}(r_{1})f_{R\gamma|\xi_{1},U_{1}}(r_{2})dr_{1}dr_{2}\Big]\\
 			&=\frac{C(n-2)}{Nh_{n}^{2}}\mathbb{E}\Big[\int E[\|\Delta W_{12}\|^{2}\nu_{12}^{2}|\Delta R_{12}'\gamma=r_{1}h_{n},\xi_{1},U_{1}]\times E[\|\Delta W_{13}\|^{2}\nu_{13}^{2}|\Delta R_{13}'\gamma=r_{2}h_{n},\xi_{1},U_{1}]\\
 			&\times K^{2}(r_{1})K^{2}(r_{2})f_{R\gamma|\xi_{1},U_{1}}(r_{1}h_{n})f_{R\gamma|\xi_{1},U_{1}}(r_{2}h_{n})dr_{1}dr_{2}\Big]\\
 			&=O\left(\frac{1}{nh_{n}^{2}}\right)
 		\end{align*}
 		Thus,
 		\begin{align*}
 			h_{n}E\left[\left(\frac{1}{N}\sum_{i<j}Q_{ij,W\nu}^{2}-E[Q_{12,W\nu}^{2}]\right)^{2}\right]=O\left(\frac{1}{Nh_{n}^{2}}\right)+O\left(\frac{1}{nh_{n}}\right)=o(1),
 		\end{align*}
 		and by Markov's inequality,
 		\begin{align*}
 			\sqrt{h_{n}}\left(\frac{1}{N}\sum_{i<j}Q_{ij,W\nu}^{2}-E[Q_{12,W\nu}^{2}]\right)=o_{p}(1).
 		\end{align*}
 		Then,
 		\begin{align*}
 			\frac{1}{v_{n}^{2}}\sum_{t=1}^{N}\left(V_{n,t}^{2}-E[V_{n,t}^{2}]\right)=O(h_{n})\times o_{p}\left(\frac{1}{\sqrt{h_{n}}}\right) =o_{p}(\sqrt{h_{n}})=o_{p}(1),
 		\end{align*}
 		which shows the stability condition.
 		
 		\subsubsection*{Step 3: Conclusion}\mbox{}\\
 		\indent By Steps 0-3, we have established that if $c_{W}'\Sigma_{W\nu,1}c_{W}>0$,
 		\begin{align*}
 			&\sqrt{n}c'S_{WW}^{-1}S_{W\nu}\\
 			&=\underbrace{\sqrt{n}L_{W\nu}}_{\to_{d}\mathcal{N}(0,c_{W}'\Sigma_{W\nu,1}c_{W})} + \underbrace{\frac{\sqrt{n}}{\sqrt{Nh_{n}}}}_{\to 0}\times\underbrace{\sqrt{Nh_{n}}(P_{W\nu}+Q_{W\nu})}_{\to_{d}\mathcal{N}(0,c_{W}'\Sigma_{W\nu,2}c_{W})}+o_{p}(\sqrt{n}/\sqrt{Nh_{n}})\to_{d}\mathcal{N}(0,c_{W}'\Sigma_{W\nu,1}c_{W}),
 		\end{align*}
 		as $n\to\infty$ by Assumption \ref{asm:band}, and if $c_{W}'\Sigma_{W\nu,1}c_{W}=0$,
 		\begin{align*}
 			\sqrt{Nh_{n}}c'S_{WW}^{-1}S_{W\nu}=\underbrace{\sqrt{Nh_{n}}L_{W\nu}}_{\to_{p}0} + \underbrace{\sqrt{Nh_{n}}(P_{W\nu}+Q_{W\nu})}_{\to_{d}\mathcal{N}(0,c_{W}'\Sigma_{W\nu,2}c_{W})}+o_{p}(1)\to_{d}\mathcal{N}(0,c_{W}'\Sigma_{W\nu,2}c_{W}),
 		\end{align*}
 		as $n\to\infty$. This completes the proof.
 	\end{proof}
 	
 	\subsubsection*{Proof of Lemma \ref{lem:hS_WW}}
 	\begin{proof}
 		By expanding $K(\Delta R_{ij}'\hat{\gamma}_{n}/h_{n})$ around $\Delta R_{ij}'\gamma$, we have
 		\begin{align*}
 			\hat{S}_{WW}&=S_{WW}+\frac{1}{Nh_{n}^{2}}\sum_{i<j}d_{ij1}d_{ij2}\Delta W_{ij}\Delta W_{ij}'\Delta R_{ij}'(\hat{\gamma}_{n}-\gamma)k(c_{ij,n}^{*})
 		\end{align*}
 		where $c_{ij,n}^{*}$ is in between $\Delta R_{ij}'\gamma$ and $\Delta R_{ij}'\hat{\gamma}_{n}$ and $k$ is the first derivative of $K$. Thus, for some constant $C>0$,
 		\begin{align*}
 			\|\hat{S}_{WW}-S_{WW}\|\leq &\underbrace{\frac{C}{N}\sum_{i<j}\|\Delta W_{ij}\|^{2}\|\Delta R_{ij}\|}_{D_{4,1}}\times h_{n}^{-2}\|\hat{\gamma}_{n}-\gamma\|
 		\end{align*}
 		Notice that $\|\Delta W_{ij}\|=\|w(X_{i1},X_{j1})-w(X_{i2},X_{j2})\|,\|R_{ij}\|=\|r(Z_{i1},Z_{j1}-r(Z_{i2},Z_{j2})\|$ are symmetric in $i$ and $j$ by the symmetry of $w$, $r$, and $\|\cdot\|$ so that $D_{4,1}$ is a second-order U-statistics. Also,
 		\begin{align*}
 			E[\|\Delta W_{12}\|^{2}\|\Delta R_{ij}\|]<\infty.
 		\end{align*}
 		by Cauchy-Schwartz with Assumption \ref{asm:unconmoments}.
 		Thus, we can apply the law of large numbers for U-statistics (\cite{Hoeffding1961}):
 		\begin{align*}
 			D_{4,1}=O_{p}(1).
 		\end{align*}
 		By the hypothesis and Assumption \ref{asm:first},
 		\begin{align*}
 			h_{n}^{-2}\|\hat{\gamma}_{n}-\gamma\|&=\frac{\|\sqrt{Nh_{n}}(\hat{\gamma}_{n}-\gamma)\|}{\sqrt{Nh_{n}^{5}}}\\
 			&=\frac{\|\sqrt{Nh_{n}}(\hat{\gamma}_{n}-\gamma)\|}{\sqrt{Nh_{n}^{2k+3}}}\times \sqrt{h_{n}^{2k-2}}\\
 			&=o_{p}(1),
 		\end{align*}
 		as $\sqrt{Nh_{n}^{2k+3}}$ is either diverging or $O(1)$, $\sqrt{h^{2k-2}}=o(1)$ for $k\geq 2$. Thus,
 		\begin{align*}
 			\|\hat{S}_{WW}-S_{WW}\|=O_{p}(1)\times o_{p}(1)=o_{p}(1).
 		\end{align*}
 		This shows that $\hat{S}_{WW}=S_{WW}+o_{p}(1)$.
 	\end{proof}
 	
 	\subsubsection*{Proof of Lemma \ref{lem:hs_Wl}}
 	\begin{proof}
 		Expanding $K(\Delta R_{ij}'\hat{\gamma}_{n}/h_{n})$ around $\Delta R_{ij}'\gamma$, for some constant $C>0$, we have
 		\begin{align*}
 			&\sqrt{Nh_{n}}\|\hat{S}_{W\lambda}-S_{W\lambda}\|\\
 			&\leq\underbrace{\frac{1}{Nh_{n}^{2}}\sum_{i<j}\|\Delta W_{ij}\|\|\Delta R_{ij}\||\lambda_{ij}||k(\Delta R_{ij}'\gamma/h_{n})|}_{D_{5,1}}\times \sqrt{Nh_{n}}\|\hat{\gamma}_{n}-\gamma\|\\
 			&+\underbrace{\frac{1}{Nh_{n}^{2}}\sum_{i<j}\|\Delta W_{ij}\|\|\Delta R_{ij}\|^{2}|\lambda_{ij}||k(\Delta R_{ij}'\gamma/h_{n})|}_{D_{5,2}}\times \frac{\sqrt{Nh_{n}}\|\hat{\gamma}_{n}-\gamma\|}{2h_{n}}\\
 			&+C\underbrace{\frac{1}{N}\sum_{i<j}\|\Delta W_{ij}\|\|\Delta R_{ij}\|^{3}|\lambda_{ij}|}_{D_{5,3}}\times \frac{\sqrt{Nh_{n}}\|\hat{\gamma}_{n}-\gamma\|^{3}}{6h_{n}^{4}}
 		\end{align*}
 		We follow the following steps to bound the right hand side.
 		\subsubsection*{Step 1 $D_{5,1}$ and $D_{5,2}$}\mbox{}\\
 		\indent Observe that
 		\begin{align*}
 			E[D_{5,1}]&=\frac{1}{h_{n}^{2}}\int E[\|\Delta W_{12}\|\|\Delta R_{12}\||\Lambda_{12}||\Delta R_{12}'\gamma=r]|r||k(r/h_{n})|f_{R\gamma}(r)dr\\
 			&=\int E[\|\Delta W_{12}\|\|\Delta R_{12}\||\Lambda_{12}||\Delta R_{12}'\gamma=rh_{n}]|r||k(r)|f_{R\gamma}(rh_{n})dr\\
 			&=O(1),
 		\end{align*}
 		where the last line holds from Assumptions \ref{asm:index}, \ref{asm:moments} and \ref{asm:kernel}. Also, writing each summand of $D_{5,1}$ by $D_{5,1,ij}$, we have
 		\begin{align*}
 			Var[D_{5,1}]&=\frac{1}{Nh_{n}^{4}}Var[D_{5,1,12}]+\frac{2(n-2)}{Nh_{n}^{4}}Cov[D_{5,1,12},D_{5,1,13}]\\
 			&\leq \frac{1}{Nh_{n}^{4}}E[D^{2}_{5,1,12}]+\frac{2(n-2)}{Nh_{n}^{4}}E[D_{5,1,12}\times D_{5,1,13}].
 		\end{align*}
 		The first term on the far right side is $O(1/(Nh_{n}^{2}))$ because
 		\begin{align*}
 			E[D_{5,1,12}^{2}]&=\int E[\|\Delta W_{12}\|^{2}\|\Delta R_{12}\|^{2}\Lambda_{12}^{2}|\Delta R_{12}'\gamma=r]r^{2}k(r/h_{n})^{2}f_{R\gamma}(r)dr\\
 			&=h_{n}^{2}\int E[\|\Delta W_{12}\|^{2}\|\Delta R_{12}\|^{2}\Lambda_{12}^{2}|\Delta R_{12}'\gamma=rh_{n}]r^{2}k(r)^{2}f_{R\gamma}(rh_{n})dr \\
 			&=O(h_{n}^{2}),
 		\end{align*}
 		where the last line holds from Assumptions \ref{asm:index}, \ref{asm:moments}, and \ref{asm:kernel}. The second term on the far right side is $O(1/n)$ because
 		\begin{align*}
 			E[D_{5,1,12}\times D_{5,1,13}]&=E\Big[\int E[\|\Delta W_{12}\|\|\Delta R_{12}\||\Lambda_{12}||\Delta R_{12}'\gamma=r_{1},\xi_{1},U_{1}]\\
 			&\times E[\|\Delta W_{13}\|\|\Delta R_{13}\||\Lambda_{13}||\Delta R_{13}'\gamma=r_{2},\xi_{1},U_{1}] \\
 			&\times |r_{1}||r_{2}||k(r_{1}/h_{n})||k(r_{2}/h_{n})|f_{R\gamma|\xi_{1},U_{1}}(r_{1})f_{R\gamma|\xi_{1},U_{1}}(r_{2})dr_{1}dr_{2}\Big]\\
 			&=h_{n}^{4}E\Big[\int E[\|\Delta W_{12}\|\|\Delta R_{12}\||\Lambda_{12}||\Delta R_{12}'\gamma=r_{1}h_{n},\xi_{1},U_{1}]\\
 			&\times E[\|\Delta W_{13}\|\|\Delta R_{13}\||\Lambda_{13}||\Delta R_{13}'\gamma=r_{2}h_{n},\xi_{1},U_{1}] \\
 			&\times |r_{1}||r_{2}||k(r_{1})||k(r_{2})|f_{R\gamma|\xi_{1},U_{1}}(r_{1}h_{n})f_{R\gamma|\xi_{1},U_{1}}(r_{2}h_{n})dr_{1}dr_{2}\Big]\\
 			&=O(h_{n}^{4}),
 		\end{align*}
 		where the last line holds from Assumptions \ref{asm:index}, \ref{asm:moments}, and \ref{asm:kernel}. Thus,
 		\begin{align*}
 			Var[D_{5,1}]=O\left(\frac{1}{Nh_{n}^{2}}\right)+O\left(\frac{1}{n}\right)=o(1),
 		\end{align*}
 		since $Nh_{n}^{2}=Nh^{2k+3}\times h_{n}^{1-2k}$ diverges by the hypothesis. Thus,
 		\begin{align*}
 			D_{5,1}=O_{p}(1).
 		\end{align*}
 		By a similar calculation, we have
 		\begin{align*}
 			D_{5,2}=O_{p}(1).
 		\end{align*}
 		
 		\subsubsection*{Step 2: $D_{5,3}$}\mbox{}\\
 		\indent First, observe that
 		\begin{align*}
 			E[D_{5,3}]=E[\|\Delta W_{12}\|\|\Delta R_{12}\|^{4}|\Lambda_{12}|]<\infty
 		\end{align*}
 		by H\"{o}lder's inequality with Assumption \ref{asm:unconmoments}.
 		Note that by construction, $\Lambda_{ij}$ is written as a function of $\xi_{i}$ and $\xi_{j}$ with symmetry with respect to $i$ and $j$, which implies that $D_{5,3}$ is a second-order U-statistics. Since each summand is non-negative and has a finite mean, we can apply the law of large numbers for U-statistics (\cite{Hoeffding1961}) to show that
 		\begin{align*}
 			D_{5,3}=O_{p}(1).
 		\end{align*}
 		
 		\subsubsection*{Step 3: Conclusion}\mbox{}\\
 		\indent Finally, by Assumption \ref{asm:first} and the hypothesis,
 		\begin{align*}
 			&\sqrt{Nh_{n}}\|\hat{\gamma}_{n}-\gamma\|=o_{p}(1),\\
 			&\frac{\sqrt{Nh_{n}}\|\hat{\gamma}_{n}-\gamma\|^{2}}{2h_{n}}=\frac{\|\sqrt{Nh_{n}}(\hat{\gamma}_{n}-\gamma)\|^{2}}{2\sqrt{Nh_{n}^{3}}}=o_{p}(1),\\
 			&\frac{\sqrt{Nh_{n}}\|\hat{\gamma}_{n}-\gamma\|^{3}}{6h_{n}^{4}}=\frac{\|\sqrt{Nh_{n}}(\hat{\gamma}_{n}-\gamma)\|^{3}}{6Nh_{n}^{5}}=o_{p}(1).
 		\end{align*}
 		Thus,
 		\begin{align*}
 			\sqrt{Nh_{n}}\|\hat{S}_{W\lambda}-S_{W\lambda}\|=o_{p}(1).
 		\end{align*}
 		This implies that
 		\begin{align*}
 			\hat{S}_{W\lambda}=S_{W\lambda}+o_{p}\left(\frac{1}{\sqrt{Nh_{n}}}\right).
 		\end{align*}
 		This completes the proof.
 	\end{proof}
 	
 	\subsubsection*{Proof of Lemma \ref{lem:hS_Wnu}}
 	\begin{proof} By expanding $K(\Delta R_{ij}'\hat{\gamma}_{n}/h_{n})$ around $\Delta R_{ij}'\gamma$, we have
 		\begin{align*}
 			&\sqrt{Nh_{n}}(\hat{S}_{W\nu}-S_{W\nu})\\
 			&=\underbrace{\frac{1}{Nh_{n}^{2}}\sum_{i<j}d_{ij1}d_{ij2}\Delta W_{ij}\Delta R_{ij}'\nu_{ij}k(\Delta R_{ij}'\gamma/h_{n})}_{D_{6,1}}\sqrt{Nh_{n}}(\hat{\gamma}_{n}-\gamma)\\
 			&+(\hat{\gamma}_{n}-\gamma)'\underbrace{\frac{1}{Nh_{n}^{2}}\sum_{i<j}d_{ij1}d_{ij2}\Delta W_{ij}\Delta R_{ij}\Delta R_{ij}'\nu_{ij}k(\Delta R_{ij}'\gamma/h_{n})}_{D_{6,2}}\sqrt{Nh_{n}}\frac{(\hat{\gamma}_{n}-\gamma)}{h_{n}}\\
 			&+\underbrace{\frac{\sqrt{Nh_{n}}}{Nh_{n}^{4}}\sum_{i<j}d_{ij1}d_{ij2}\Delta W_{ij}\nu_{ij}k(c_{ij,n}^{*}/h_{n})\left(\Delta R_{ij}'(\hat{\gamma}_{n}-\gamma)\right)^{3}}_{D_{6,3}}
 		\end{align*}
 		We bound each component by the following steps.
 		
 		\subsubsection*{Step 1: $D_{6,1}$ and $D_{6,2}$}\mbox{}\\
 		\indent Note that $E[D_{6,1}]=E[D_{6,2}]=0$ by the conditional mean independence of $\nu_{ij}$. Write $D_{6,1,ij}$ as each summand of $D_{6,1}$. Observe that, by the similar calculation as above,
 		\begin{align*}
 			Var[\|D_{6,1}\|]\leq \frac{1}{Nh_{n}^{4}}E[\|D_{6,1,12}\|^{2}]+\frac{2(n-2)}{Nh_{n}^{4}}E[\|D_{6,1,12}\|\times \|D_{6,1,13}\|].
 		\end{align*} 
 		The first term on the right hand side is $O(1/(Nh_{n}^{3}))$ since
 		\begin{align*}
 			E[\|D_{6,1,12}\|^{2}]&\leq \int E[\|\Delta W_{12}\|^{2}\|\Delta R_{12}\|^{2}\nu_{12}^{2}|\Delta R_{12}'\gamma=r]k(r/h_{n})^{2}f_{R\gamma}(r)dr\\
 			&=h_{n}\int E[\|\Delta W_{12}\|^{2}\|\Delta R_{12}\|^{2}\nu_{12}^{2}|\Delta R_{12}'\gamma=rh_{n}]k(r)^{2}f_{R\gamma}(rh_{n})dr\\
 			&=O(h_{n}),
 		\end{align*}
 		where the last line holds from Assumptions \ref{asm:index}, \ref{asm:moments}, and \ref{asm:kernel}. The second term on the right hand side is $O(1/(nh_{n}^{2}))$ since
 		\begin{align*}
 			E[\|D_{6,1,12}\|\times \|D_{6,1,13}\|]&\leq 
 			E\Big[\int E[\|\Delta W_{12}\|\|\Delta R_{12}\||\nu|_{12}|\Delta R_{12}'\gamma=r_{1},\xi_{1},U_{1}]\\
 			&\times E[\|\Delta W_{13}\|\|\Delta R_{13}\||\nu|_{13}|\Delta R_{13}'\gamma=r_{2},\xi_{1},U_{1}]\\
 			&\times |k(r_{1}/h_{n})||k(r_{2}/h_{n})|f_{R\gamma|\xi_{1},U_{1}}(r_{1})f_{R\gamma|\xi_{1},U_{1}}(r_{2})dr_{1}dr_{2}\Big]\\
 			&=h_{n}^{2}E\Big[\int E[\|\Delta W_{12}\|\|\Delta R_{12}\||\nu|_{12}|\Delta R_{12}'\gamma=r_{1}h_{n},\xi_{1},U_{1}]\\
 			&\times E[\|\Delta W_{13}\|\|\Delta R_{13}\||\nu|_{13}|\Delta R_{13}'\gamma=r_{2}h_{n},\xi_{1},U_{1}]\\
 			&\times |k(r_{1})||k(r_{2})|f_{R\gamma|\xi_{1},U_{1}}(r_{1}h_{n})f_{R\gamma|\xi_{1},U_{1}}(r_{2}h_{n})dr_{1}dr_{2}\Big]\\
 			&=O(h_{n}^{2}),
 		\end{align*} 
 		where the last line holds from Assumptions \ref{asm:index}, \ref{asm:moments}, and \ref{asm:kernel}. Hence,
 		\begin{align*}
 			Var[\|D_{6,1}\|]=O\left(\frac{1}{Nh_{n}^{3}}\right)+O\left(\frac{1}{nh_{n}^{2}}\right)=o(1),
 		\end{align*}
 		since both $Nh_{n}^{3}=Nh_{n}^{2k+3}\times h_{n}^{-2k}$ and $nh_{n}^{2}\sim \sqrt{Nh_{n}^{4}}=\sqrt{Nh_{n}^{2k+3}}\times \sqrt{h_{n}^{-2k+1}}$ diverge under the hypothesis. This shows that
 		\begin{align*}
 			D_{6,1} = o_{p}(1).
 		\end{align*}
 		A similar calculation shows that
 		\begin{align*}
 			D_{6,2}=o_{p}(1),
 		\end{align*}
 		as well.
 		
 		\subsubsection*{Step 2: $D_{6,3}$}\mbox{}\\
 		\indent Observe that, for some constant $C>0$
 		\begin{align*}
 			\|D_{6,3}\|\leq C\underbrace{\frac{1}{N}\sum_{i<j}\|\Delta W_{ij}\|\|\Delta R_{ij}\|^{3}|\nu_{ij}|}_{D_{6,4}}\times \frac{\sqrt{Nh_{n}}\|\hat{\gamma}_{n}-\gamma\|^{3}}{Nh_{n}^{4}}.
 		\end{align*}
 		Observe that
 		\begin{align*}
 			E[D_{6,4}]=E[\|\Delta W_{12}\|\|\Delta R_{12}\|^{3}|\nu_{12}|]<\infty,
 		\end{align*}
 		by Cauchy-Schwartz with Assumption \ref{asm:moments}. Also, by writing each summand of $D_{6,4}$ as $D_{6,4,ij}$, we have
 		\begin{align*}
 			Var[D_{6,4}]\leq \frac{E[D_{6,4,12}^{2}]}{N}+\frac{2(n-2)}{N}E[D_{6,4,12}\times D_{6,4,13}].
 		\end{align*}
 		Since
 		\begin{align*}
 			E[D_{6,4,12}^{2}]&=E[\|\Delta W_{12}\|^{2}\|\Delta R_{12}\|^{6}\nu_{12}^{2}]<\infty\\
 			E[D_{6,4,12}\times D_{6,4,13}]&=E[\|\Delta W_{12}\|\|\Delta W_{13}\|\|\Delta R_{12}\|^{3}\|\Delta R_{13}\|^{3}|\nu_{12}||\nu_{13}|]<\infty
 		\end{align*}
 		by H\"{o}lder's inequality with Assumption \ref{asm:unconmoments}, 
 		\begin{align*}
 			Var[D_{6,4}]=O\left(\frac{1}{N}\right)+O\left(\frac{1}{n}\right)=o(1).
 		\end{align*}
 		This shows that
 		\begin{align*}
 			D_{6,4}=O_{p}(1).
 		\end{align*}
 		Hence, by the previous calculation for the term involving $\hat{\gamma}_{n}-\gamma$,
 		\begin{align*}
 			\|D_{6,3}\|=O_{p}(1)\times o_{p}(1)=o_{p}(1).
 		\end{align*}
 		
 		\subsubsection*{Step 3: Conclusion}\mbox{}\\
 		\indent By the above steps and the hypothesis on $\hat{\gamma}_{n}-\gamma$,
 		\begin{align*}
 			\sqrt{Nh_{n}}\|\hat{S}_{W\nu}-S_{W\nu}\|=o_{p}(1).
 		\end{align*}
 		This implies that
 		\begin{align*}
 			\hat{S}_{W\nu}=S_{W\nu}+o_{p}\left(\frac{1}{\sqrt{Nh_{n}}}\right).
 		\end{align*}
 		This completes the proof.
 	\end{proof}
 
 	\subsubsection*{Proof of Lemma \ref{lem:tSigma1}}
 	\begin{proof}
 			Define
 		\begin{align*}
 			S_{ij,1}&\equiv 2d_{ij1}d_{ij2}K_{h_{n}}(\Delta R_{ij}'\gamma)\Delta W_{ij} \nu_{ij},\\
 			S_{ij,2}&\equiv 2d_{ij1}d_{ij2}K_{h_{n}}(\Delta R_{ij}'\gamma)\Delta W_{ij} \lambda_{ij},\\
 			S_{ij,3}&\equiv 2d_{ij1}d_{ij2}K_{h_{n}}(\Delta R_{ij}'\gamma)\Delta W_{ij}\Delta W_{ij}'(\beta-\hat{\beta}_{n}).
 		\end{align*}
 		Since 
 		\begin{align*}
 			\Delta\hat{\epsilon}_{ij}=\Delta W_{ij}'(\beta-\hat{\beta}_{n})+\lambda_{ij}+\nu_{ij},
 		\end{align*}
 		we have
 		\begin{align*}
 			S_{ij}=S_{ij,1}+S_{ij,2}+S_{ij,3}.
 		\end{align*}
 		Thus,
 		\begin{align*}
 			&\tilde{\Sigma}_{W\nu,1}\\
 			&=\underbrace{{n\choose 3}^{-1}\sum_{i<j<k}\underbrace{\frac{1}{3}(S_{ij,1}S_{ik,1}'+S_{ij,1}S_{jk,1}'+S_{ik,1}S_{jk,1})}_{D_{7,ijk}}}_{D_{7}}+\mathcal{O}_{7},
 		\end{align*}
 		where $\mathcal{O}_{7}$ is the remainder term.
 		
 		We first show that $c'D_{7}c\to_{p}c'\Sigma_{W\nu,1}c$. Note that
 		\begin{align*}
 			E[c'D_{7}c]&=E[c'S_{12,1}S_{13,1}c]\\
 			&=\frac{4}{h_{n}^{2}}\int E[d_{121}d_{122}d_{131}d_{132}c'\Delta W_{12}\Delta W_{13}'c\nu_{12}\nu_{13}|\Delta R_{12}'\gamma=s_{1},\Delta R_{13}'\gamma=s_{2}]\\
 			&\times K(s_{1}/h_{n})K(s_{2}/h_{n})f_{R\gamma,2}(s_{1},s_{2})ds_{1}ds_{2}\\
 			&=4\int E[d_{121}d_{122}d_{131}d_{132}c'\Delta W_{12}\Delta W_{13}'c\nu_{12}\nu_{13}|\Delta R_{12}'\gamma=s_{1}h_{n},\Delta R_{13}'\gamma=s_{2}h_{n}]\\
 			&\times K(s_{1})K(s_{2})f_{R\gamma,2}(s_{1}h_{n},s_{2}h_{n})ds_{1}ds_{2}\\
 			&\to c'\Sigma_{W\nu,1}c,
 		\end{align*}
 		as $n\to\infty$ by the dominated convergence theorem under Assumptions \ref{asm:index}, \ref{asm:moments}, and \ref{asm:kernel}. Define the third order U-statistics
 		\begin{align*}
 			U_{n,1} = {n\choose 3}^{-1}\sum_{i<j<k}p_{n}(\boldsymbol{\xi}_{i},\boldsymbol{\xi}_{j},\boldsymbol{\xi}_{k}),
 		\end{align*}
 		where $\boldsymbol{\xi}_{i}=(\xi_{i},U_{i})$ and
 		\begin{align*}
 			p_{n}(\boldsymbol{\xi}_{i},\boldsymbol{\xi}_{j},\boldsymbol{\xi}_{k})=E[c'D_{7,ijk}c|\boldsymbol{\xi}_{i},\boldsymbol{\xi}_{j},\boldsymbol{\xi}_{k}]
 		\end{align*}
 		By the calculation of \cite{Graham2019} in Appendix B, 
 		\begin{align*}
 			E[(c'D_{7}c-U_{n,1})^{2}]&= {n\choose 3}^{-1}E\left[(c'D_{7,123}c-E[c'D_{7,123}c|\boldsymbol{\xi}_{1},\boldsymbol{\xi}_{2},\boldsymbol{\xi}_{3}])^{2}\right]\\
 			&+{n\choose 3}^{-2}\times 3{n\choose 2}{n-2\choose 2}\times E\left[c'D_{7,123}c-E[c'D_{7,123}c|\boldsymbol{\xi}_{1},\boldsymbol{\xi}_{2},\boldsymbol{\xi}_{3}]\right]\\ &\times E\left[c'D_{7,124}c-E[c'D_{7,124}c|\boldsymbol{\xi}_{1},\boldsymbol{\xi}_{2},\boldsymbol{\xi}_{4}]\right]\\
 			&=O\left(\frac{E[(c'D_{7,123}c)^{2}]}{n^{3}}\right).
 		\end{align*}
 		Observe that
 		\begin{align*}
 			E[(c'D_{7,123}c)^{2}]&=\frac{1}{9}\left(3E[(c'S_{12,1}c\times c'S_{13,1}c)^{2}]+6E[(c'S_{12,1}c)^{2}\times c'S_{13,1}c\times c'S_{23,1}c]\right)\\
 			&=O\left(\frac{1}{h_{n}^{2}}\right),
 		\end{align*}
 		since for some positive constant $C>0$,
 		\begin{align*}
 			E[(c'S_{12,1}c\times c'S_{13,1}c)^{2}]&\leq\frac{C}{h_{n}^{4}}\int E[\|\Delta W_{12}\|^{2}\|\Delta W_{13}\|^{2}\nu_{12}^{2}\nu_{13}^{2}|\Delta R_{12}'\gamma=s_{1},\Delta R_{13}'\gamma=s_{2}]\\
 			&\times K^{2}(s_{1}/h_{n})K^{2}(s_{2}/h_{n})f_{R\gamma,2}(s_{1},s_{2})ds_{1}ds_{2}\\
 			&=\frac{C}{h_{n}}\int E[\|\Delta W_{12}\|^{2}\|\Delta W_{13}\|^{2}\nu_{12}^{2}\nu_{13}^{2}|\Delta R_{12}'\gamma=s_{1}h_{n},\Delta R_{13}'\gamma=s_{2}h_{n}]\\
 			&\times K^{2}(s_{1})K^{2}(s_{2})f_{R\gamma,2}(s_{1}h_{n},s_{2}h_{n})ds_{1}ds_{2}\\
 			&=O\left(\frac{1}{h_{n}^{2}}\right),
 		\end{align*}
 		as $n\to\infty$ with the last line coming from Assumption \ref{asm:index}, \ref{asm:moments}, and \ref{asm:kernel}, and,
 		\begin{align*}
 			&E[(c'S_{12,1}c)^{2}\times c'S_{13,1}c\times c'S_{23,1}c] \\
 			&=E[(c'S_{12,1}c)^{2}\times E[c'S_{13,1}c|\xi_{1},U_{1}]\times E[c'S_{23,1}c|\xi_{2},U_{2}]]\\
 			&=E[E[(c'S_{12,1}c)^{2}\times c'S_{13,1}c|\xi_{1},U_{1}]\times E[c'S_{23,1}|\xi_{2},U_{2}]]\\
 			&=E[(c'S_{12,1}c)^{2}\times c'S_{13,1}c]\times E[c'S_{12,1}c]\\
 			&\leq O(1)\times \frac{C}{h_{n}^{3}}\int E[\|\Delta W_{12}\|^{2}\nu_{12}^{2}|\Delta R_{12}'\gamma=s_{1},\xi_{1},U_{1}]
 			\times
 			E[\|\Delta W_{13}\||\nu_{13}||\Delta R_{13}'\gamma=s_{2},\xi_{1},U_{1}]\\
 			&\times K^{2}(s_{1}/h_{n})K(s_{2}/h_{n})f_{R\gamma|\xi_{1},U_{1}}(s_{1})
 			f_{R\gamma|\xi_{1},U_{1}}(s_{2})ds_{1}ds_{2}ds_{3}\\
 			&=O(1)\times \frac{C}{h_{n}}\int E[\|\Delta W_{12}\|^{2}\nu_{12}^{2}|\Delta R_{12}'\gamma=s_{1}h_{n},\xi_{1},U_{1}]
 			\times
 			E[\|\Delta W_{13}\||\nu_{13}||\Delta R_{13}'\gamma=s_{2}h_{n},\xi_{1},U_{1}]\\
 			&\times K^{2}(s_{1})K(s_{2})f_{R\gamma|\xi_{1},U_{1}}(s_{1}h_{n})
 			f_{R\gamma|\xi_{1},U_{1}}(s_{2}h_{n})ds_{1}ds_{2}ds_{3}\\
 			&=O\left(\frac{1}{h_{n}}\right),
 		\end{align*}
 		where the first to third lines follow from the conditional independence of $S_{ij,1}$, the random sampling of $\xi_{i}$, and the conditional independence and exchangeability of $U_{i}$ under Assumption \ref{asm:data}, and the last line follows from Assumptions \ref{asm:index}, \ref{asm:moments}, and \ref{asm:kernel}.
 		Observe that, by conditional independence of $S_{ij,1}$ and $S_{ik,1}$ given $\xi_{i},U_{i}$ and $S_{ij,1}=S_{ji,1}$, one can show that
 		\begin{align*}
 			E[c'D_{7,123}c\times c'D_{7,124}c]&=\frac{1}{9}\big\{2E[(c'S_{12,1}c)^{2}\times c'S_{13,1}c\times c'S_{14,1}c]\\
 			&+2E[(c'S_{12,1}c)^{2}\times c'S_{13,1}c]\times  E[c'S_{13,1}c]
 			+5E[c'S_{12,1}c\times c'S_{13,1}c]^{2}\big\}\\
 			&=\left(\frac{1}{h_{n}}\right),
 		\end{align*}
 		where the last line holds since 
 		\begin{align*}
 			&E[(c'S_{12,1}c)^{2}\times c'S_{13,1}c\times c'S_{14,1}c]\\
 			&\leq \frac{C}{h_{n}^{4}}E\bigg[\int E[\|\Delta W_{12}\|^{2}\nu_{12}^{2}|\Delta R_{12}'\gamma=s_{1},\xi_{1},U_{1}]\times
 			E[\|\Delta W_{12}\|\nu_{12}|\Delta R_{12}'\gamma=s_{2},\xi_{1},U_{1}]\\
 			&\times E[\|\Delta W_{14}\|\nu_{14}|\Delta R_{14}'\gamma=s_{3},\xi_{1},U_{1}]
 			\times K^{2}(s_{1}/h_{n})K(s_{2}/h_{n})K(s_{3}/h_{n})\\
 			&\times f_{R\gamma|\xi_{1},U_{1}}(s_{1})f_{R\gamma|\xi_{1},U_{1}}(s_{2})
 			f_{R\gamma|\xi_{1},U_{1}}(s_{3})ds_{1}ds_{2}ds_{3}\bigg]\\
 			&=\frac{C}{h_{n}}E\bigg[\int E[\|\Delta W_{12}\|^{2}\nu_{12}^{2}|\Delta R_{12}'\gamma=s_{1}h_{n},\xi_{1},U_{1}]\times
 			E[\|\Delta W_{12}\|\nu_{12}|\Delta R_{12}'\gamma=s_{2}h_{n},\xi_{1},U_{1}]\\
 			&\times E[\|\Delta W_{14}\|\nu_{14}|\Delta R_{14}'\gamma=s_{3}h_{n},\xi_{1},U_{1}]
 			\times K^{2}(s_{1})K(s_{2})K(s_{3})\\
 			&\times f_{R\gamma|\xi_{1},U_{1}}(s_{1}h_{n})f_{R\gamma|\xi_{1},U_{1}}(s_{2}h_{n})
 			f_{R\gamma|\xi_{1},U_{1}}(s_{3}h_{n})ds_{1}ds_{2}ds_{3}\bigg]\\
 			&=O\left(\frac{1}{h_{n}}\right),
 		\end{align*}
 		where the last line holds by Assumptions \ref{asm:index}, \ref{asm:moments}, \ref{asm:kernel}, 
 		\begin{align*}
 			&E[(c'S_{12,1}c)^{2}\times c'S_{13,1}c]\\
 			&\leq \frac{C}{h_{n}^{3}}E\bigg[\int E[\|\Delta W_{12}\|^{2}\nu_{12}^{2}|\Delta R_{12}'\gamma=s_{1},\xi_{1},U_{1}]\\
 			& \times E[\|\Delta W_{13}\||\nu_{13}| |\Delta R_{13}'\gamma=s_{2},\xi_{1},U_{1}]K^{2}(s_{1}/h_{n})K(s_{2}/h_{n})f_{R\gamma|\xi_{1},U_{1}}(s_{1})f_{R\gamma|\xi_{1},U_{1}}(s_{2})ds_{1}ds_{2}\bigg]\\
 			&\leq \frac{C}{h_{n}}E\bigg[\int E[\|\Delta W_{12}\|^{2}\nu_{12}^{2}|\Delta R_{12}'\gamma=s_{1}h_{n},\xi_{1},U_{1}]\\
 			&\times E[\|\Delta W_{13}\||\nu_{13}| |\Delta R_{13}'\gamma=s_{2}h_{n},\xi_{1},U_{1}]K^{2}(s_{1})K(s_{2})f_{R\gamma|\xi_{1},U_{1}}(s_{1}h_{n})f_{R\gamma|\xi_{1},U_{1}}(s_{2}h_{n})ds_{1}ds_{2}\bigg]\\
 			&=O\left(\frac{1}{h_{n}}\right),
 		\end{align*}
 		where the last equality holds from Assumptions \ref{asm:index}, \ref{asm:moments}, and \ref{asm:kernel}, and
 		\begin{align*}
 			&E[c'S_{12}c\times c'S_{13}c]\\
 			&\leq \frac{C}{h_{n}^{2}}E\bigg[\int E[\|\Delta W_{12}\||\nu_{12}||\Delta R_{12}'\gamma=s_{1},\xi_{1},U_{1}]\\
 			& \times E[\|\Delta W_{13}\||\nu_{13}| |\Delta R_{13}'\gamma=s_{2},\xi_{1},U_{1}]K(s_{1}/h_{n})K(s_{2}/h_{n})f_{R\gamma|\xi_{1},U_{1}}(s_{1})f_{R\gamma|\xi_{1},U_{1}}(s_{2})ds_{1}ds_{2}\bigg]\\
 			&\leq CE\bigg[\int E[\|\Delta W_{12}\||\nu_{12}||\Delta R_{12}'\gamma=s_{1}h_{n},\xi_{1},U_{1}]\\
 			&\times E[\|\Delta W_{13}\||\nu_{13}| |\Delta R_{13}'\gamma=s_{2}h_{n},\xi_{1},U_{1}]K(s_{1})K(s_{2})f_{R\gamma|\xi_{1},U_{1}}(s_{1}h_{n})f_{R\gamma|\xi_{1},U_{1}}(s_{2}h_{n})ds_{1}ds_{2}\bigg]\\
 			&=O(1),
 		\end{align*}
 		where the last equality holds from Assumptions \ref{asm:index}, \ref{asm:moments}, and \ref{asm:kernel}. 
 		Thus,
 		\begin{align*}
 			E\left[(c'D_{7}c-U_{n,1})^{2}\right]=O\left(\frac{1}{n^{3}h_{n}^{2}}\right)=o(1).
 		\end{align*}
 		Thus, $c'D_{1}$ is well approximated by $U_{n}$. Also, since $nh_{n}^{2}\to\infty$ with the stated assumption on $h_{n}$,
 		\begin{align*}
 			E\left[\left(p_{n}(\boldsymbol{\xi}_{i},\boldsymbol{\xi}_{j},\boldsymbol{\xi}_{k})\right)^{2}\right]=O\left( E[(c'D_{7,123}c)^{2}]\right)=O\left(\frac{n}{nh_{n}^{2}}\right)=o(1)\times O(n),
 		\end{align*} 
 		and by Lemma A.3 of \cite{Ahn1993}, we have
 		\begin{align*}
 			U_{n}=E[U_{n,1}]+o_{p}(1).
 		\end{align*}
 		This shows that
 		\begin{align*}
 			c'D_{7}c&= E[U_{n,1}] +\underbrace{c'D_{7}c-U_{n,1}}_{=o_{p}(1)} + \underbrace{U_{n,1}-E[U_{n,1}]}_{=o_{p}(1)}\\
 			&=E[c'D_{7}c] +o_{p}(1)\\
 			&=c'\Sigma_{W\nu,1}c +o_{p}(1).
 		\end{align*}
 		This completes $c'D_{7}c\to_{p}c'\Sigma_{W\nu,1}c$ as $n\to\infty$. 
 		
 		The remainder term $\mathcal{O}_{7}$ with each term involving either $S_{ij,2}$ or (and) $S_{ij,3}$ is of smaller order than $D_{7}$ since $S_{ij,2}$ and $S_{ij,3}$ involve $\|\hat{\beta}_{n}-\beta\|=O_{p}(1/\sqrt{n})$ and $\lambda_{ij}\sim h_{n}$ for large $n$; By computing in a similar way as before, we can establish that $E[|c'\mathcal{O}_{7}c|]=o(1)$ and $Var[c'\mathcal{O}_{7}c]=o(1)$ so that $|c'\mathcal{O}_{7}c|\to_{p}0$. Hence,
 		\begin{align*}
 			|c'\tilde{\Sigma}_{\nu,1}c- c'\Sigma_{W\nu,1}c|\leq |c'D_{1}c-c'\Sigma_{W\nu,1}c|+|c'\mathcal{O}_{7}c|=o_{p}(1),
 		\end{align*}
 		which completes the proof for Lemma \ref{lem:tSigma1}.
 	\end{proof}
 	\subsubsection*{Proof of Lemma \ref{lem:Sigma1er}}
 	\begin{proof}
 		Define
 		\begin{align*}
 			\hat{S}_{ij,1}&=\frac{2}{h_{n}^{2}}d_{ij1}d_{ij2}k\left(\frac{c_{ij,n}^{*}}{h_{n}}\right)\Delta W_{ij}\Delta R_{ij}'\Delta\hat{\epsilon}_{ij}
 		\end{align*}
 		where $c_{ij,n}^{*}$ is in between $\Delta R_{ij}'\gamma$ and $\Delta R_{ij}'\hat{\gamma}_{n}$. In the following argument, we treat $\Delta \hat{\epsilon}_{ij}$ in $\hat{S}_{ij,1}$ as $\nu_{ij}$ because only the existence of higher moments is important and bounding the terms involving $\nu_{ij}$ suffices. By the expression for $\Delta \hat{\epsilon}_{ij}$, we have
 		\begin{align*}
 			\hat{S}_{ij}&=S_{ij,1}+S_{ij,2}+S_{ij,3}+\hat{S}_{ij,1}(\hat{\gamma}_{n}-\gamma).
 		\end{align*}
 	    By the proof of \ref{lem:tSigma1}, we know that
 		\begin{align*}
 			{n\choose 3}^{-1}\sum_{i<j<k}\frac{1}{3}(S_{ij,p}S_{ik,p}'+S_{ij,p}S_{jk,p}'+S_{ik,p}S_{jk,p}')&=o_{p}(1),
 		\end{align*}
 		for $p=2,3$. Then,
 		\begin{align*}
 			\|\hat{\Sigma}_{W\nu,1}-\tilde{\Sigma}_{W\nu,1}\|&\leq \sum_{p=1}^{3}\underbrace{{n\choose 3}^{-1}\sum_{i<j<k}\underbrace{\frac{h_{n}^{2}}{3}(\|S_{ij,p}\|\|\hat{S}_{ik,1}\|+\|S_{ij,p}\|\|\hat{S}_{jk,1}\|+\|S_{ik,p}\|\|\hat{S}_{jk,1}\|)}_{D_{8,1,ijk}^{p}}}_{D_{8,1}^{p}}\frac{\|\hat{\gamma}_{n}-\gamma\|}{h_{n}^{2}}\\
 			&+\underbrace{\sum_{p=1}^{3}{n\choose 3}^{-1}\sum_{i<j<k}\frac{h_{n}^{2}}{3}(\|\hat{S}_{ij,1}\|\|S_{ik,p}\|+\|\hat{S}_{ij,1}\|\|S_{jk,p}\|+\|\hat{S}_{ik,1}\|\|S_{jk,p}\|)}_{D_{8,2}}\frac{\|\hat{\gamma}_{n}-\gamma\|}{h_{n}^{2}}\\
 			&+\underbrace{{n\choose 3}^{-1}\sum_{i<j<k}\frac{h_{n}^{4}}{3}(\|\hat{S}_{ij,1}\|\|\hat{S}_{ik,1}\|+\|\hat{S}_{ij,1}\|\|\hat{S}_{jk,1}\|+\|\hat{S}_{ik,1}\|\|\hat{S}_{jk,1}\|)}_{D_{8,3}}\frac{\|\hat{\gamma}_{n}-\gamma\|^{2}}{h_{n}^{4}}
 		\end{align*}
 		
 		For $D_{8,1}^{p}$, it suffices to bound $D_{8,1}^{1}$ as the similar calculation applies to the other terms. By Assumption \ref{asm:kernel}, for some constant $C>0$,
 		\begin{align*}
 			\|S_{ij,1}\|&\leq \frac{2}{h_{n}}\underbrace{\|\Delta W_{ij}\||\nu_{ij}||K(\Delta R_{ij}'\gamma/h_{n})|}_{g_{ij,1}},\\
 			\|\hat{S}_{ij,1}\|&\leq \frac{C}{h_{n}^{2}}\underbrace{\|\Delta W_{ij}\|\|\Delta R_{ij}\||\nu_{ij}|}_{g_{ij,2}}.
 		\end{align*}
 		Thus,
 		\begin{align*}
 			D_{8,1}^{1}\leq C{n\choose 3}^{-1}\sum_{i<j<k}\underbrace{(g_{ij,1}g_{ik,2}+g_{ij,1}g_{jk,2}+g_{ik,1}g_{jk,2})}_{g_{ijk,12}}.
 		\end{align*}
 		Observe that
 		\begin{align*}
 			E[D_{8,1}^{1}]&\leq \frac{1}{h_{n}}E[g_{12,1}g_{13,2}]\\
 			&=\frac{1}{h_{n}}E\Big[\int E[\|\Delta W_{12}\||\nu_{12}||\Delta R_{12}'\gamma=r,\xi_{1},U_{1}]K(r/h_{n})f_{R\gamma|\xi_{1},U_{1}}(r)dr\\
 			&\times E[\|\Delta W_{13}\|\|\Delta R_{13}\||\nu_{13}||\xi_{1},U_{1}]\Big]\\
 			&=E\Big[\int E[\|\Delta W_{12}\||\nu_{12}||\Delta R_{12}'\gamma=rh_{n},\xi_{1},U_{1}]K(r)f_{R\gamma|\xi_{1},U_{1}}(rh_{n})dr\\
 			&\times E[\|\Delta W_{13}\|\|\Delta R_{13}\||\nu_{13}||\xi_{1},U_{1}]\Big]\\
 			&=O(1),
 		\end{align*}
 		where the last equality follows from Assumptions \ref{asm:index}, \ref{asm:moments}, \ref{asm:unconmoments}, and \ref{asm:kernel}. For variance, the leading term involves covariances between variables with one common node, which has $n\times {n-1\choose 4}$ elements (up to some constant scale):
 		\begin{align*}
 			Var[D_{8,1}^{1}]=O\Bigg(\frac{1}{h_{n}^{2}}\times {n\choose 3}^{-2}\times  n\times {n-1\choose 4}\times E[D_{8,1,123}^{1}\times D_{8,1,145}^{1}]\Bigg)=O_{p}\left(\frac{1}{nh_{n}^{2}}\right)=o(1),
 		\end{align*}
 		since $nh_{n}^{2}\sim n^{(2k-1)/(2k+3)}$ diverges for $k\geq2 $ and
 		\begin{align*}
 			E[D_{8,1,123}^{1}\times D_{8,1,145}^{1}]&\leq E[g_{123,12}\times g_{145,12}]\\
 			&\leq E[g_{123,12}^{2}]\\
 			&\leq CE[\|\Delta W_{12}\|^{2}\|\Delta W_{13}\|^{2}\|\Delta R_{13}\|^{2}\nu_{12}^{2}\nu_{13}^{2}]\\
 			&=O(1),
 		\end{align*}
 		by Cauchy-Schwartz under Assumption \ref{asm:unconmoments}. Thus, $D_{8,1}=O_{p}(1)$. Similarly, $D_{8,2}=O_{p}(1)$ and $D_{8,3}=O_{p}(1)$ hold. 
 		
 		Notice that
 		\begin{align*}
 			\frac{\|\hat{\gamma}_{n}-\gamma\|}{h_{n}^{2}}=\frac{\sqrt{Nh_{n}}\|\hat{\gamma}_{n}-\gamma\|}{\sqrt{Nh_{n}^{5}}}=o_{p}(1),
 		\end{align*}
 		since $1/\sqrt{Nh_{n}^{5}}$ diverges by the hypothesis, and similarly,
 		\begin{align*}
 			\frac{\|\hat{\gamma}_{n}-\gamma\|^{2}}{h_{n}^{4}}=\frac{(\sqrt{Nh_{n}}\|\hat{\gamma}_{n}-\gamma\|)^{2}}{Nh_{n}^{5}}=o_{p}(1).
 		\end{align*}
 		Hence,
 		\begin{align*}
 			\|\hat{\Sigma}_{W\nu,1}-\tilde{\Sigma}_{W\nu,1}\|\leq O_{p}(1)\times o_{p}(1)+o_{p}(1)=o_{p}(1),
 		\end{align*}
 		which completes the proof for Lemma \ref{lem:Sigma1er}.
 	\end{proof}
 	\subsubsection*{Proof of Lemma \ref{lem:negligible_partial}}
 	\begin{proof}
 		We only show $nh_{n}c_{W}'\hat{\Sigma}=o_{p}(1)$ as the other case follows by taking transpose. We also write $c_{W}$ as $c$ for short. The statement is proved by showing that Lemmas \ref{lem:tSigma1} and \ref{lem:Sigma1er} hold even after re-scaled by $n^{1-\alpha/2}h_{n}$. First, we show that $c'\Sigma_{W\nu,1}c=0$ implies that $c'\Sigma_{W\nu,1}=0$.
 		\subsubsection*{Step 1: The implication of $c'\Sigma_{W\nu,1}c=0$}\mbox{}\\
 			\indent Note that
 			\begin{align*}
 				\Sigma_{W\nu,1}&=f_{R\gamma,2}(0,0)Pr(d_{121}d_{122}d_{131}d_{132}=1|\Delta R_{12}'\gamma=\Delta R_{13}'\gamma=0)\\
 				&\times E[\Delta W_{12}\Delta W_{13}'\nu_{12}\nu_{13}|\Delta R_{12}'\gamma=\Delta R_{13}'\gamma=0].
 			\end{align*}
 			By Assumption \ref{asm:index}, $f_{R\gamma,2}(0,0)>0$. Also, by the conditional independence of $d_{12t}$ and $d_{13s}$ under Assumption \ref{asm:data}, 
 			\begin{align*}
 			&Pr(d_{121}d_{122}d_{131}d_{132}=1|\Delta R_{12}'\gamma=\Delta R_{13}'\gamma=0) \\
 			&=E\left[\left(Pr(d_{122}d_{122}=1|\Delta R_{12}'\gamma=0,\xi_{1},U_{1})\right)^{2}\right|\Delta R_{12}'\gamma=0].
 			\end{align*}
 			Since $Pr(d_{121}d_{122}=1|\Delta R_{12}'\gamma=0)>0$ is implied by Assumption \ref{asm:inv}, it must be that
 			\begin{align*}
 				Pr(d_{121}d_{122}d_{131}d_{132}=1|\Delta R_{12}'\gamma=\Delta R_{13}'\gamma=0)>0,
 			\end{align*}
 			as otherwise $Pr(d_{121}d_{122}=1|\Delta R_{12}'\gamma=0,\xi_{1},U_{1})$ is constant at $0$, which contradicts with the locally positive probability of $d_{121}d_{122}=1$. Thus, $c'\Sigma_{W\nu,1}c=0$ is equivalent to
 			\begin{align*}
 				&E[\Delta c'W_{12}\Delta W_{13}'c\nu_{12}\nu_{13}|\Delta R_{12}'\gamma=\Delta R_{13}'\gamma=0] \\
 				&=E\left[\left(E[c'\Delta W_{12}\nu_{12}|\Delta R_{12}'\gamma=0,\xi_{1},U_{1}]\right)^{2}|\Delta R_{12}'\gamma=0\right]\\
 				&=0,
 			\end{align*}
 			which, in turn, is equivalent to (by the mean independence of $\nu_{12}$),
 			\begin{align*}
 				E[c'\Delta W_{12}\nu_{12}|\Delta R_{12}'\gamma=0, \xi_{1}, U_{1}] = 0,
 			\end{align*}
 			almost surely. Thus, $c'\Sigma_{W\nu,1}c=0$ implies that
 			\begin{align*}
 				c'\Sigma_{W\nu,1}&=f_{R\gamma,2}(0,0)Pr(d_{121}d_{122}d_{131}d_{132}=1|\Delta R_{12}'\gamma=\Delta R_{13}'\gamma=0)\\
 				&\times E\left[E[c'\Delta W_{12}\nu_{12}|\Delta R_{12}'\gamma=0,\xi_{1},U_{1}]E[\Delta W_{13}'\nu_{13}|\Delta R_{13}'\gamma,\xi_{1},U_{1}]|\Delta R_{12}'\gamma=\Delta R_{13}'\gamma=0\right]\\
 				&=0. 
 			\end{align*}
 		\subsubsection*{Step 2: $nh_{n}c'\tilde{\Sigma}_{W\nu,1}=o_{p}(1)$}\mbox{}\\
 			\indent Remember that 
 			\begin{align*}
 				\tilde{\Sigma}_{W\nu,1}=D_{7}+\mathcal{O}_{7}.
 			\end{align*}
 			For $D_{7}$, from the calculation in the proof of Lemma \ref{lem:S_Wnu}, we have that 
 			\begin{align*}
 				E[c'D_{7}] = E[c'S_{12,1}S_{13,1}']=O(h_{n}^{k}).
 			\end{align*}
 			, which shows $nh_{n}E[c'D_{7}]=O(nh_{n}^{k+1})=o(1)$ under the hypothesis.
 			For any non-zero vector $a\in\mathbb{R}^{q_{w}}$, redefining $U_{n}$ and $U_{n,1}$ with the kernel $p_{n}(\boldsymbol{\xi}_{i},\boldsymbol{\xi}_{j},\boldsymbol{\xi}_{k})=E[c'D_{7,ijk}a|\boldsymbol{\xi}_{i},\boldsymbol{\xi}_{j},\boldsymbol{\xi}_{k}]$, we can repeat the calculation in the proof of Lemma \ref{lem:tSigma1} to get
 			\begin{align*}
 				E[(nh_{n}c'D_{7}a-nh_{n}U_{n,1})^{2}]=n^{2}h_{n}^{2}\times O\left(\frac{1}{n^{3}h_{n}^{2}}\right) =o(1).
 			\end{align*}
 			Also, $E[nh_{n}p_{n}(\boldsymbol{\xi}_{i},\boldsymbol{\xi}_{j},\boldsymbol{\xi}_{k})]=E[nh_{n}c'D_{7}]=o(1)$ and
 			\begin{align*}
 				E[(nh_{n}p_{n}(\boldsymbol{\xi}_{i},\boldsymbol{\xi}_{j}, \boldsymbol{\xi}_{l}))^{2}]=O(n),
 			\end{align*}
 			so that by Lemma A.3 of \cite{Ahn1993},
 			\begin{align*}
 				nh_{n}U_{n}= nh_{n}E[U_{n,1}] +o_{p}(1)=o_{p}(1).
 			\end{align*}
 			This shows that, since $a$ is arbitrary,
 			\begin{align*}
 				nh_{n}c'D_{7}=o_{p}(1).
 			\end{align*}
 		
 			For the remainder term $\mathcal{O}_{7}$, this should again be of smaller order than $nh_{n}c'D_{7}$ since $\beta-\hat{\beta}_{n}=O_{p}(1/\sqrt{n})$ and $\lambda_{ij}=\Delta R_{ij}'\gamma \Lambda_{ij}$ is locally $O(h_{n}^{k+1})$ under the smoothing kernel and smoothness conditions on the density. For example, one of the elements in $\mathcal{O}_{7}$ is given by
 			\begin{align*}
 				&nh_{n}{n\choose 3}\sum_{i<j<k}\frac{1}{3}(S_{ij,2}S_{ik,2}'+S_{ij,2}S_{jk,2}'+S_{ik,2}S_{jk,2})\\
 				&\leq {n\choose 3}\sum_{i<j<k}\frac{4}{3h_{n}^{2}}\Bigg(|\Delta W_{ij}\|^{2}\|\Delta W_{ik}\|^{2}|K(\Delta R_{ij}'\gamma/h_{n})||K(\Delta R_{ik}'\gamma/h_{n})|\\
 				&+\|\Delta W_{ij}\|^{2}\|\Delta W_{jk}\|^{2}|K(\Delta R_{ij}'\gamma/h_{n})||K(\Delta R_{jk}/h_{n})|\\
 				&+\|\Delta W_{ik}'\|^{2}\|\Delta W_{jk}\|^{2}|K(\Delta R_{ik}/h_{n})||K(\Delta R_{jk}/h_{n})|\Bigg)nh_{n}\|\beta-\hat{\beta}_{n}\|^{2}\\
 				&=O_{p}(h_{n})=o_{p}(1),
 			\end{align*}  
 			where the last line can be shown by the same calculation as before to show $O_{p}(1)$ for the summation part and $\|\beta-\hat{\beta}_{n}\|^{2}=O_{p}(1/n)$ from Theorem \ref{thm:normality}. Similarly, we can show the negligibility of the elements in $\mathcal{O}_{7}$. This finishes the step 2. 
 			
 		\subsubsection*{Step 3: $n^{1-\alpha/2}h_{n}\|\hat{\Sigma}_{W\nu,1}-\tilde{\Sigma}_{W\nu,1}\|=o_{p}(1)$}\mbox{}\\
 		\indent By the proof of Lemma \ref{lem:Sigma1er}, we have
 		\begin{align*}
 			n^{1-\alpha/2}h_{n}\|\hat{\Sigma}_{W\nu,1}-\tilde{\Sigma}_{W\nu,1}\|\leq O_{p}(1)\Bigg(\frac{n^{1-\alpha/2}\|\hat{\gamma}_{n}-\gamma\|}{h_{n}}+\frac{n^{1-\alpha/2}\|\hat{\gamma}_{n}-\gamma\|^{2}}{h_{n}^{3}}\Bigg).
 		\end{align*}
 		Observe that
 		\begin{align*}
 			\frac{n^{1-\alpha/2}\|\hat{\gamma}_{n}-\gamma\|}{h_{n}}&=O\Bigg(\frac{\sqrt{Nh_{n}}\|\hat{\gamma}_{n}-\gamma\|}{\sqrt{n^{\alpha}h_{n}^{3}}}\Bigg)=o_{p}(1),\\
 			\frac{n^{1-\alpha/2}\|\hat{\gamma}_{n}-\gamma\|^{2}}{h_{n}^{3}}&=O\Bigg(\frac{\|\sqrt{Nh_{n}}(\hat{\gamma}_{n}-\gamma)\|^{2}}{\sqrt{n^{2+\alpha}h_{n}^{7}}}\Bigg)=o_{p}(1),
 		\end{align*}
 		by Assumption \ref{asm:first} and
 		\begin{align*}
 			n^{\alpha}h_{n}^{3}&\sim n^{(\alpha(2k+3)-6)/(2k+3)}\to\infty,\\
 			n^{2+\alpha}h_{n}^{7}&\sim n^{((2k+3)\alpha+4k-8)/(2k+3)}\to\infty,
 		\end{align*}
 		for $\alpha\in[6/(2k+3),1)$. Hence,
 		\begin{align*}
 			n^{1-\alpha/2}h_{n}\|\hat{\Sigma}_{W\nu,1}-\tilde{\Sigma}_{W\nu,1}\|=o_{p}(1).
 		\end{align*}
 		
 		Steps 1-3 complete the proof of Lemma \ref{lem:negligible_partial}.
 	\end{proof}
 
 	\subsubsection*{Proof of Lemma \ref{lem:negligible_full}}
 	\begin{proof}
 		Write $c_{W}$ as $c$ for short. It suffices to show that $nh_{n}\|c'\hat{\Sigma}_{W\nu,1}c-c'\tilde{\Sigma}_{W\nu,1}c\|=o_{p}(1)$ as we already show, in the proof of Lemma \ref{lem:negligible_partial}, that $nh_{n}c'\tilde{\Sigma}_{W\nu,1}=o_{p}(1)$, which implies $nh_{n}c'\tilde{\Sigma}_{W\nu,1}c=o_{p}(1)$. To save the space, in the following argument, we treat $\Delta \hat{\epsilon}_{ij}$ as $\nu_{ij}$; The other terms are similarly bounded using the properties of $\beta-\hat{\beta}_{n}$ and $\lambda_{ij}$.
 		
 		Re-define
 		\begin{align*}
 			S_{ij}&=\frac{2}{h_{n}}d_{ij1}d_{ij2}K\left(\frac{\Delta R_{ij}'\gamma}{h_{n}}\right)c'\Delta W_{ij}\nu_{ij}\\
 			\hat{S}_{ij}&=\frac{2}{h_{n}^{2}}d_{ij1}d_{ij2}k\left(\frac{\Delta R_{ij}'\gamma}{h_{n}}\right)c'\Delta W_{ij}\Delta R_{ij}'\nu_{ij}\\
 			\hat{S}_{ij,2}&=\frac{1}{h_{n}^{3}}d_{ij1}d_{ij2}k\left(\frac{c_{ij,n}^{*}}{h_{n}}\right)\Delta R_{ij}c'\Delta W_{ij}\Delta R_{ij}'\nu_{ij},
 		\end{align*}
 		where $c_{ij,n}^{*}$ is in between $\Delta R_{ij}'\gamma$ and $\Delta R_{ij}'\hat{\gamma}_{n}$. We have that
 		\begin{align*}
 			\hat{S}_{ij}=S_{ij}+\hat{S}_{ij,1}(\hat{\gamma}_{n}-\gamma)+(\hat{\gamma}_{n}-\gamma)'\hat{S}_{ij,2}(\hat{\gamma}_{n}-\gamma).
 		\end{align*}
 		Note that, for some constant $C>0$, $\|\hat{S}_{ij,1}\|\leq Ch_{n}^{-2} g_{ij,2}$ as before and
 		\begin{align*}
 			\|\hat{S}_{ij,2}\|\leq\frac{C}{h_{n}^{3}}\underbrace{\|\Delta W_{ij}\|\|\Delta R_{ij}\|^{2}|\nu_{ij}|}_{g_{ij,3}}.
 		\end{align*}
 		
 		Observe that 
	 	\begin{align*}
	 		&c'\hat{\Sigma}_{W\nu,1}c-c'\tilde{\Sigma}_{W\nu,1}c\\
			&\leq \frac{(\hat{\gamma}_{n}-\gamma)'}{\sqrt{h_{n}}}\underbrace{{n \choose 3}^{-1}\sum_{i<j<k}\underbrace{\frac{\sqrt{h_{n}}}{3}(S_{ij}\hat{S}_{ik,1}'+S_{ij}\hat{S}_{jk,1}'+S_{ik}\hat{S}_{jk,1}')}_{D_{10,1,ijk}}}_{D_{9,1}}\\
		    &+\underbrace{{n \choose 3}^{-1}\sum_{i<j<k}\frac{\sqrt{h_{n}}}{3}(\hat{S}_{ij,1}S_{ik}+\hat{S}_{ij,1}S_{jk}+\hat{S}_{ik,1}S_{jk})}_{D_{10,2}}\frac{(\hat{\gamma}_{n}-\gamma)}{\sqrt{h_{n}}}\\
			&+\frac{(\hat{\gamma}_{n}-\gamma)'}{h_{n}^{3/2}}\underbrace{{n \choose 3}^{-1}\sum_{i<j<k}\frac{h_{n}^{3}}{3}(S_{ij}\hat{S}_{ik,2}'+S_{ij}\hat{S}_{jk,2}'+S_{ik}\hat{S}_{jk,2}')}_{D_{10,3}}\frac{(\hat{\gamma}_{n}-\gamma)}{h_{n}^{3/2}}\\
			&+\frac{(\hat{\gamma}_{n}-\gamma)'}{h_{n}^{3/2}}\underbrace{{n \choose 3}^{-1}\sum_{i<j<k}\frac{h_{n}^{3}}{3}(\hat{S}_{ij,2}S_{ik}+\hat{S}_{ij,2}S_{jk}+\hat{S}_{ik,2}S_{jk})}_{D_{9,4}}\frac{(\hat{\gamma}_{n}-\gamma)}{h_{n}^{3/2}}\\
			&+C^{2}\underbrace{{n \choose 3}^{-1}\sum_{i<j<k}\frac{1}{3}(g_{ij,2}g_{ik,3}+g_{ij,2}g_{jk,3}+g_{ik,2}g_{jk,3})}_{D_{10,5}}\frac{\|\hat{\gamma}_{n}-\gamma\|^{3}}{h_{n}^{5}}\\
			&+C^{2}\underbrace{{n \choose 3}^{-1}\sum_{i<j<k}\frac{1}{3}(g_{ij,3}g_{ik,2}+g_{ij,3}g_{jk,2}+g_{jk,3}g_{ik,2})}_{D_{10,6}}\frac{\|\hat{\gamma}_{n}-\gamma\|^{3}}{h_{n}^{5}}\\
			&+\underbrace{{n \choose 3}^{-1}\sum_{i<j<k}\frac{h_{n}^{2}}{3}(\|\hat{S}_{ij,1}\|\|\hat{S}_{ik,1}\|+\|\hat{S}_{ij,1}\|\|\hat{S}_{jk,1}\|+\|\hat{S}_{jk,1}\|\|\hat{S}_{ik,1}\|)}_{D_{10,7}}\frac{\|\hat{\gamma}_{n}-\gamma\|^{2}}{h_{n}^{2}}\\
			&+C^{2}\underbrace{{n \choose 3}^{-1}\sum_{i<j<k}\frac{1}{3}(g_{ij,3}g_{ik,3}+g_{ij,3}g_{jk,3}+g_{jk,3}g_{ik,3})}_{D_{10,8}}\frac{\|\hat{\gamma}_{n}-\gamma\|^{4}}{h_{n}^{6}}.
	 	\end{align*}
	 	
	 	First we stochastically bound $D_{10,1}$ and $D_{10,2}$. For any vector $a\in\mathbb{R}^{q_{r}}$ and some constant $C>0$,
	 	\begin{align*}
	 		E[a'D_{10,1}]&=\frac{\sqrt{h_{n}}}{h_{n}^{3}}E\left[E[c'S_{12}|\xi_{1},U_{1}]E[a'\hat{S}_{13}c|\xi_{1},U_{1}]\right]\\
		 	&\leq \frac{1}{h_{n}^{3/2}}\left\{E\left[\int 	E[d_{121}d_{122}c'\Delta W \nu_{12}|\Delta R_{12}=s_{1}h_{n},\xi_{1},U_{1}]^{2}K(s_{1})f_{R\gamma|\xi_{1},U_{1}}(s_{1}h_{n})\right]\right\}^{1/2}\\
		 	&\times CE[\|\Delta W_{12}\|^{2}\|\Delta 	R_{13}\|^{2}\nu_{13}^{2}]^{1/2}\\
		 	&=O(h_{n}^{(2k-1)/2})=o(1),
	 	\end{align*}
	 	where the first line hold from the conditional independence, the inequality follows from Cauchy-Schwartz and Assumption \ref{asm:kernel}, and the final line holds from the implication from $c'\Sigma_{W\nu,1}c=0$ and Assumptions \ref{asm:index}, \ref{asm:moments}, \ref{asm:unconmoments}, and \ref{asm:kernel}. Repeating the calculation in the proof for Lemma \ref{lem:tSigma1} and adjusting for covariances with one index in common,
	 	\begin{align*}
	 		Var[a'D_{10,1}]=O\left(\frac{1}{n^{2}h_{n}^{3}}\right)+O\left(\frac{h_{n}E[S_{12}a'\hat{S}_{13,1}'S_{14}a'\hat{S}_{15,1}']}{n}\right)=o(1),
	 	\end{align*}
	 	where the last equality holds from the stated assumption on $h_{n}$ for $k\geq 1$ and
	 	\begin{align*}
	 		&E[S_{12}a'\hat{S}_{13,1}'S_{14}a'\hat{S}_{15,1}']\\
	 		&=\frac{16}{h_{n}^{2}}E\bigg[E[d_{121}d_{122}c'\Delta W_{12}\nu_{12}|\Delta R_{12}'\gamma=h_{n}s_{1},\boldsymbol{\xi}_{1}]\\
	 		&\times E[d_{131}d_{132}c'\Delta W_{13}a '\Delta R_{13}\nu_{13}|\Delta R_{13}'\gamma=h_{n}s_{2},\boldsymbol{\xi}_{1}]\\
	 		&\times E[d_{141}d_{142}c'\Delta W_{14}\nu_{14}|\Delta R_{14}'\gamma=h_{n}s_{3},\boldsymbol{\xi}_{1}]\\
	 		&\times E[d_{151}d_{152}c'\Delta W_{15}a '\Delta R_{15}\nu_{15}|\Delta R_{15}'\gamma=h_{n}s_{4},\boldsymbol{\xi}_{1}]\\
	 		&\times K(s_{1})k(s_{2})K(s_{3})k(s_{4})\Pi_{i=1}^{4}f_{R\gamma|\xi_{1},U_{1}}(s_{i}h_{n})\bigg]\\
	 		&=O\left(\frac{1}{h_{n}^{2}}\right)
	 	\end{align*}
	 	by Assumptions \ref{asm:index}, \ref{asm:moments}, and \ref{asm:kernel}. Thus, $D_{1,1}=o_{p}(1)$. Similarly, we have $D_{1,2}=o_{p}(1)$. 
	 	
	 	Next, we stochastically bound $D_{10,3}$ and $D_{9,4}$. For any finite $a,b\in\mathbb{R}^{q_{r}}$, for some $C>0$, 
	 	\begin{align*}
	 		&E[a'D_{10,3}b]\\
	 		&\leq\frac{C}{h_{n}}E[|S_{12}|g_{13,3}]\\
	 		&=CE\bigg[\int E[d_{121}d_{122}|c'\Delta W_{12}||\nu_{12}||\Delta R_{12}'\gamma=h_{n}s_{1},\boldsymbol{\xi}_{1}]\\
	 		&\times E[g_{13}|\boldsymbol{\xi}_{1}]\\
	 		&\times K(s_{1})f_{R\gamma|\xi_{1},U_{1}}(s_{1}h_{1})f_{R\gamma|\xi_{1},U_{1}}(s_{2})\bigg]\\
	 		&=O(1),
	 	\end{align*}
	 	where the last equality holds from Assumptions \ref{asm:index}, \ref{asm:moments}, \ref{asm:unconmoments} and \ref{asm:kernel}. The variance is calculated similarly as before:
	 	\begin{align*}
	 		Var[a'D_{10,3}b]=O\left(\frac{1}{n^{2}h_{n}}\right)+O\left(\frac{h_{n}}{n}\right)=o(1).
	 	\end{align*}
	 	Thus, $D_{10,3}=O_{p}(1)$. Similarly, $D_{10,4}=O_{p}(1)$.
	 	
	 	$D_{10,5}, D_{10,6}$, and $D_{10,8}$ are all $O_{p}(1)$ by the similar computation as in Lemma \ref{lem:tSigma1}.
	 	
	 	$D_{9.7}$ is stochastically bounded as follows. Observe that 
	 	\begin{align*}
	 		E[D_{10,7}] &= \frac{4}{h_{n}^{2}}E[\|\hat{S}_{12,1}\|\|\hat{S}_{13,1}\|]\\
	 		&\leq 4\int E[|c'\Delta W_{12}|c'\Delta W_{12}|\|\Delta R_{12}\|\Delta R_{13}\||\nu_{12}||\nu_{13}||\Delta R_{12}'\gamma=h_{n}s_{1},\Delta R_{13}'\gamma=h_{n}s_{2}]\\
	 		&\times k(s_{1})k(s_{2})f_{R\gamma,2}(h_{n}s_{1},h_{n}s_{2})\\
	 		&=O(1),
	 	\end{align*}
	 	where the last line holds from Assumption \ref{asm:index}, \ref{asm:moments}, and \ref{asm:kernel}. The variance is calculated similarly as before:
	 	\begin{align*}
	 		Var[D_{10,7}]=O\left(\frac{1}{n^{2}h_{n}^{2}}\right)+O\left(\frac{1}{n}\right)=o(1).
	 	\end{align*}
	 	Thus, $D_{10,7}=O_{p}(1)$.
	 	
	 	Finally, the above implies
	 	\begin{align*}
	 		&nh_{n}|c'\hat{\Sigma}_{W\nu,1}c-c'\tilde{\Sigma}_{W\nu,1}c|\\
	 		&\leq O_{p}(1)\times \left(\frac{nh_{n}\|\hat{\gamma}_{n}-\gamma\|}{h_{n}} + \frac{nh_{n}\|\hat{\gamma}_{n}-\gamma\|^{2}}{h_{n}^{3}}+\frac{nh_{n}\|\hat{\gamma}_{n}-\gamma\|^{3}}{h_{n}^{5}}+\frac{nh_{n}\|\hat{\gamma}_{n}-\gamma\|^{4}}{h_{n}^{6}}\right)
	 		&=o_{p}(1),
	 	\end{align*}
	 	because
	 	\begin{align*}
	 		\frac{nh_{n}\|\hat{\gamma}-\gamma\|}{h_{n}}&=O(\sqrt{Nh_{n}}\|\hat{\gamma}_{n}-\gamma\|)=o_{p}(1),\\
	 		\frac{nh_{n}\|\hat{\gamma}_{n}-\gamma\|^{2}}{h_{n}^{3}}&=O\left(\frac{(\sqrt{Nh_{n}}\|\hat{\gamma}_{n}-\gamma\|)^{2}}{nh_{n}^{3}}\right)=o_{p}(1),\\
	 		\frac{nh_{n}\|\hat{\gamma}_{n}-\gamma\|^{3}}{h_{n}^{5}}&=O\left(\frac{(\sqrt{Nh_{n}}\|\hat{\gamma}_{n}-\gamma\|)^{3}}{n^{2}h_{n}^{11/2}}\right)=o_{p}(1),\\
	 		\frac{nh_{n}\|\hat{\gamma}_{n}-\gamma\|^{4}}{h_{n}^{6}}&=O\left(\frac{(\sqrt{Nh_{n}}\|\hat{\gamma}_{n}-\gamma\|)^{4}}{n^{3}h_{n}^{7}}\right)=o_{p}(1),
	 	\end{align*}
	 	by Assumption \ref{asm:first}, $nh_{n}^{3}=O(n^{(2k-3)/(2k+3)})$, $n^{2}h_{n}^{11/2}=O(n^{(4k-5)/(2k+3)})$, and $n^{3}h_{n}^{7}=O(n^{(6k-5)})$ all diverging for $k\geq 2$. This completes the proof for Lemma \ref{lem:negligible_full}.
 	\end{proof}
\end{document}